\documentclass[
	reprint,
	superscriptaddress,
	showkeys,
	aps,
	pra,
	longbibliography,
	nofootinbib
]{revtex4-2}

\usepackage[utf8]{inputenc}

\usepackage[hyperref]{xcolor}
	\definecolor{goethe-blau}{cmyk}{1.0,0.2,0.0,0.4}
	\definecolor{hellgrau}{cmyk}{0.04,0.04,0.05,0.02}
	\definecolor{sandgrau}{cmyk}{0.12,0.09,0.13,0.0}
	\definecolor{dunkelgrau}{cmyk}{0.25,0.25,0.30,0.75}
	\definecolor{emo-rot}{cmyk}{0.04,1.0,0.8,0.07}
	\definecolor{purple}{cmyk}{0.08,1.0,0.3,0.36}
	\definecolor{senfgelb}{cmyk}{0.01,0.25,1.0,0.05}
	\definecolor{gruen}{cmyk}{0.62,0.4,0.87,0.09}
	\definecolor{magenta}{cmyk}{0.08,0.86,0.12,0.12}
	\definecolor{orange}{cmyk}{0.0,0.7,1.0,0.04}
	\definecolor{sonnengelb}{cmyk}{0.0,0.12,0.95,0.0}
	\definecolor{helles-gruen}{cmyk}{0.4,0.17,0.81,0.07}
	\definecolor{lichtblau}{cmyk}{0.8,0.0,0.06,0.04}

\usepackage[
	colorlinks,
	pdfpagelabels,
	breaklinks,
	pdfstartview=FitH,
	bookmarksopen=true,
	bookmarksnumbered=true,
	bookmarksopenlevel=2,
	plainpages=false,
	hypertexnames=false,
	citecolor=emo-rot,
	linkcolor=goethe-blau,
	urlcolor=purple,
	pdftitle={{Numerical fluid dynamics for FRG flow equations: Zero-dimensional QFTs as numerical test cases. III. Shock and rarefaction waves in RG flows reveal limitations of the N infinity limit in O(N)-type models}},
	pdfauthor={{Martin J. Steil and Adrian Koenigstein}},
	pdfkeywords={Functional Renormalization Group, numerical fluid dynamics, large-N expansion, saddle-point expansion, shocks, rarefaction waves, entropy}
]{hyperref}

\usepackage{amsmath} 
\usepackage{amssymb} 
\usepackage{bm} 

\usepackage{bookmark} 
\usepackage{graphicx} 
\usepackage{placeins} 

\usepackage{booktabs} 

\allowdisplaybreaks 

\begin{document}


\title{
	Numerical fluid dynamics for FRG flow equations:\texorpdfstring{\\}{ }Zero-dimensional QFTs as numerical test cases.\texorpdfstring{\\}{ }III. Shock and rarefaction waves in RG flows reveal limitations\texorpdfstring{\\}{ }of the \texorpdfstring{$N \rightarrow \infty$}{N infinity} limit in \texorpdfstring{$O(N)$}{O(N)}-type models
}

\author{Martin J. Steil}
	\email{msteil@theorie.ikp.physik.tu-darmstadt.de}
	\affiliation{
		Technische Universit\"at Darmstadt, Department of Physics, Institut f\"ur Kernphysik, Theoriezentrum,\\
		Schlossgartenstra{\ss}e 2, D-64289 Darmstadt, Germany
	}

\author{Adrian Koenigstein}
	\email{koenigstein@th.physik.uni-frankfurt.de}	
	\affiliation{
		Institut f\"ur Theoretische Physik, Goethe University,\\
		Max-von-Laue-Stra{\ss}e 1, D-60438 Frankfurt am Main, Germany
	}

\date{\today}

\begin{abstract}
	Using an $O(N)$-symmetric toy model QFT in zero space-time dimensions we discuss several aspects and limitations of the $\frac{1}{N}$-expansion. We demonstrate, how slight modifications in a classical UV action can lead the $\frac{1}{N}$-expansion astray and how the infinite-$N$ limit may alter fundamental properties of a QFT. Thereby we present the problem of calculating correlation functions from two totally different perspectives: First, we explicitly analyze our model within an $\frac{1}{N}$-saddle-point expansion and show its limitations. Secondly, we picture the same problem within the framework of the Functional Renormalization Group. Applying novel analogies between (F)RG flow equations and numerical fluid dynamics from parts I and II of this series of publications, we recast the calculation of expectation values of our toy model into solving a highly non-linear but exact advection(-diffusion) equation. In doing so, we find that the applicability of the $\frac{1}{N}$-expansion to our toy model is linked to freezing shock waves in field space in the FRG-fluid dynamic picture, while the failure of the $\frac{1}{N}$-expansion in this context is related to the annihilation of two opposing shock waves in field space.
\end{abstract}

\maketitle

\tableofcontents

\section{Introduction}
\label{chap:introduction}

	In all kinds of applications of (quantum-)statistical physics, like particle or solid state physics, strongly interacting systems uncovered fundamental shortcomings of perturbative methods for the calculation of expectation values (observables) from path integrals or partition functions, which are usually not exactly solvable \cite{Weinberg:1996kr,Peskin:1995ev,Strocchi:2013awa}. Consequently, various alternative non-perturbative approaches were developed within the last decades. Amongst others, these comprise ``brute force'' lattice Monte-Carlo simulations \cite{Philipsen:2012nu,Ding:2015ona,Guenther:2017grd}, Complex-Langevin equations \cite{Attanasio:2020spv,Berger:2019odf}, Dyson-Schwinger equations \cite{Fischer:2018sdj,Dyson:1949ha,Schwinger:1951ex,Schwinger:1951hq}, the Functional Renormalization Group \cite{Berges:2000ew,Pawlowski:2005xe,Dupuis:2020fhh}, holographic methods \cite{Maldacena:1997re,Witten:1998qj} \textit{etc}..\\
	
	Another rather old ``non-perturbative'' approach is the so-called $\frac{1}{N}$-expansion -- sometimes also denoted as large-$N$ expansion, the 't~Hooft limit, or in some contexts the mean-field approximation. This method relies on a systematic expansion of characteristic quantities of the theory, like expectation values, correlation functions, and observables, in powers of $\frac{1}{N}$. Here, $N$ is the number of different kinds of interacting degrees of freedom of the theory (particle or field types, spins, molecules, color charges \textit{etc.}), which is considered to be large in this context ($1 \lll N$). Hence, extensive quantities need to be rescaled by appropriate powers of $N$ in advance to allow for a meaningful $\frac{1}{N}$-expansion. Although involving an expansion in a small, dimensionless parameter, namely $\frac{1}{N}$, the method is considered to be non-perturbative, because it is also applicable to systems of strong interactions, where an expansion in couplings is doomed to fail. In consequence, various great successes and precise predictions trace back to this method, see, \textit{e.g.},\ Refs.~\cite{tHooft:1973alw,Veneziano:1979ec,Witten:1979vv,Witten:1979kh,Gross:1974jv,Maldacena:1997re,DAttanasio:1997yph,Keitel:2011pn,Grossi:2019urj} or the review \cite{Moshe:2003xn} -- in some cases maintaining predictive power even for systems, where $N$ is surprisingly small. However, also the large-$N$ expansion and especially retaining only its zeroth order contribution -- the infinite-$N$ limit -- comes with some limitations and certain fundamental characteristics of a (quantum) field theoretical or statistical model, like the convexity of the $\tfrac{1}{N}$-rescaled effective action, may be altered.\\

	In order to elucidate some of these aspects and interesting consequences, we study the large-$N$ expansion and the infinite-$N$ limit within two totally different setups. On the one hand, we perform a conventional saddle-point expansion of the path integral (partition function) by assuming that $N$ is large (or even infinite) \cite{Arfken:2005,Keitel:2011pn}. On the other hand, we study the same problem within the Functional Renormalization Group approach, also considering large and/or (in)finite $N$, \textit{cf.}\ Refs.~\cite{Tetradis:1995br,Litim:1995ex,DAttanasio:1997yph,Litim:2002cf,Fejos:2012rz,Litim:2016hlb,Yabunaka:2017uox,Yabunaka:2018mju,Yabunaka:2021fow,Grossi:2019urj,Grossi:2021ksl} for the infinite-$N$ limit in the FRG framework.
	
	To keep our discussion as simple as possible we work with a sober and exactly solvable toy model: the zero-dimensional $O(N)$~model -- an ultra local and strongly coupled QFT in a single space-time point\footnote{The zero-dimensional $O(N)$~model is also referred to as $O(N)$-vector model and can be seen as the high-temperature limit of a quantum mechanical system \cite{Moroz:2011thesis}. It was also considered as a statistical model for the formation of polymers \cite{Nishigaki:1990sk}.}. The property of being exactly solvable (in terms of conventional one-dimensional integrals) promotes the model to a perfect testing ground for methods of (quantum\nobreakdash-)statistical physics and we are by far not the first scientists, who are using this toy model for this purpose, \textit{cf.} Refs.~\cite{Fl_rchinger_2010,Moroz:2011thesis,Keitel:2011pn,Strocchi:2013awa,Kemler:2013yka,Pawlowski:talk,Rentrop_2015,Rosa:2016czs,Liang:2017whg,SkinnerScript,Millington:2019nkw,Alexander:2019cgw,Catalano:2019,Millington:2020Talk,Millington:2021ftp,Kades:2021hir}. Even a lot of aspects of the large-/infinite-$N$ limit have been discussed already within this setup, \textit{cf.} Refs.~\cite{Hikami:1978ya,Bessis:1980ss,DiVecchia:1990ce,Nishigaki:1990sk,Schelstraete:1994sc,Zinn-Justin:1998hwu,Keitel:2011pn} -- especially for quartic actions (potentials).\\

	Within this work, we use the zero-dimensional $O(N)$~model to highlight the following aspects:
		\begin{enumerate}
			\item	Considering a rather simple -- but non-analytic -- one-parameter family of classical actions (potentials) we demonstrate that there is a narrow line between a straightforward applicability of the large-$N$ saddle-point expansion and a total failure of this method. In our zero-dimensional pedagogical and tailor made example, this point of failure is easy to detect. However, it may serve as a warning for applications of the large-$N$ limit and the corresponding saddle-point expansion of the path integral in higher-dimensional scenarios, where it is not necessarily easy to judge, if all requirements for a meaningful $\tfrac{1}{N}$-expansion are fulfilled.
			
			\item	Switching perspectives to the FRG formalism, we make use of the fact that the corresponding RG flow equation is exact for the zero-dimensional $O(N)$~model. Being ``exact'' in this context means that truncating the flow equation is not necessary (for finite and infinite $N$) since the partial differential equation (PDE) for the RG flow can be solved numerically. To do so, we apply the novel fluid dynamic reformulation of this RG flow equation in terms of a highly non-linear advection-diffusion equation, which was demonstrated and discussed in parts I and II of this series of publications \cite{Koenigstein:2021syz,Koenigstein:2021rxj} and Refs.~\cite{Grossi:2019urj,Grossi:2021ksl,Stoll:2021ori}. Within this fluid dynamic framework, we show that the RG flow in the infinite-$N$ limit is purely advection driven, while diffusive contributions enter only at finite $N$. This also generalizes to higher space-time dimensions.
			
			As a direct consequence, depending on the classical action (potential) -- UV initial condition, the infinite-$N$-RG flows tend to form or sustain non-analyticities of different kinds, \textit{e.g.}, shock and rarefaction waves or jump discontinuities, \textit{cf.}\ Ref.~\cite{Grossi:2019urj,Grossi:2021ksl}. 
				
			We find that for our toy model with non-analytic classical action, shock and rarefaction waves are present and the problem presents at the UV initial scale involves two Riemann problems. But we do not stop by turning the calculation of ordinary $N$-dimensional integrals with spherical symmetry for expectation values into a fluid dynamical problem. We also demonstrate that the (non\nobreakdash-)applicability of the large-$N$ saddle point expansion translates into the (collision) freezing of interacting shock waves in these fluid dynamic RG flows.
				
			\item	Still working in the FRG-fluid dynamic framework, we also demonstrate that the inclusion of the radial $\sigma$-mode, thus switching from infinite-$N$ to arbitrary but finite $N$, totally changes the physics of the system. In the infinite-$N$ limit, we explicitly show by numerical calculations that convexity and smoothness are not necessarily realized for $\tfrac{1}{N}$-rescaled IR potentials, which effectively violates the Coleman-Mermin-Wagner-Hohenberg theorem \cite{Coleman:1973ci,Mermin:1966,Hohenberg:1967}, respectively a special zero-dimensional version of the theorem \cite{Moroz:2011thesis,Koenigstein:2021syz}. Interestingly, as soon as $N$ is finite, the highly non-linear diffusive contribution of the radial $\sigma$ mode unavoidably restores convexity and smoothness of the $\tfrac{1}{N}$-rescaled IR potentials. Hence, the large-$N$ expansion with finite $N$ and the infinite-$N$ limit (only retaining the zeroth order of the $\tfrac{1}{N}$-expansion) may lead to two fundamentally different results. A qualitative similar result is found in a parallel work \cite{Stoll:2021ori} by the authors and their collaborators in the context of the Gross-Neveu model  \cite{Gross:1974jv} in 1+1 space-time dimension.
			
			\item	As a last aspect, we also highlight further direct consequences of our fluid dynamic interpretation of RG flows. Utilizing the method of characteristics \cite{polyanin2016handbook,LeVeque:2002,Delgado2006Aug} and the Rankine-Hugoniot condition \cite{Rankine:1870,Hugoniot:1887}, we directly trace the locations of shock and rarefaction waves in the field space derivative of the scale-dependent potential during the RG flows. Applications of the aforementioned methods in (F)RG studies can be found in, \textit{e.g.}, Refs.~\cite{Tetradis:1995br,Litim:1995ex,Aoki:2014,Aoki:2017rjl,Grossi:2019urj}.
			
			Interacting shock and rarefaction waves, but also diffusive processes go hand in hand with the rise of entropy in fluid dynamic problems -- as it is well-known from our everyday life. Remarkably, this is also observed in our RG flows, where we were able to identify a numerical entropy function. This entropy production manifests the irreversibility of RG flows and the corresponding semi-group property of RG transformations.
		\end{enumerate}
	
	At this point, we remark that our work was partially influenced by the excellent publication \cite{Grossi:2019urj} on the infinite-$N$ limit of the FRG flow equations of the $O(N)$~model in three Euclidean space-time dimensions and the interpretation of these RG flows as advection equations, which can develop different kind of discontinuities \cite{Grossi:2019urj}. The application of the method of characteristics in the large-$N$ limit of FRG flow equation predates the explicit identification and detailed understanding of infinite-$N$ FRG flow equations as advection equations and goes back (to the best of our knowledge) to Refs.~\cite{Tetradis:1995br,Litim:1995ex}. The authors of Ref.~\cite{Grossi:2019urj}, E.~Grossi and N.~Wink, were also involved in the first two parts of this series of publications \cite{Koenigstein:2021syz,Koenigstein:2021rxj} and also worked together with F.~Ihssen and J.~M.~Pawlowski on calculations in the quark meson model in the infinite-$N$ limit \cite{Grossi:2021ksl,Ihssen2020}, which was also based on a fluid dynamic interpretation of RG flows.

	In addition, we thank the referee for pointing out that there are recent works, which also deal with the shortcomings of the standard infinite-$N$ limit in the context of FRG and link this to non-analytic structures in the fixed-point potential \cite{Yabunaka:2018mju,Yabunaka:2021fow}.
	An interesting future prospect is certainly to draw connection between our works in the fluid-dynamic framework and these results.

\section{The zero dimensional \texorpdfstring{$O(N)$}{O(N)}-model}
	In this work we consider a purely bosonic, zero-dimensional QFT consisting of $N$ real scalar ``fields'' ${\vec{\phi} = ( \phi_1 , \, \phi_2 , \, \ldots, \, \phi_N )}$, which transform according to 
		\begin{align}
			\vec{\phi} \mapsto \vec{\phi}^{\, \prime} = O \, \vec{\phi} \, ,
		\end{align}
	where $O \in O(N)$. Due to the absence of space and time dimensions, the fields $\vec{\phi}$ are strictly speaking not fields but merely plain numbers, without any space-time dependence. Space-time derivatives and integrals do not exist. The most general action $\mathcal{S} ( \vec{\phi} \, )$ of such an ultra-local model, which is invariant under $O(N)$ rotations, is given by the ordinary function
		\begin{align}
			\mathcal{S}( \vec{\phi} \, ) = U ( \vec{\phi} \, ) = U ( \rho ) \, ,
		\end{align}
	with the $O(N)$ invariant
		\begin{align}
			\rho \equiv \tfrac{1}{2} \, \vec{\phi}^{\, 2} \, ,	\label{eq:invariant}
		\end{align}
	and the scalar self-interaction potential $U$, which merely needs to be bounded from below and grow at least linearly in $\rho$ to have well-defined expectation values. All non-vanishing correlation functions, \textit{e.g.}, the two-point function
		\begin{align}
			\langle \phi_i \, \phi_j \rangle = \, & \tfrac{1}{N} \, \delta_{i j} \, \langle \vec{\phi}^{\, 2} \rangle \, ,	\label{eq:on-model_relation_2pfi}
		\end{align}
	can be expressed in terms of ordinary $N$-dimensional integrals,
		\begin{align}
			\big\langle ( \vec{\phi}^{\, 2} )^n \big\rangle \equiv \frac{1}{\mathcal{Z}_0} \int_{- \infty}^{\infty} \mathrm{d}^N \phi \, ( \vec{\phi}^{\, 2} )^n \, \mathrm{e}^{- U ( \vec{\phi} \, )} \, ,	\label{eq:expectation_values}
		\end{align}
	with the normalization
		\begin{align}
			\mathcal{Z}_0 \equiv \int_{- \infty}^{\infty} \mathrm{d}^N \phi \, \mathrm{e}^{- U ( \vec{\phi} \, )} \, .
		\end{align}
	Using (hyper\nobreakdash-)spherical coordinates -- the $O(N)$-invariant -- these expectation values of $( \vec{\phi}^{\, 2} )^n$ can be computed in terms of one-dimensional integrals
		\begin{align}
			\langle ( \vec{\phi}^{\, 2} )^n \rangle = \frac{2^n \int_0^\infty \mathrm{d}\rho \, \rho^\frac{(N - 2)}{2} \,  \rho^n \, \mathrm{e}^{-U ( \rho )}}{\int_0^\infty \mathrm{d} \rho \, \rho^\frac{(N - 2)}{2} \, \mathrm{e}^{- U ( \rho )}}	\, .	\label{eq:ON_expectation_value}
		\end{align}
	Connected and $1$PI-correlation functions are related to correlation functions, see, \textit{e.g.}, Eqs.~(70)-(75) in Ref.~\cite{Koenigstein:2021syz} for explicit expressions of the first three non-vanishing connected and $1$PI-correlation functions of the zero dimensional $O(N)$~model or Ref.~\cite{Keitel:2011pn}.\\
	
	For related works on this toy model QFT, we again refer to Refs.~\cite{Fl_rchinger_2010,Moroz:2011thesis,Keitel:2011pn,Strocchi:2013awa,Kemler:2013yka,Pawlowski:talk,Rentrop_2015,Rosa:2016czs,Liang:2017whg,SkinnerScript,Millington:2019nkw,Alexander:2019cgw,Millington:2020Talk,Hikami:1978ya,Bessis:1980ss,DiVecchia:1990ce,Nishigaki:1990sk,Schelstraete:1994sc,Zinn-Justin:1998hwu,Keitel:2011pn,Koenigstein:2021syz,Koenigstein:2021rxj,Catalano:2019}.

\subsection{Free theory}
\label{subsec:free_theory}

	For later reference, we recapitulate some results for the massive non-interacting free theory. The action of the corresponding $O(N)$~model is given by
		\begin{align}
			U_0 ( \rho ) \equiv m^2 \rho \, ,
		\end{align}
	with the positive non-zero ``mass'' $m$. The expectation values \eqref{eq:ON_expectation_value} can be computed analytically in terms of Gamma functions resulting in
		\begin{align}
			\big\langle ( \vec{\phi}^{\, 2} )^0 \big\rangle = \, & \big\langle 1 \big\rangle = 1 \, ,	\vphantom{\bigg(\bigg)}
			\\
			\big\langle ( \vec{\phi}^{\, 2} )^n \big\rangle = \, & \tfrac{N + 2 n - 2}{m^2} \, \big\langle ( \vec{\phi}^{\, 2} )^{n-1} \big\rangle \, ,	\vphantom{\bigg(\bigg)}
		\end{align}
	for $n > 1$. For the $1$PI-correlation functions this result implies
		\begin{align}
			&	\Gamma^{(2)} = m^2 \, ,	&&	\text{and}	&&	\forall n \neq 2 \quad \Gamma^{(n)} = 0 \, ,	\label{eq:free_Gamma}
		\end{align}
	where we used the short-hand notation $\Gamma^{(n)} \equiv \, \Gamma^{(n)}_{\varphi_i \ldots \varphi_i}$ of Refs.~\cite{Koenigstein:2021syz,Keitel:2011pn}. In their interpretation as interaction vertices this result for $\Gamma^{(n)}$ is rather intuitive for a ``massive non-interacting'' theory, which has only a non-vanishing $1$PI two-point function, because the underlying probability distribution is Gaussian.

\subsection{Reformulation for large-\texorpdfstring{$N$}{N}}

	For computations at large $N$ and in the limit ${N \rightarrow \infty}$ the rescaling
		\begin{align}
			&	\rho \mapsto y = \tfrac{1}{N} \, \rho \, ,	&&	U ( \rho ) \mapsto V ( y ) = \tfrac{1}{N} \, U ( \rho ) \, ,	\label{eq:rescalings}
		\end{align}
	has proven particularly useful, see, \text{e.g.}, Refs.~\cite{Keitel:2011pn,Grossi:2019urj}, because both $y$ and $V ( y )$ are of $\mathcal{O} ( N^0 )$. The expression \eqref{eq:ON_expectation_value} for $\langle( \vec{\phi}^{\, 2} )^n\rangle$ reads
		\begin{align}
			\langle ( \vec{\phi}^{\, 2} )^n \rangle = \, & \frac{2^n N^n \int_0^\infty \mathrm{d} y \, y^\frac{(N - 2)}{2} \,  y^n \, \mathrm{e}^{-N V ( y )}}{\int_0^\infty \mathrm{d} y \, y^\frac{(N - 2)}{2} \, \mathrm{e}^{-N V ( \rho )}} \, ,	\vphantom{\Bigg(\Bigg)}	\label{eq:ON_expectation_value_largeN}
			\\
			= \, & \frac{2^n N^n \int_0^\infty \mathrm{d}y \, y^{n - 1} \, \mathrm{e}^{- N \, [ V ( y ) - \frac{1}{2} \ln ( y ) ]}}{\int_0^\infty \mathrm{d} y \, y^{-1} \, \mathrm{e}^{- N \, [ V ( y ) - \frac{1}{2} \ln ( y ) ]}} \, ,	\vphantom{\Bigg(\Bigg)}	\nonumber
		\end{align}
	in terms of $y$ and $V ( y )$ and we note $\langle ( \vec{\phi}^{\, 2} )^n \rangle=\mathcal{O} ( N^n )$. For certain potentials $V(y)$ the involved integrals
		\begin{align}
			I_n^N [ V ] \equiv \int_0^\infty \mathrm{d} y \, y^{n - 1} \, \mathrm{e}^{- N \, [ V ( y ) - \frac{1}{2} \ln ( y ) ]}	\label{eq:In_largeN}
		\end{align}
	can be solved in terms of known functions, see, \textit{e.g.} Refs.~\cite{Keitel:2011pn,Koenigstein:2021syz,Keitel:2011pn} as well as Sub.Sec.~\ref{subsec:RP}, and/or they can be computed in the limit ${N \rightarrow \infty}$ by means of a saddle-point expansion, see Sec.~\ref{sec:saddle_point} and App.~\ref{sec:saddle_point_app}.

\subsection{An instructive toy model}
\label{subsec:RP}

\begin{figure}
			\centering
			\includegraphics{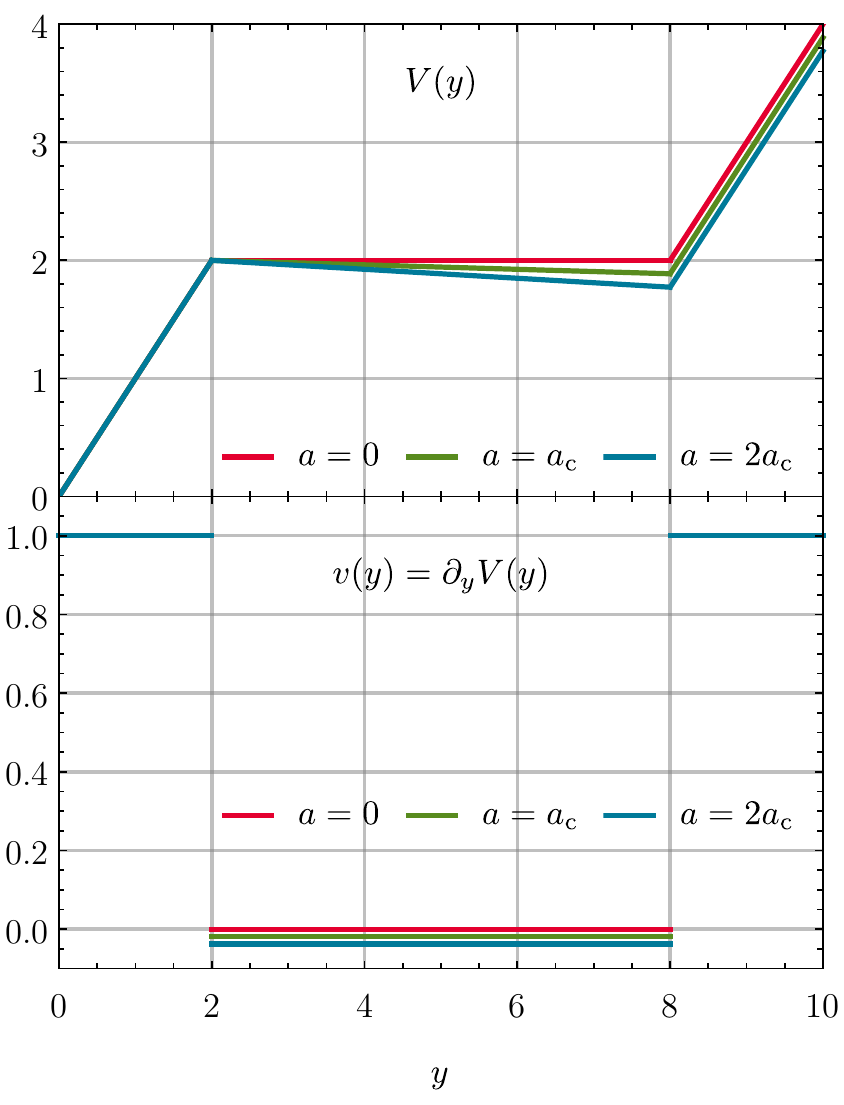}
			\caption{\label{fig:Vofy}%
				The potential $V ( y )$ from Eq.~\eqref{eq:RP_Vofy} (upper panel) and its derivative $v ( y ) = \partial_y V ( y )$ from Eq.~\eqref{eq:RP_vofy} (lower panel) for different values of the parameter $a$ and where $a_\mathrm{c}$ is given by Eq.~\eqref{eq:a_critical}.
			}
		\end{figure}

	In this subsection we present an explicit $O(N)$~model, respectively its $\frac{1}{N}$-rescaled self-interaction potential $V(y)$, which turns out to be a rather instructive toy model when studied at large and infinite $N$. We consider a family of piecewise linear potentials
		\begin{align}
			V ( y ) =
			\begin{cases}
				y							&	\text{for} \quad 0 \leq y \leq 2 \, ,	\vphantom{\bigg(\bigg)}
				\\
				- a \, y + 2 \, ( a + 1 )	&	\text{for} \quad 2 < y \leq 8 \, ,	\vphantom{\bigg(\bigg)}
				\\
				y - 6 \, ( a + 1 )			&	\text{for} \quad 8 < y \, ,	\vphantom{\bigg(\bigg)}
			\end{cases}	\label{eq:RP_Vofy}
		\end{align}
	with a parameter $a\geq0$. The first derivative of $V ( y )$ presents as a simple piecewise constant function in the $\tfrac{1}{N}$-rescaled invariant $y$
		\begin{align}
		v ( y ) = \partial_y V ( y ) =
		\begin{cases}
			1	&	\text{for} \quad 0 \leq y \leq 2 \, ,	\vphantom{\bigg(\bigg)}
			\\
			- a	&	\text{for} \quad 2 < y \leq 8 \, ,	\vphantom{\bigg(\bigg)}
			\\
			1	&	\text{for} \quad 8 < y \, ,	\vphantom{\bigg(\bigg)}
		\end{cases}	\label{eq:RP_vofy}
	\end{align}
	which is very similar to the one studied in Ref.~\cite{Grossi:2019urj}. The potential \eqref{eq:RP_Vofy} and its $y$-derivative \eqref{eq:RP_vofy} are plotted in Fig.~\ref{fig:Vofy} for illustrative purposes\footnote{In Sec.~\ref{sec:FRG}, Fig.~\ref{fig:Vofx}, we also plot the potential and its derivative as functions of the rescaled field $x$, where $\tfrac{1}{2} \, x^2 \equiv y\equiv \tfrac{1}{2N} \, \vec{\phi}^{\, 2}$, which might be a more familiar variable choice for some readers.}. In the context of conservation equations and fluid dynamics in general, initial value problems with piecewise constant initial conditions involving a single discontinuity are refereed to as Riemann problems and canonical examples can be found in the textbooks \cite{Lax1973,Ames:1992,LeVeque:1992,LeVeque:2002,Hesthaven2007,Toro2009,RezzollaZanotti:2013}. When considering Eq.~\eqref{eq:RP_vofy} as the initial condition of a conservation equation, see App.~\ref{app:RPandEntropy}, we are faced with two Riemann problems (one at $y=2$ and one at $y=8$) at the UV initial scale.
	
	This model has several interesting properties:
		\begin{enumerate}
			\item	The expectation values of Eq.~\eqref{eq:ON_expectation_value_largeN} can be evaluated in terms of known functions. In the limit ${N \rightarrow \infty}$ the $1$PI-correlation functions can be computed analytically for all $a\geq0$. We will discover within this section that there are two distinct parameter regimes, which are particularly interesting when studying this problem within the saddle-point and FRG frameworks.
			
			\item	For certain parameters $a$, which are smaller than some critical value $a_\mathrm{c}$, the $1$PI-correlation functions $\Gamma^{(n)}$ -- the underlying expectation values \eqref{eq:ON_expectation_value_largeN} -- can be computed by means of a saddle-point expansion. For $a \geq a_\mathrm{c}$ the saddle-point expansion is not applicable. This is discussed in detail in the following Sec.~\ref{sec:saddle_point}.
			
			\item	The model under consideration presents initially as two Riemann problems in the FRG (fluid dynamic) framework. The two distinct parameter regimes, ${0 \leq a \leq a_\mathrm{c}}$ and $a > a_\mathrm{c}$, present with qualitatively different FRG flows. The interpretation involving Riemann problems, its numerical solution, and its consequences are discussed in detail in Sec.~\ref{sec:FRG}.
		\end{enumerate}

	For now, we turn to the computation of the correlation functions of the model under consideration. Solutions in terms of known functions for the necessary integrals \eqref{eq:In_largeN} for the potential \eqref{eq:RP_Vofy} are presented in App.~\ref{sec:RP_integrals}.
	
	In the limit ${N \rightarrow \infty}$ the direct computations of App.~\ref{sec:RP_integrals} revealed two distinct regimes in parameter space separated by
		\begin{align}
			a_\mathrm{c} = \tfrac{1}{4} - \tfrac{1}{3} \ln ( 2 ) \approx 0.018951 \, .	\label{eq:a_critical}
		\end{align}
	For the infinite-$N$ limit of the expectation values \eqref{eq:ON_expectation_value} we find,
		\begin{align}
			\lim_{N \rightarrow \infty} \tfrac{1}{N^n} \, \big\langle ( \vec{\phi}^{\, 2} )^n \big\rangle =
			\begin{cases}
				1 \, ,		&	\text{for} \quad 0 \leq a \leq a_\mathrm{c} \, ,	\vphantom{\bigg(\bigg)}
				\\
				16^n \, ,	&	\text{for} \quad a_\mathrm{c} < a \, ,	\vphantom{\bigg(\bigg)}
			\end{cases}
		\end{align}
	For the corresponding $1$PI-correlation functions this implies in the limit ${N \rightarrow \infty}$
		\begin{align}
			\Gamma^{(2)} =	\label{eq:two-point_exact}
			\begin{cases}
			1 \, ,				&	\text{for} \quad 0 \leq a \leq a_\mathrm{c} \, ,	\vphantom{\bigg(\bigg)}
			\\
			\tfrac{1}{16} \, ,	&	\text{for} \quad a_\mathrm{c} < a \, ,	\vphantom{\bigg(\bigg)}
		\end{cases}
		\end{align}
	as well as for all $a \geq 0$ 
		\begin{align}
			\forall n \neq 2 \quad \Gamma^{(n)} = 0 \, .	\label{eq:n-point_exact}
		\end{align}
	Thus, in the limit ${N \rightarrow \infty}$ and in terms of $1$PI-vertices the current model under consideration presents as a massive non-interacting theory for all $a \geq 0$, \textit{cf.} Eq.~\eqref{eq:free_Gamma}. The situation for $0\leq a<a_\mathrm{c}$ and the corresponding ``mass'' as well as the origin of the critical value $a_\mathrm{c}$ can be understood intuitively in the context of the saddle-point expansion discussed in Sec.~\ref{sec:saddle_point}. The situations for $a =a_\mathrm{c}$ and $a >a_\mathrm{c}$ are more involved and not accessible with a saddle-point expansion. However, a study in the FRG framework is possible and rather instructive as we will demonstrate in Sec.~\ref{sec:FRG}. In terms of correlation functions the theory undergoes a first-order phase transition at $a_\mathrm{c}$ when varying the external parameter $a$, \textit{cf.} Sub.Sec.~III~C of Ref.~\cite{Grossi:2019urj} and references therein.\\ 
	
	For finite $N$ higher order $n$-point functions do not vanish and the theory is of ``interactive type'', but in the scope of this paper we nevertheless mainly focus on $\Gamma^{(2)}$ -- especially when it comes to numerical computations.\\
	
	In Tab.~\ref{tab:Gamma2N} we summarize several (exact) reference values for $\Gamma^{(2)}$ for later use.
		\begin{table}[b]
			\caption{\label{tab:Gamma2N}%
				Reference values for $\Gamma^{(2)} = N ( \langle \vec{\phi}^{\, 2} \rangle )^{- 1}$ for selected $N$ and $a$ computed with the expressions \eqref{eq:InNV_a0} and \eqref{eq:InNV} as well as their large $N$ asymptotics. The exact analytical results are in some cases rather lengthy and therefore we present in those cases only six decimal digits for readability.
			}
			\begin{ruledtabular}
				\begin{tabular}{l c c c}
					$N$			&	$a = 0$		&	$a = a_\mathrm{c}$	&	$a = 2 a_\mathrm{c}$
					\\
					\colrule\addlinespace[0.25em]
					$2$			&	$0.356907$	&	$0.327332$ 			&	$0.299162$
					\\
					$32$		&	$0.962306$	&	$0.475285$			&	$0.087158$
					\\
					$\infty$	&	 $1$		&	$1$					&	$0.0625$
				\end{tabular}
			\end{ruledtabular}
		\end{table}

\section{The saddle-point expansion at large-\texorpdfstring{$N$}{N}}
\label{sec:saddle_point}

\begin{figure}
			\centering
			\includegraphics{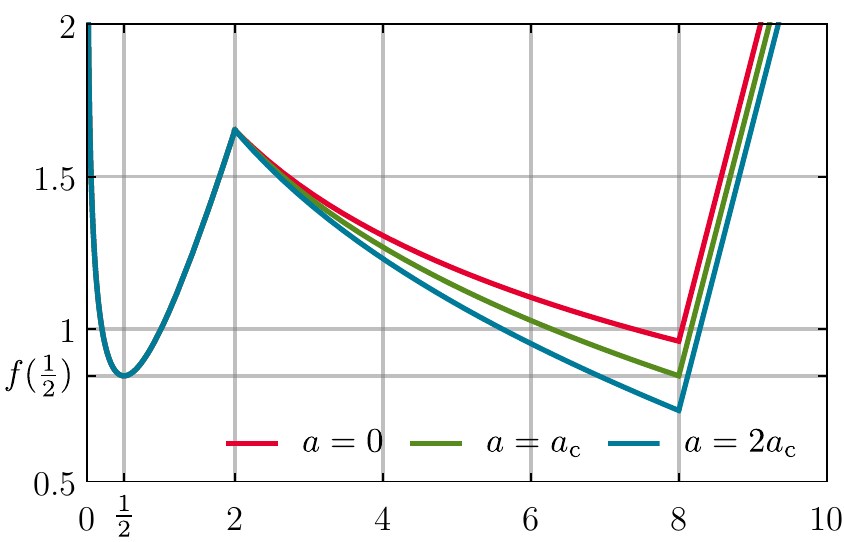}
			\caption{\label{fig:sp_fofy}%
				The function ${f ( y ) = V ( y ) - \tfrac{1}{2} \ln ( y )}$ for the potential \eqref{eq:RP_Vofy} for selected values of the parameter $a$ with ${a_\mathrm{c} = \tfrac{1}{4} - \tfrac{1}{3} \ln ( 2 ) \approx 0.018951}$. The local minima of $f ( y )$ are located at $y_0 = \tfrac{1}{2}$ and $y_{0, 2} = 8$, where $y_0$ ($y_{0, 2}$) presents as the unique global minimum for $0 \leq a < a_\mathrm{c}$ ($a > a_\mathrm{c}$). At $a = a_\mathrm{c}$ both minima coincide and present both as global minima of $f ( y )$. The non-analyticity of $f ( y )$ in $y_{0, 2} = 8$ inherited from the piecewise definition of $V ( y )$ is clearly visible in the plot.
			}
		\end{figure}

	In this section we will analyze the instructive toy model of Sub.Sec.~\ref{subsec:RP} within a saddle-point approximation for large $N$. In App.~\ref{sec:saddle_point_app} we discuss the large $N$ saddle-point expansion of integrals like \eqref{eq:In_largeN} concluding in the asymptotic series \eqref{eq:SPapp_series} for $\langle ( \vec{\phi}^{\, 2} )^n \rangle$. To apply the series \eqref{eq:SPapp_series} to the interaction potential \eqref{eq:RP_Vofy} of the model under consideration, we first have to compute the global minimum $y_0$ of the exponents of the integrands in Eq.~\eqref{eq:ON_expectation_value_largeN},
		\begin{align}
			f ( y ) = V ( y ) - \tfrac{1}{2} \ln ( y ) \, ,\label{eq:fofy}
		\end{align}
	and check for analyticity of $f ( y )$ and $g ( y ) = y^{n - 1}$ around $y_0$. The function $f ( y )$ for the model under consideration is plotted in Fig.~\ref{fig:sp_fofy} for different parameters $a$.

	There is always a minimum on the first section (${0 \leq y \leq 2}$) of the piecewise linear potential
		\begin{align}
			0 \overset{!}{=} \, & \partial_y f(y) \big|_{y = y_0} =	\vphantom{\bigg(\bigg)}
			\\
			= \, & \big[ \partial_y V ( y ) - \tfrac{1}{2 y} \big] \big|_{y = y_0} =	\vphantom{\bigg(\bigg)}	\nonumber
			\\
			= \, & 1 - \tfrac{1}{2 y_0} \, .	\vphantom{\bigg(\bigg)}	\nonumber
		\end{align}
	It follows that
		\begin{align}
			&	y_0 = \tfrac{1}{2} \, ,	&&	V ( y_0 ) = \tfrac{1}{2} \, ,	&&	f ( y_0 ) = \tfrac{1}{2} \, \big[ 1 - \ln \big( \tfrac{1}{2} \big) \big] \, ,
		\end{align}
	and for the second and third derivatives, we find
		\begin{align}
			&	\partial_y^2 V ( y ) \big|_{y = y_0} = 0 \, ,	&&	\partial_y^2 f ( y ) \big|_{y = y_0} = 2 \, ,	\vphantom{\bigg(\bigg)}
			\\
			&	\partial_y^3 V ( y ) \big|_{y = y_0} = 0 \, ,	&&	\partial_y^3 f ( y ) \big|_{y = y_0} = - 8 \, .	\vphantom{\bigg(\bigg)}
		\end{align}
	We note that $V ( y )$ and therefore also $f ( y )$ are smooth, thus $C^\infty$, and analytic around $y_0 = \tfrac{1}{2}$. Also $g ( y ) = y^{n - 1}$ is analytic and $C^\infty$ around $y_0 = \tfrac{1}{2}$. We can therefore use the asymptotic series \eqref{eq:SPapp_series} to compute the non-vanishing expectation values,
		\begin{align}
			\tfrac{1}{N} \, \langle \vec{\phi}^{\, 2}  \rangle = \, & 1 \, ,	\vphantom{\bigg(\bigg)}
			\\
			\tfrac{1}{N^2} \, \langle ( \vec{\phi}^{\, 2} )^2 \rangle = \, & 1 + \tfrac{2}{N} \, ,	\vphantom{\bigg(\bigg)}
			\\
			\tfrac{1}{N^3} \, \langle ( \vec{\phi}^{\, 2} )^3 \rangle = \, & 1 + \tfrac{6}{N} + \tfrac{8}{N^2} \, ,	\vphantom{\bigg(\bigg)}
			\\
			\vdots \, \,  &	\vphantom{\bigg(\bigg)}
		\end{align}
	and the corresponding $1$PI-correlation functions
		\begin{align}
			&	\Gamma^{(2)} = 1 \, ,	&&	\forall n \neq 2 \quad \Gamma^{(n)} = 0 \, .
		\end{align}
	Both are exact results (without taking any limits) and we find that $\tfrac{1}{N^n} \, \langle ( \vec{\phi}^{\, 2} )^n \rangle = 1 + \mathcal{O} ( N^{- 1} )$, while the maximal correction to $1$ is always of $\mathcal{O} ( N^{- ( n - 1 )})$. Considering the corresponding $\Gamma^{(2n)}$ we recover the $1$PI-correlation functions of a free massive theory, see Eq.~\eqref{eq:free_Gamma} with $m^2 = 1$, which -- as an exact and $N$-independent result -- also holds trivially in leading order in the limit ${N \rightarrow \infty}$. This is a rather unsurprising result since the $\tfrac{1}{N}$-rescaled potential $V ( y )$ manifests as a linear potential with slope $1$ -- corresponding to a non-interacting theory with $m^2 = 1$ -- for ${0 \leq y \leq 2}$.\\
	
	The previous large-$N$ saddle-point approximation is however limited to potentials \eqref{eq:RP_Vofy} with $0 \leq a < a_\mathrm{c}$. For $a\geq a_\mathrm{c}$ the function ${f ( y ) = V ( y ) - \tfrac{1}{2} \ln ( y )}$ develops a global minimum at $y_{0,2} = 8$, which becomes the unique global minimum for $a>a_\mathrm{c}$ while at $a=a_\mathrm{c}$ both $y_0$ and $y_{0,2}$ are global minima, see Fig.~\ref{fig:sp_fofy}. For $a\geq a_\mathrm{c}$ the saddle-point expansion breaks down since at $a=a_\mathrm{c}$ the function $f(y)$ has no unique minimum and for $a>a_\mathrm{c}$ the function $f(y)$ is non-analytic in its global minimum (the ``expansion point'') $y_{0,2}=8$. The value of $a_\mathrm{c}$ and the related qualitatively distinct scenarios were established in Sub.Sec.~\ref{subsec:RP}. In the corresponding exact computations of App.~\ref{sec:RP_integrals} the threshold ${a_\mathrm{c}= \tfrac{1}{4} - \tfrac{1}{3} \ln ( 2 ) \approx 0.018951}$ appears when considering the limit ${N \rightarrow \infty}$ of rather complicated symbolic expressions. On the other hand, within the framework of the saddle-point expansion the value of $a_\mathrm{c}$ can be derived and understood in a very instructive way as the breakdown point of the saddle-point expansion,
		\begin{align}
			f \big( y_0 = \tfrac{1}{2} \big) \overset{!}{=} \, & f ( y_{0,2} = 8 )	\vphantom{\bigg(\bigg)}	\label{eq:saddle_minima}
			\\
			\tfrac{1}{2} - \tfrac{1}{2} \ln \big( \tfrac{1}{2} \big) = \, & 8 - 6 \, ( a_\mathrm{c} + 1 ) - \tfrac{1}{2} \ln ( 8 ) \, ,	\vphantom{\bigg(\bigg)}	\nonumber
		\end{align}
	which is solved by
		\begin{align}
			a_\mathrm{c} = \, & \tfrac{1}{4} - \tfrac{1}{3} \ln ( 2 ) \approx 0.018951 \, .
		\end{align}
	For $a$ below $a_\mathrm{c}$ the model presents as a free massive theory in its saddle-point and the analytical results in the limit ${N \rightarrow \infty}$ of Sub.Sec.~\ref{subsec:RP} make perfect sense.\\
	
	In this paper we are not interested in a quantitative review of the large-$N$ saddle-point expansion beyond the limit ${N \rightarrow \infty}$. For such a discussion in the context of zero-dimensional $O(N)$~models we refer the interested reader to the excellent and pedagogical Ref.~\cite{Keitel:2011pn}.\\
	
	At and beyond the critical value $a_\mathrm{c}$ -- at and beyond the corresponding first-order phase transition -- the saddle-point expansion is no longer applicable and alternative methods are required for the computation of correlation functions. Apart from the direct symbolic computations of Sub.Sec.~\ref{subsec:RP} the FRG is a potent tool for computations at arbitrary finite and infinite $N$, as we will demonstrate in the next section. This becomes in particular interesting, when exact reference results are no longer accessible and when it is hard to judge if a possible expansion point for the $\frac{1}{N}$-expansion is (non\nobreakdash-)analytic. This is the case for a lot of higher-dimensional models (also involving fermionic degrees of freedom \textit{cf.} \cite{Stoll:2021ori}).

\section{FRG and fluid dynamics}
\label{sec:FRG}

	This section is dedicated to the FRG analysis of our zero-dimensional $O(N)$-symmetric toy model for the piecewise UV initial potential \eqref{eq:RP_Vofy} in terms of a fluid-dynamic problem. However, we do not provide a detailed introduction to the FRG formalism at this point. Instead we only recapitulate some key aspects of (zero-dimensional) FRG and almost directly start off with the RG flow equation of the zero-dimensional $O(N)$~model with arbitrary action $\mathcal{S} ( \vec{\phi} \, ) = U ( \vec{\phi} \, )$. For detailed reviews and introductions to the FRG and generic applications, we refer to Refs.~\cite{Berges:2000ew,Pawlowski:2005xe,Kopietz:2010zz,Rosten:2010vm,Gies:2006wv,Delamotte:2007pf,Dupuis:2020fhh,Gies:2006wv,PawlowskiScript}. For a pedagogical introduction into the FRG in zero dimensions, we refer to part I of this series of publications \cite{Koenigstein:2021syz} or Ref.~\cite{Keitel:2011pn}.

\subsection{The RG flow equation}
	
	In Sec.~II of Ref.~\cite{Koenigstein:2021syz} we motivated the overall technical idea of the FRG: By introducing an artificial parameter dependent ``mass-like'' term, the regulator\footnote{%
		In contrast to FRG calculations in non-zero spacetime dimensions all regulators in zero-dimensions are equivalent up to a reparameterization.
		Therefore the explicit choice of regulator does not alter any results.%
	}, \textit{e.g.},
		\begin{align}
			&	r ( t ) = \Lambda \, \mathrm{e}^{-t} \, ,	&&	t \in [ 0, \infty ) \, ,	\label{eq:roft}
		\end{align}
	in the generating functional for correlation functions
		\begin{align}
			\mathcal{Z}_t [ \vec{J} \, ] \equiv \int_{- \infty}^{\infty} \mathrm{d}^N \phi \, \mathrm{e}^{- \frac{1}{2} \, r ( t ) \, \vec{\phi}^{\, 2} - U ( \vec{\phi} \, ) + \vec{J} \cdot \vec{\phi}} \, ,	\label{eq:partition_fucntion}
		\end{align}
	we are able to continuously deform partition functions for arbitrary theories into Gaussian-type partition functions and vice versa by simply changing the so-called RG time $t$ and choosing a large UV cutoff $\Lambda$. For $t = 0$ the generating functional \eqref{eq:partition_fucntion} is completely dominated by the artificial mass $r ( t )$ and almost perfectly Gaussian (as long as $\Lambda$ is much larger than the typical model scales). For $t \rightarrow \infty$ Eq.~\eqref{eq:partition_fucntion} transforms back into the original problem. The same holds true for the corresponding expectation values. However, Gaussian type integrals can easily be solved, which lead to the idea to describe the continuous deformation via an evolution equation for $\mathcal{Z}_t [ \vec{J} \, ]$ in terms of a PDE and to initialize this PDE at the Gaussian point $t = 0$ (the UV). At the end of our discussion in Sec.~II of \cite{Koenigstein:2021syz} we demonstrated, that it is more convenient to formulate this process on the level of the generating functional for the $1$PI-correlation functions, namely the effective action,
		\begin{align}
			\Gamma [ \vec{\varphi} \, ] \equiv \, & \underset{\vec{J}}{\mathrm{sup}} \{ \vec{J} \cdot \vec{\varphi} - \ln \mathcal{Z} [ \vec{J} \, ] \} \, ,	\label{eq:effective_action}
		\end{align}
	It was shown, that there exists an evolution equation for a scale (RG-time) dependent version of $\Gamma [ \vec{\varphi} \, ]$, namely the scale-dependent effective average action $\bar{\Gamma}_t [ \vec{\varphi} \, ]$. This evolution equation is called the Exact Renormalization Group (ERG) equation,
		\begin{align}
			\partial_t \bar{\Gamma}_t [ \vec{\varphi} \, ] = \, & \mathrm{Tr} \big[ \big( \tfrac{1}{2} \, \partial_t R_t \big) \, \big( \bar{\Gamma}_t^{(2)} [ \vec{\varphi} \, ] + R_t \big)^{-1} \big] \, .	\label{eq:wetterich-equation}
		\end{align}
	Here, $\bar{\Gamma}_t^{(2)} [ \vec{\varphi} \, ]$ is the full scale- and field-dependent two-point function, while $R_t = r ( t ) \, \openone_{N \times N}$ is the regulator function that is diagonal in field space. The trace is exclusively a field space trace for our model.
	
	Equation \eqref{eq:wetterich-equation} is the zero-dimensional $N$-boson version of the much more general functional ERG equation, which can be applied to arbitrary QFTs. The most general version of Eq.~\eqref{eq:wetterich-equation} is a modern implementation of K.~G.~Wilson's concept of the RG \cite{Wilson:1971bg,Wilson:1971dh,Wilson:1979qg} originally developed by U.~Ellwanger, T.~R.~Morris, and C.~Wetterich \cite{Ellwanger:1993mw,Morris:1993qb,Wetterich:1991be,Wetterich:1992yh} and was successfully applied to a broad range of problems in quantum statistical physics -- included all kinds of strongly correlated systems, \textit{cf.}\ Refs.~\cite{Berges:2000ew,Pawlowski:2005xe,Kopietz:2010zz,Rosten:2010vm,Gies:2006wv,Delamotte:2007pf,Dupuis:2020fhh,Gies:2006wv,PawlowskiScript} and references therein.
	
	Equation \eqref{eq:wetterich-equation} is usually initialized with the classical action in the UV at $t = 0$, $\bar{\Gamma}_{t = 0} [ \vec{\varphi} \, ] = \mathcal{S} [ \vec{\varphi} \, ]$, while after the evolution to $t \rightarrow \infty$, the full quantum effective action is approached in the IR, $\bar{\Gamma}_{t \rightarrow \infty} [ \vec{\varphi} \, ] = \Gamma [ \vec{\varphi} \, ]$.
	
	Usually it is far to complicated or even impossible to solve the functional ERG equation in higher-dimensional theories directly and the evolution in theory space needs to be truncated to some subspace by using an ansatz for $\bar{\Gamma}_t [ \vec{\varphi} \, ]$. This is done by utilizing the symmetries of the system and suitable projection prescriptions, see \textit{e.g.}, Refs.~\cite{Adams:1995cv,Berges:2000ew,Pawlowski:2017gxj,Pawlowski:2014zaa,Papp:1999he,Schaefer:1999em,Cichutek:2020bli,Eser:2018jqo,Eser:2019pvd,Divotgey:2019xea,Canet:2002gs,Delamotte:2007pf,Otto:2019zjy,Otto:2020hoz}. For our zero-dimensional scenario the ERG equation~\eqref{eq:wetterich-equation} manifests directly as a PDE for the effective average action which is a only a function and not a functional in zero space-time dimensions. For the zero dimensional $O(N)$ model it is possible to solve the PDE~\eqref{eq:wetterich-equation} prescribing the RG flow in its full generality without the need for truncations of $\bar{\Gamma}_t [ \vec{\varphi} \, ]$ \cite{Koenigstein:2021syz,Koenigstein:2021rxj,Keitel:2011pn,Steil:partIV}. For the zero-dimensional $O(N)$ model, the most general form of the effective average action reads
		\begin{align}
			\bar{\Gamma}_t [ \vec{\varphi} \, ] = U ( t, \vec{\varphi} \, ) \, ,
		\end{align}
	were $U ( t, \vec{\varphi} \, )$ is the scale dependent effective potential. We merely take advantage of the $O(N)$ symmetry and, \textit{w.l.o.g.}, choose the field vector $\vec{\varphi} = ( \sigma, 0, \ldots, 0 )$ in the functional equation \eqref{eq:wetterich-equation}. We end up with an RG flow equation for the scale-dependent effective potential $U ( t, \sigma )$,
		\begin{align}
			& \partial_t U ( t, \sigma ) =	\vphantom{\bigg(\bigg)}	\label{eq:rg_flow_U}
			\\
			= \, & (N - 1) \, \frac{\tfrac{1}{2} \, \partial_t r ( t )}{r ( t ) + \frac{1}{\sigma} \, \partial_\sigma U ( t, \sigma )} + \frac{\tfrac{1}{2} \, \partial_t r ( t )}{r ( t ) + \partial_\sigma^2 U ( t, \sigma )}	\nonumber
		\end{align}
	that is initialized with the classical potential ${U ( t = 0, \sigma ) = \mathcal{S} ( \sigma ) = U ( \sigma )}$.

\subsection{The RG and (numerical) fluid dynamics}
	
	It turns out that it is usually not appropriate to solve the PDE \eqref{eq:rg_flow_U} in terms of $U ( t, \sigma )$ directly. In Refs.~\cite{Koenigstein:2021syz,Koenigstein:2021rxj,Grossi:2019urj,Grossi:2021ksl,Stoll:2021ori} it is demonstrated that by taking a derivative \textit{w.r.t.}\ $\sigma$ of Eq.~\eqref{eq:rg_flow_U}, one can transform the RG flow into a typical conservation law for $u ( t, \sigma ) \equiv \partial_\sigma U ( t, \sigma )$,
		\begin{align}
			& \partial_t u ( t, \sigma ) =	\vphantom{\bigg(\bigg)}	\label{eq:frg_flow_x_without_rescale}
			\\
			= \, & \frac{\mathrm{d}}{\mathrm{d} \sigma} \bigg[ (N - 1) \, \frac{\tfrac{1}{2} \, \partial_t r ( t )}{r ( t ) + \frac{1}{\sigma} \, u ( t, \sigma )} + \frac{\tfrac{1}{2} \, \partial_t r ( t )}{r ( t ) + \partial_\sigma u ( t, \sigma )} \bigg]	\vphantom{\bigg(\bigg)}		\nonumber
		\end{align}
	Conservation laws like this are well-known from (numerical) fluid dynamics. More precisely, the conservation law at hand can be classified as an advection-diffusion equation for the ``fluid'' $u ( t, \sigma )$ in one temporal and one spatial dimension. Hereby, $t \in [ 0, \infty )$ plays the role of a manifestly positive effective time coordinate, while field space $\sigma \in ( - \infty, + \infty)$ is interpreted as an effective infinite spatial domain. Formally, we identify the highly non-linear, explicitly position-dependent advection flux,
		\begin{align}
			F [ t, \sigma, u ] = \, & - \frac{(N - 1 ) \,  \tfrac{1}{2} \, \partial_t r ( t )}{r ( t ) + \frac{1}{\sigma} \, u ( t, \sigma )} = -%
			\begin{gathered}%
				\includegraphics{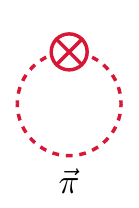}%
			\end{gathered} \, ,	\label{eq:advection_flux}%
		\end{align}
	which originates from the pion\footnote{In this article, we use the terminology of high-energy physics for the zero-dimensional counterparts of the Nambu-Goldstone modes \cite{Nambu:1960tm,Goldstone:1961eq,Goldstone:1962es}, which are called Anderson-Bogoliubov modes \cite{Anderson:1958:PhysRev.110.827,Anderson:1958:PhysRev.112.1900,Bogoljubov:1958} in condensed matter physics. Of course, there is no real notion of massless modes in zero-dimensions, because there is no particle propagation. The same holds true for the ``massive'' radial $\sigma$ mode.} loops and the highly non-linear diffusion flux\footnote{A particularly important feature of the diffusion flux is the non-linearity of the diffusion coefficient $D [ t, \partial_\sigma u ]$, which is obtained as the prefactor of $\partial_\sigma^2 u$, when executing the $\sigma$-derivative in Eq.~\eqref{eq:frg_flow_x_without_rescale}. The inverse dependence on the difference between regulator $r ( t )$ and gradient $\partial_\sigma u$ effectively ensures the convexity of the effective potential $U ( t \rightarrow \infty, \sigma )$, as is also discussed in Refs.~\cite{Koenigstein:2021syz,Koenigstein:2021rxj,Stoll:2021ori} and Sub.Sec.~\ref{subsec:FRGfiniteN}.}
		\begin{align}
			Q [ t, \partial_\sigma u ] = \, & \frac{\tfrac{1}{2} \, \partial_t r ( t )}{r ( t ) + \partial_\sigma u ( t, \sigma )} =
			\begin{gathered}
				\includegraphics{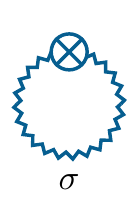}
			\end{gathered} \, ,	\label{eq:diffusion_flux}
		\end{align}
	that stems from the $\sigma$ loop contribution to the RG flow -- the ``radial mode''.
	
	 Non-linear advection-diffusion equations tend to form different kinds of non-analyticities and solutions to these problems usually only exist in a weak-formulation of the PDE \cite{Ames:1992,LeVeque:1992,LeVeque:2002,Hesthaven2007,RezzollaZanotti:2013}. In our everyday life, we are familiar with phenomena like thunder storms, supersonic aircraft, uncontrolled decompression in pipeline systems \textit{etc.}. The theoretical description of such phenomena in a fluid dynamic setup involves discontinuities like shocks, rarefaction waves \textit{etc.}. However, it was found that similar phenomena can also occur in RG flows \cite{Aoki:2014,Aoki:2017rjl,Grossi:2019urj,Grossi:2021ksl,Koenigstein:2021syz,Koenigstein:2021rxj,Stoll:2021ori}. The discussion of shocks and rarefaction waves in RG flows along field space direction and their interpretation and interplay with the failure of the large-$N$ expansion is a central aspect of this work.\\

	In order to solve highly non-linear PDEs like Eq.~\eqref{eq:frg_flow_x_without_rescale} numerically including an adequate resolution of potential non-analytical behavior, it is best to directly address those, who gained most experience on similar types of problems: experts from numerical and computational fluid dynamics! Therefore, we decided to choose a highly potent and well-established discretization scheme, in order to tackle Eq.~\eqref{eq:frg_flow_x_without_rescale} and its variations (see below) numerically. We employ two semi-discrete finite volume schemes using a so-called \textit{Monotonic Upstream-centered Scheme for Conservation Laws} (MUSCL) reconstruction. Most computations are performed using the \textit{KT scheme} developed by A.~Kurganov and E.~Tadmor in Ref.~\cite{KTO2-0}. For computations in the rescaled invariant $y$ we use a variation of the KT scheme: the \textit{KNP scheme} presented by A.~Kurganov, S. Noelle, and G. Petrova in Ref.~\cite{KTO2-1}. For a detailed discussion on especially the KT numerical scheme, as well as its explicit implementation in the context of RG flows, we refer to Ref.~\cite{Koenigstein:2021syz}. In Ref.~\cite{Koenigstein:2021syz} and Ref.~\cite{Stoll:2021ori}, we also discuss at length and test the explicit implementation of the extremely important spatial (field space) boundary conditions for RG flow equations of type \eqref{eq:frg_flow_x_without_rescale}, which is not repeated at this point.

\subsection{The \texorpdfstring{$\tfrac{1}{N}$}{1/N}-rescaled RG flow equation}

	Now that we have introduced the RG flow equation for the derivative of the effective potential ${u ( t, \sigma ) = \partial_\sigma U ( t, \sigma )}$ in Eq.~\eqref{eq:frg_flow_x_without_rescale} and briefly recapitulated its relation to numerical fluid dynamics from Refs.~\cite{Koenigstein:2021syz,Koenigstein:2021rxj,Grossi:2019urj,Grossi:2021ksl,Stoll:2021ori}, we have to slightly modify this PDE to facilitate the large- and infinite-$N$ studies of this publication. To this end, we make use of the rescalings \eqref{eq:rescalings} of extensive quantities,
		\begin{align}
			&	\sigma \mapsto x = \tfrac{1}{\sqrt{N}} \, \sigma \, ,	&&	U ( t, \sigma ) \mapsto V ( t, x ) = \tfrac{1}{N} \,  U ( t, \sigma ) \, .
		\end{align}
	and additionally introduce $v ( t, x ) \equiv \partial_x V ( t, x )$,
		\begin{align}
			u ( t, \sigma ) \mapsto v ( t, x ) = \tfrac{1}{\sqrt{N}} \, u ( t, \sigma ) \, .
		\end{align}
	On the level of the RG flow equation, this results in a slight modification of the prefactors of the fluxes \eqref{eq:advection_flux} and \eqref{eq:diffusion_flux},	
		\begin{align}
			& \partial_t v ( t, x ) =	\vphantom{\bigg(\bigg)}	\label{eq:frg_flow_x}
			\\
			= \, & \frac{\mathrm{d}}{\mathrm{d} x} \bigg[ \frac{N - 1}{N} \, \frac{\tfrac{1}{2} \, \partial_t r ( t )}{r ( t ) + \frac{1}{x} \, v ( t, x )} + \frac{1}{N} \, \frac{\tfrac{1}{2} \, \partial_t r ( t )}{r ( t ) + \partial_x v ( t, x )} \bigg] \, ,	\vphantom{\bigg(\bigg)}	\nonumber
		\end{align}
	which makes it easily possible to take the infinite-$N$ limit and to compare RG flows for infinite and finite values of $N$. Already at this point we find that increasing $N$ makes the problem more and more advection driven. In the limit $N \rightarrow \infty$ the diffusion flux vanishes completely and we are left over with the infinite-$N$ flow equation,
		\begin{align}
			\partial_t v ( t, x ) = \, & \frac{\mathrm{d}}{\mathrm{d} x} \bigg[ \frac{\tfrac{1}{2} \, \partial_t r ( t )}{r ( t ) + \frac{1}{x} \, v ( t, x )}\bigg] \, . \label{eq:frg_flow_Ninf_x}
		\end{align}
	This PDE presents as an advective hyperbolic conservation law and is very similar to its higher-dimensional counterpart \cite{Tetradis:1995br,Litim:1995ex,Grossi:2019urj}.
	
	Of course, we can also formulate the RG flow equation \eqref{eq:frg_flow_x} as a fluid dynamic problem in the $\tfrac{1}{N}$-rescaled invariant $y = \tfrac{1}{2} \, x^2$,
		\begin{align}
			\partial_t v ( t, y ) = \, & \frac{\mathrm{d}}{\mathrm{d} y} \bigg[ \frac{N - 1}{N} \, \frac{\tfrac{1}{2} \, \partial_t r ( t )}{r ( t ) + v ( t, y )} +	\vphantom{\bigg(\bigg)}	\label{eq:frg_flow_y}
			\\
			&  + \frac{1}{N} \, \frac{\tfrac{1}{2} \, \partial_t r ( t )}{r ( t ) + v(t,y) +2 y \, \partial_y v ( t, y )} \bigg] \, ,	\vphantom{\bigg(\bigg)}	\nonumber
		\end{align}
	as is done in Refs.~\cite{Grossi:2019urj,Grossi:2021ksl}. This PDE might even appear more natural to readers, who are familiar with common FRG literature in higher space-time dimensions, \textit{cf.} Refs.~\cite{Tetradis:1995br,Litim:1995ex,Grossi:2019urj}. Overall the structure of the equation keeps its conservative form in terms of an advection-diffusion equation\footnote{This generalizes in $x$ and $y$ to arbitrary dimensions and also to the fixed-point form of the RG flow equation \cite{Koenigstein:2021syz,Koenigstein:2021rxj,Stoll:2021ori}. Regarding fixed-points in the infinite-$N$ limit for the $O(N)$~model in the FRG context we refer the interested reader to Refs.~\cite{Litim:2016hlb,Litim:2018pxe} for a detailed discussion of the situation in $d=3$ dimensions.}.
	
	The main difference is that the advective contribution lost its unpleasant position dependence, which is now found in the second formerly diffusive contribution. The diffusive term has changed more drastically and can no longer exclusively be interpreted as a non-linear diffusion flux. In Refs.~\cite{Koenigstein:2021syz,Koenigstein:2021rxj,Stoll:2021ori} we argue at length, that, due to several reasons, we currently believe that a formulation in $x$ instead of $y$ is favorable as soon as we allow for diffusive contributions to the RG flow -- hence at finite $N$. In Sec.~IV~D of Ref.~\cite{Koenigstein:2021syz} we discuss the difficulties arising when attempting to formulate the inevitable spatial boundary condition at $y = 0$, when using the (rescaled) invariant $y$. An oversimplified argument is that there is no physical meaning of negative values of $y$, which makes a correct formulation of a boundary condition, that correctly captures possible influx due to diffusion, extremely challenging -- if not impossible. In a formulation in $x$, this is not a problem at all, since negative $x$ formally exist and anti-symmetric boundary conditions can be used for $u ( t, x )$ at $x = 0$ \cite{Koenigstein:2021syz,Koenigstein:2021rxj,Stoll:2021ori}. Additionally, it is understandable that a sober split of advection and diffusion fluxes is no longer possible in $y$, by simply executing the total $y$-derivative on the \textit{r.h.s.}\ of Eq.~\eqref{eq:frg_flow_y} for the last term. Hence, as long as $N$ is finite, one has to live with the challenging $x$-dependence in the advection flux of the PDE \eqref{eq:frg_flow_x}, which can however be handled by suitable discretizations, as demonstrated at length in part I of this series of publications \cite{Koenigstein:2021syz}.

	However, in the infinite-$N$ limit the second term of the PDE \eqref{eq:frg_flow_y} vanishes and the problem again reduces to a hyperbolic non-linear advection equation -- without any explicit position dependences,
		\begin{align}
			\partial_t v ( t, y ) = \, & \frac{\mathrm{d}}{\mathrm{d} y} \bigg[ \frac{\tfrac{1}{2} \, \partial_t r ( t )}{r ( t ) + v ( t, y )} \bigg]\equiv - \tfrac{\mathrm{d}}{\mathrm{d} y} \, G [ t, v ] \, . \label{eq:frg_flow_Ninf_y}
		\end{align}
	Because the newly defined advection flux $G [ t, v ]$ has manifestly negative sign\footnote{For all $y\in\mathbb{R}^+$ and $t\in\mathbb{R}^+$ $G [ t, v ] < 0$, since $\partial_t r ( t ) <0$, holds for all well-defined initial conditions, which realize $r ( t ) + v ( t, y ) > 0$ at $t=0$. The latter inequality is guaranteed dynamically at $t > 0$ by the flow equation as long as it is realized in the UV at the initial scale $t = 0$, \textit{cf.}\ part I of this series of publications \cite{Koenigstein:2021syz}.}, there can not be any influx at $y = 0$ into the spatial domain $y \in [ 0, \infty )$ of the problem resolving the conceptual issues with the $y = 0$ boundary and allowing practical computations in the rescaled invariant $y$.
	
	Though, as is explained in detail in App.~\ref{app:flow_equation_in_rho} there is another remaining pitfall in this formulation: Using equidistant discretizations of the computational domain in $y$ implies a very low spatial resolution at small field values $x=\tfrac{1}{\sqrt{N}} \, \sigma$. This becomes relevant for computations close to $a_\mathrm{c}$ within our testing scenario or in generic higher-dimensional models in their symmetric phase.\\
	
	In summary, for the scope of this work, we will stick to version \eqref{eq:frg_flow_x} and \eqref{eq:frg_flow_Ninf_x} of the RG flow equations in the main text and results using Eq.~\eqref{eq:frg_flow_Ninf_y} are only presented in App.~\ref{app:flow_equation_in_rho}.
	
\subsection{UV initial condition}
	
	As explained above, the RG flows for $V$ or $v$ respectively need to be initialized with the UV potential or rather its spatial derivative.\\
	
	For the infinite-$N$ RG flow equation \eqref{eq:frg_flow_Ninf_y} formulated in $y$, the UV initial condition for $v ( t, y )$ is given by Eq.~\eqref{eq:RP_vofy}. On the level of the fluid-dynamic reformulation of the RG flow in $y$, \textit{cf.} App.~\ref{app:RPandEntropy}, Eq.~\eqref{eq:frg_flow_Ninf_y} together with the initial condition \eqref{eq:RP_vofy} present as two Riemann problems \cite{LeVeque:1992,LeVeque:2002,Hesthaven2007,RezzollaZanotti:2013,Grossi:2019urj} involving jump discontinuities at $y=2$ and $y=8$, \textit{cf.} App.~\ref{app:RPandEntropy}.\\
		\begin{figure}
			\centering
			\includegraphics{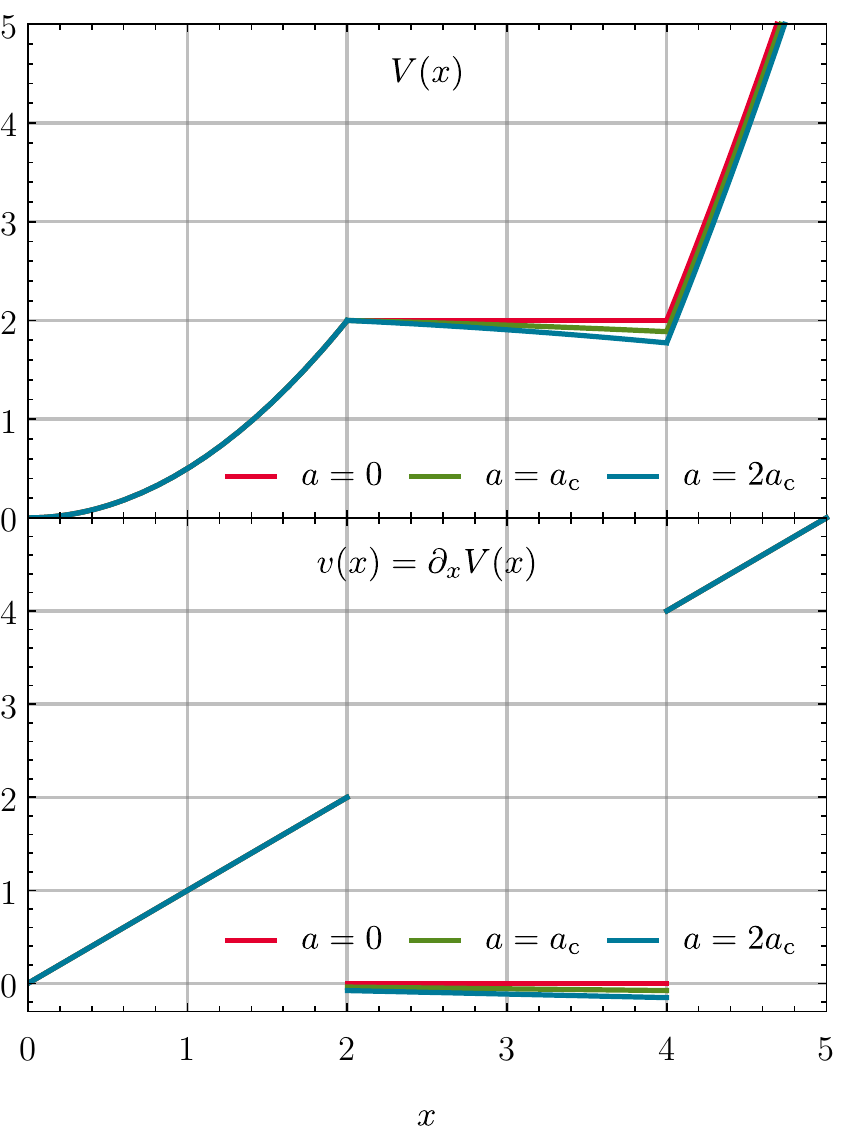}
			\caption{\label{fig:Vofx}%
				The potential $V ( x )$ from Eq.~\eqref{eq:RP_Vofx} (upper panel) and its $x$-derivative $v ( x ) = \partial_x V ( x )$ from Eq.~\eqref{eq:RP_vofx} (lower panel) for different values of the parameter $a$ and where $a_\mathrm{c}$ is given by Eq.~\eqref{eq:a_critical}.
			}
		\end{figure}
	
	For the RG flow equations \eqref{eq:frg_flow_x} and \eqref{eq:frg_flow_Ninf_x} the initial conditions \eqref{eq:RP_Vofy} and \eqref{eq:RP_vofy} have to be transformed to the variable $x$. For the one parameter family of UV potentials, this reads
		\begin{align}
			V ( x ) =
			\begin{cases}
				\tfrac{1}{2} \, x^2							&	\text{for} \quad |x| \leq 2 \, ,	\vphantom{\bigg(\bigg)}
				\\
				- a \, \tfrac{1}{2} \, x^2 + 2 \, ( a + 1 )	&	\text{for} \quad 2 < |x| \leq 4 \, ,	\vphantom{\bigg(\bigg)}
				\\
				\tfrac{1}{2} \, x^2 - 6 \, ( a + 1 )		&	\text{for} \quad 4 < |x| \, .	\vphantom{\bigg(\bigg)}
			\end{cases}\label{eq:RP_Vofx}
		\end{align}
	Hence, our UV potential is actually a piecewise quadratic function of $x = \tfrac{1}{\sqrt{N}} \, \sigma$, while its $x$-derivative is given by the piecewise linear function
		\begin{align}
			v ( x ) = \partial_x V ( x ) =
			\begin{cases}
				x		&	\text{for} \quad |x| \leq 2 \, ,	\vphantom{\bigg(\bigg)}
				\\
				- a \, x	&	\text{for} \quad 2 < |x| \leq 4 \, ,	\vphantom{\bigg(\bigg)}
				\\
				x		&	\text{for} \quad 4 < |x| \, .	\vphantom{\bigg(\bigg)}
			\end{cases} \label{eq:RP_vofx}
		\end{align}
	For illustrative purpose, we plot $v ( x )$ and $V ( x )$ in Fig.~\ref{fig:Vofx} for selected values of $a$.

\subsection{RG flows at infinite \texorpdfstring{$N$}{N} -- shocks and rarefaction waves in advective flows}
\label{subsec:FRGlargeN}

	Next, we turn to the results for the RG flows for Eq.~\eqref{eq:RP_vofx} in the limit $N \rightarrow \infty$. Before presenting the numerical results, which are obtained by a numerical solution of the PDE \eqref{eq:frg_flow_Ninf_x} with the KT scheme \cite{KTO2-0}, we use the so-called \textit{method of characteristics} to discuss analytic results for solutions of the purely hyperbolic conservation law \eqref{eq:frg_flow_Ninf_x}. This helps to better understand the underlying processes in the fluid-dynamical framework and the results of our numeric calculations.
	
	In the FRG framework the method of characteristics was used by N. Tetradis and D.~Litim in Refs. \cite{Tetradis:1995br,Litim:1995ex} to obtain analytical solutions to FRG flow equations of the $O(N)$~model in dimensions $d > 0$ in the infinite-$N$ limit. K.-I. Aoki, S.-I. Kumamoto, D. Sato, and M. Yamada also used the method of characteristics and the Rankine-Hugoniot condition in their studies~\cite{Aoki:2014,Aoki:2017rjl} of weak solutions and dynamical symmetry breaking. Unfortunately (or luckily for the authors of Refs.~\cite{Grossi:2019urj,Grossi:2021ksl,Koenigstein:2021syz,Koenigstein:2021rxj,Steil:partIV,Stoll:2021ori}), their otherwise remarkable work lacks the fluid dynamical interpretation and with it an instructive way to understand characteristic curves in this context. The latter was put forward in the context of the FRG for the first time in Ref.~\cite{Grossi:2019urj}.

\subsubsection{Characteristic curves}

\begin{figure}
			\centering
			\includegraphics{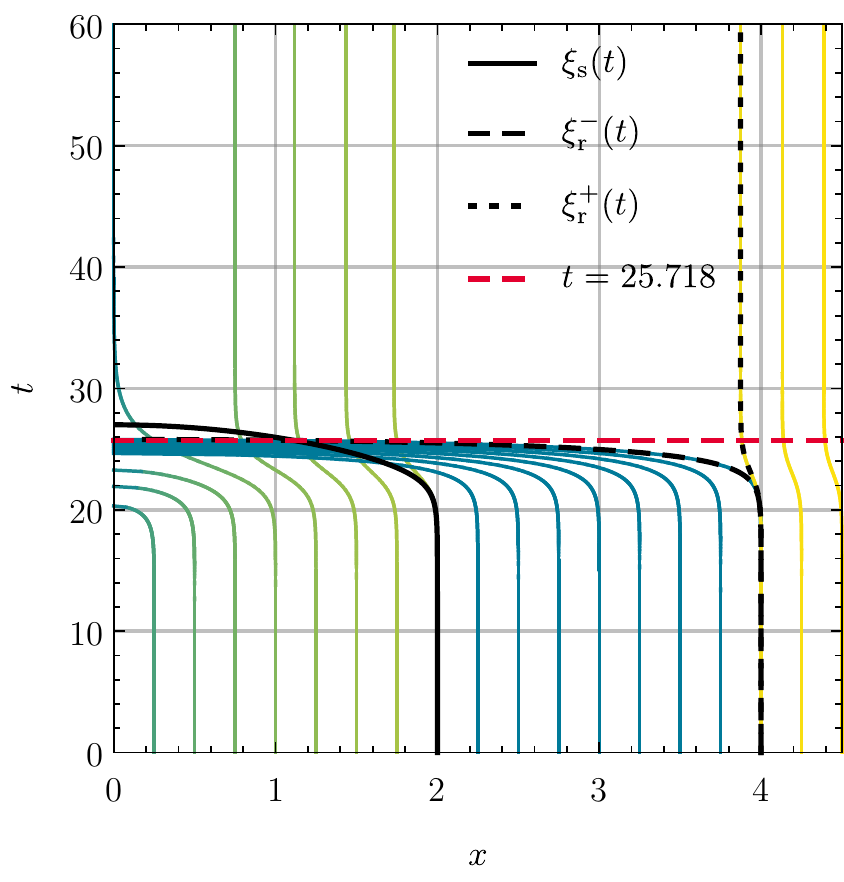}
			\caption{\label{fig:frg_largeNChar}%
				Selected characteristic curves $(t,x(t))$, see Eq.~\eqref{eq:MoC_xoftau}, for $a=0$ and $\Lambda=10^{10}$ in {blue}, {green} and {yellow}, shock position $\xi_\mathrm{s} ( t )$, see Eq.~\eqref{eq:xioft}, as solid black line, and the tips of the rarefaction fan $\xi_\mathrm{r}^\mp ( t )$, see Eqs.~\eqref{eq:rarefaction_minus} and \eqref{eq:rarefaction_plus} originating at $( t = 0, \xi_\mathrm{r}^\mp ( 0 ) = 4 )$ as dashed black lines. The changing color on the characteristic curves indicates the change of $v(t,x(t))$ along them, see Eq.~\eqref{eq:MoC_vxoftau}, where {blue} corresponds to $v ( t, x ( t ) ) = 0$ and {yellow} corresponds to $v ( t, x ( t ) ) = 4.5$. The shock wave and the rarefaction fan collide at $( t, x ) \approx ( 25.718, 1.115 )$ (the time is marked with the {red-dashed} line) rendering the expressions $\xi_\mathrm{r}^\pm ( t )$ and $\xi_\mathrm{s} ( t )$ as well as the characteristics that intersect with the shock and rarefaction wave invalid for later times.
			}
		\end{figure}

	In Fig.~\ref{fig:frg_largeNChar} we plot the characteristic curves of the fluid. These are defined as those (parametric) curves $( t, x ( t ) )$ in the domain $[ 0, \infty ) \times ( - \infty, + \infty )$ of the problem, where the ratio $\frac{v ( t, x ( t ))}{x ( t )}$ stays constant\footnote{If formulated in terms of the $\frac{1}{N}$-rescaled invariant $y$, these are the (parametric) curves $( t, y ( t ) )$ on $[ 0, \infty ) \times [ 0, \infty )$, where $v ( t, y ( t ))$ is constant, see, \textit{e.g.}, Ref.~\cite{Grossi:2019urj}. Both formulations can be transformed into each other by simple coordinate transformations, see App.~\ref{app:method_of_characteristics}.}. In App.~\ref{app:method_of_characteristics} we derive these implicit analytic solutions for the PDE \eqref{eq:frg_flow_Ninf_x} with initial condition \eqref{eq:RP_vofx} in great detail and the explicit solutions for $x ( t )$ and $v ( t, x ( t ))$ are given by Eqs.~\eqref{eq:MoC_xoftau} and \eqref{eq:MoC_vxoftau}. If needed, $x ( t )$ and $v ( t, x ( t ))$ can be used to reconstruct the full solution of the PDE, $v ( t, x )$, for $t \in [ 0, \infty )$ and $x \in ( - \infty, + \infty )$, which usually needs to be done numerically since the involved expressions can usually not be inverted analytically. Though, this method only works as long as the solution $v ( t, x )$ is not multi-valued, which means that it is valid until any characteristics intersect at some point $x$ in position space  (here field space). Once the analytical solution becomes multi-valued, the physical solution exists only in a weak sense, see, \textit{e.g.} Refs.~\cite{Ames:1992,LeVeque:1992,LeVeque:2002,Hesthaven2007,RezzollaZanotti:2013} for details. Intersecting characteristics correspond to the formation of a shock wave, since several fluid elements are approaching the same point in the spatial domain at different velocities \cite{Ames:1992,LeVeque:1992,LeVeque:2002,Hesthaven2007,RezzollaZanotti:2013}. The movement of this shock wave, its (parametric) curve $( t, \xi_\mathrm{s} ( t ) )$ is described by the Rankine-Hugoniot shock condition \cite{Rankine:1870,Hugoniot:1887}. A derivation is presented in App.~\ref{app:rankine-hugoniot_condition_and_shock_position}. On the other hand, there might also be positions in field space that ``separate'' the characteristic curves into distinct regimes and that are the origin of infinitely many characteristic curves. These are so-called \textit{rarefaction waves}, which each cause a \textit{rarefaction fan} of infinitely many characteristic curves. As their name suggests, they are associated to points $x$ (in field space), where fluid elements are moving apart from each other and cause a rarefaction of the fluid (in physical fluids corresponding to a reduction of density as a direct opposite of a compression wave). A rarefaction fan can be described by the spatially closest characteristic curves that are moving to the left ($-$) and right ($+$) apart from each other, $( t, \xi^\mp_\mathrm{r} ( t ) )$.\\
	
	Now we are equipped with the vocabulary to efficiently interpret and analyze Fig.~\ref{fig:frg_largeNChar}. \textit{W.l.o.g.}\ we choose the initial condition \eqref{eq:RP_vofx} with $a = 0$. (The plots and the discussion for different choices of $a$ are qualitatively very similar.) Furthermore, we only restrict our plot of the characteristics and parts of the discussion to positive $x$. For negative $x$ the dynamics is perfectly mirrored about the $t$ axis in Fig.~\ref{fig:frg_largeNChar}.
	
	The initial condition $v ( t = 0, x ) = v ( x )$ corresponds to the initial values of $v ( t, x ( t ) )$ on the characteristic curves at $t = 0$ along the $x$-axis. The color-coding indicates the value of $v ( t, x ( t ) )$ according to Eq.~\eqref{eq:MoC_ode_y_3} along the curves $( t, x ( t ) )$, where {blue} corresponds to $v ( t, x ( t ) ) = 0$ and {yellow} corresponds to $v ( t, x ( t ) ) = 4.5$. 
	
	Firstly and in general, we observe that all characteristic curves only move towards smaller $| x |$, while $v ( t, x ( t ) )$ only decreases (increases) along each characteristic curve at positive (negative) $x$. This implies that the fluid $v ( t, x)$ only moves towards $x = 0$. This can already be seen from the manifestly (positive) negative sign of the local fluid velocity $\partial_v F [ t, x, v ]$ for (negative) positive $x$, \textit{cf.} Eqs.~\eqref{eq:advection_flux} and \eqref{eq:frg_flow_Ninf_y} or our discussion in Ref.~\cite{Koenigstein:2021syz}. Hence, we find that right moving waves of the fluid from negative $x$ and left moving waves of the fluid from positive $x$ annihilate in $x = 0$, which is also manifestly encoded in the anti-symmetry $v ( t, x ) = - v ( t, - x)$.
	
	Secondly we observe that the fluid elements, which start off in the interval $2 < | x | < 4$, move faster towards $x = 0$ than the fluid elements, that start at $| x | < 2$. As soon as the former try to overtake the latter, the solution gets multi-valued and a shock forms. Actually, this happens already at $t = 0$, but we can also see how more and more characteristics ``join'' and ``accelerate'' the shock wave. The movement of the shock wave, $( t, \xi_\mathrm{s} ( t ))$ is described analytically by Eq.~\eqref{eq:xioft} and depicted as a black solid line in Fig.~\ref{fig:frg_largeNChar}.
	
	Thirdly, there is another important phenomenon going on about $| x | = 4$. We find that fluid elements at $| x | < 4$ are traveling fast towards $x = 0$, while the characteristic curves that start at $| x | > 4$ move slower towards $x = 0$ and that only for a very short period of (RG) time, before the characteristic curve closest to $|x| = 4$ freezes at ${| x | \simeq \sqrt{15} \approx 3.873}$. This effectively causes a rarefaction wave in $v ( t, x )$, which is described analytically by
		\begin{align}
			\xi_\mathrm{r}^- ( t ) = \, & \pm \sqrt{16 - \frac{1}{\Lambda \, \mathrm{e}^{- t} - a} + \frac{1}{\Lambda - a} } \, ,	\vphantom{\Bigg(\Bigg)}	\label{eq:rarefaction_minus}
			\\
			v_\mathrm{r}^- ( t ) = \, & - a \, \xi_\mathrm{r}^- ( t ) \, ,	\vphantom{\Bigg(\Bigg)}
			\\
			\xi_\mathrm{r}^+ ( t ) = \, & \pm \sqrt{16 - \frac{1}{\Lambda \, \mathrm{e}^{- t} + 1} + \frac{1}{\Lambda + 1} } \, ,	\vphantom{\Bigg(\Bigg)}	\label{eq:rarefaction_plus}
			\\
			v_\mathrm{r}^+ ( t ) = \, & \xi_\mathrm{r}^+ ( t ) \, ,	\vphantom{\Bigg(\Bigg)}
		\end{align}
	where $v_\mathrm{r}^\mp ( t )$ are the values of of the fluid at the edges of the rarefaction fan. The rarefaction fan is marked in Fig.~\ref{fig:frg_largeNChar} by black-dashed lines that are analytically described by Eqs.~\eqref{eq:rarefaction_minus} and \eqref{eq:rarefaction_plus}. The rarefaction wave also forms already at $t = 0$.\\
	
	Interestingly, there is a (RG) time and field space position $( t, x ) \simeq ( 25.718, 1.115 )$, where the rarefaction fan catches up the shock wave (indicated by the {red-dashed} horizontal line). Up to this point, our analytical solutions for the shock $\xi_\mathrm{s} ( t )$ and the left tip of the rarefaction wave $\xi_\mathrm{r}^- ( t )$ are valid and we could in principle even integrate backwards in (RG) time and reconstruct the UV potential. However, when the shock and the rarefaction wave meet and interact, some highly non-linear dynamics is going on and we can no longer trust our analytical solutions. At later (RG) times, we totally have to rely on adequate numerical solutions.
	
	Interestingly, it is exactly this complicated non-analytic dynamics, which makes the RG flow manifestly irreversible and produces some abstract form of entropy, see App.~\ref{app:flow_equation_in_rho} and especially Fig.~\ref{fig:frg_largeN_TVD} as well as Refs.~\cite{Koenigstein:2021syz,Koenigstein:2021rxj}, because information about the UV initial potential is unavoidably lost\footnote{One might also argue that an infinite number of new couplings or interaction vertices is generated at this point.}. Actually, this is the dynamics that fundamentally encodes the irreversibility of RG transformations on the level of the PDE, \textit{cf.}\ also Refs.~\cite{Wilson:1979qg,Zumbach:1994vg,Zamolodchikov:1986gt} for similar discussions.
	
	However, most remarkably in the context of the infinite-$N$ limit: We find numerically that it is the complicated interplay between the shock and rarefaction waves (at positive and negative $x$), which either causes the shock waves to freeze at some non-zero $|x|$ or to crash into each other and annihilate in $x = 0$, depending on the choice of $a$ -- smaller, equal, or greater than $a_\mathrm{c}$. This means that the (non\nobreakdash-)applicability of the large-$N$ saddle point expansion of Sec.~\ref{sec:saddle_point}, which was caused by a (non\nobreakdash-)analytic ``expansion point'' -- the underlying first-order phase transition, translates into freezing or the annihilation shock waves in field space in RG flow equations. For further details on the relation between first-order phase transitions and the interaction/freezing of shock and rarefaction waves we refer the interested reader to Sub.Sec.~III~C of Ref.~\cite{Grossi:2019urj}. However, to proceed with this discussion and to understand this interrelation, we have to leave the sure ground of analytical solutions and turn to high precision numerical computations of this challenging dynamics.
	
\subsubsection{Numerical results at infinite \texorpdfstring{$N$}{N}}\label{subsubsec:infiniteNflows}

	Next, we apply the KT scheme \cite{KTO2-0} from numerical fluid dynamics to the problem posed by the PDE \eqref{eq:frg_flow_Ninf_x} with initial condition \eqref{eq:RP_vofx}. All details on the numeric implementation are provided in Ref.~\cite{Koenigstein:2021syz}. The corresponding (numerical) parameters are either incorporated in the figures or their corresponding captions. Additionally, we discuss the choice of some of our (numeric) parameters and some aspects of the implementation in App.~\ref{app:FRGnumerics}.\\
	
\begin{figure}
			\centering
			\includegraphics{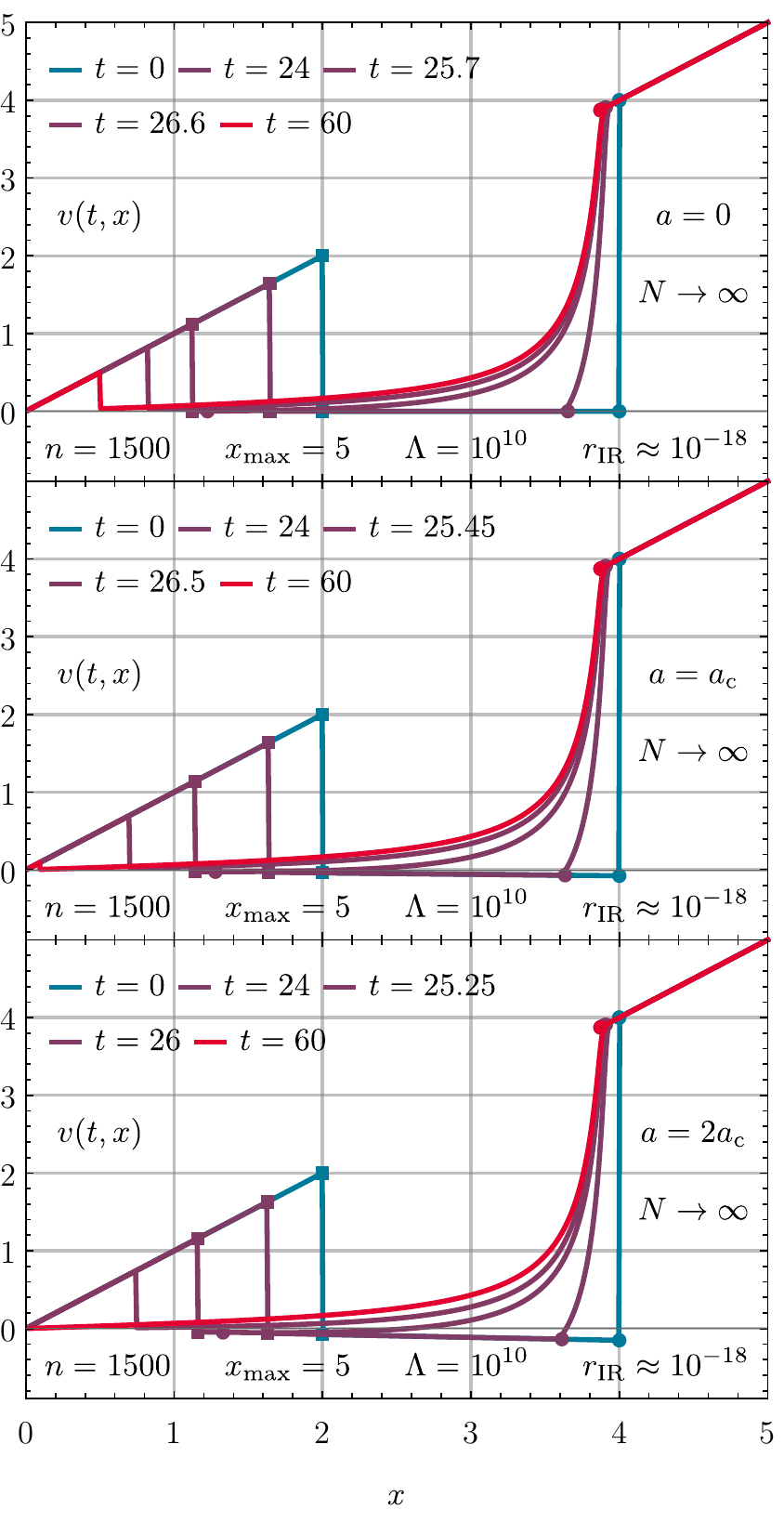}
			\caption{\label{fig:frg_largeN_flows}%
				The RG flow of the derivative of the rescaled effective potential $v ( t, x )$ for the zero dimensional $O ( N )$-model in the limit $N \rightarrow \infty$ for the initial condition \eqref{eq:RP_vofx} with $a = 0$, $a = a_\mathrm{c}$ and $a = 2 a_\mathrm{c}$ in the upper, middle and lower panel respectively. {Blue} curves represent the UV initial conditions at $t=0$, {red} curves correspond to the IR potentials at $t = 60$ and the {violet} curves are at intermediate, selected RG times $t$ chosen around the respective collision of the shock $\xi_\mathrm{s} ( t ) $ with the left tip of the rarefaction fan $\xi_\mathrm{r}^{-}(t)$. The squares mark the shock $( \xi_\mathrm{s} ( t ), v ( t, \xi_\mathrm{s} ( t )^\pm ) )$, while the disks mark the tips of the rarefaction fan $( \xi_\mathrm{r}^\pm ( t ), v ( t, \xi_\mathrm{r}^\pm ( t ) ) )$. The left tip of the rarefaction fan and the shock are only marked up to the RG time when they meet since the underlying analysis based on the method of characteristics and Rankine-Hugoniot condition breaks down after their collision.
			} 
		\end{figure}

	We obtain the following numeric results for the RG flows of $v ( t, x )$: In Fig.~\ref{fig:frg_largeN_flows} we plot the RG flow of $v ( t, x )$ from the UV initial condition \eqref{eq:RP_vofx} (see Fig.~\ref{fig:Vofx}) at $t = 0$ to the IR at $t \rightarrow \infty$. Of course, for practical (numerical) calculations one has to stop the integration at some finite $t$ in the IR\footnote{Zero-dimensional RG is exceptional and it is actually not needed to introduce a numerical IR cutoff $r_\mathrm{IR}$ \cite{Keitel:2011pn,Koenigstein:2021syz}, which can be seen by simple reparametrization of the RG time. Nevertheless, we use an extremely small IR cutoff, to be as close as possible to higher dimensional applications of the FRG-fluid dynamic framework.}. Here we chose $t = 60$, which corresponds to an IR cutoff $r_\mathrm{IR} \approx 10^{-18}$, which is 18 orders of magnitude below model scales (which are considered to be of order one in $\tfrac{1}{N}$-rescaled quantities). Our UV cutoff $\Lambda$ was chosen to be ten orders of magnitude above model scales to guarantee RG consistency \cite{Braun:2018svj,Koenigstein:2021syz} to a sufficient level. In total, we are integrating over 28 orders of magnitude in the regulator scale and corresponding tests for UV-cutoff independence are presented in App.~\ref{app:FRGnumerics}\footnote{Again, zero-dimensional QFTs are special. Due to their ultra locality, they are extremely coupled in field space, which makes an RG flow over several orders of magnitude unavoidable to reach a freeze out of all dynamics. In higher-dimensional calculations, we expect this problem to be less severe, due to increasing importance of momentum-dependences of vertices and an increasing phase space, see also our discussion in Ref.~\cite{Koenigstein:2021syz}. However, in our parallel work on the $(1 + 1)$-dimensional Gross-Neveu model \cite{Stoll:2021ori}, we find that also in a realistic QFT an integration over several orders of magnitude is needed to capture all relevant physical effects.}.\\
	
	Figure~\ref{fig:frg_largeN_flows} shows RG flows for $v ( t, x )$ for different values of $a$. In the upper panel $a = 0$ and therefore clearly below $a_\mathrm{c}$, such that this RG flow corresponds to the situation, where the $\tfrac{1}{N}$-expansion is applicable. The middle panel shows the RG flow exactly at the threshold $a = a_\mathrm{c}$, where the exponent \eqref{eq:fofy} has two degenerate minima \eqref{eq:saddle_minima}, with one being a non-analytic point, preventing a saddle point expansion. The bottom panel in Fig.~\ref{fig:frg_largeN_flows} corresponds to a situation, where $a > a_\mathrm{c}$ and the saddle-point expansion again fails as it is not applicable to this initial condition.
	
	As already mentioned at the end of the previous subsection, we find that the different situations within the saddle-point expansion are realized by freezing or colliding and annihilating shock waves, caused by the interplay with the rarefaction wave. This is clearly seen in Fig.~\ref{fig:frg_largeN_flows}, where the position $\xi_\mathrm{s} ( t )$ of the shock wave is marked with squares and the positions $\xi_\mathrm{r}^- ( t )$ and $\xi_\mathrm{r}^+ ( t )$ of the tips of the rarefaction fan are marked with disks -- up to the RG time, where they meet and interact rendering the analytic expressions invalid.
	
	Explicitly, we find that for $a = 0$ (upper panel Fig.~\ref{fig:frg_largeN_flows}) the opposing shock waves ultimately freeze at ${| x | = |\xi_\mathrm{s} ( t = 60 )| \approx 0.496}$. We obtained this value using computations at different numerical spatial resolutions $\Delta x$ by varying the number of volume cells $n$ while keeping the computational extend fixed to $x\in[0,5]$. The explicit value of $| x | \approx 0.496$ has been extracted from the fit
		\begin{align}
			|\xi_\mathrm{s} ( t \rightarrow \infty )| \approx |\xi_\mathrm{s} ( t = 60 )| =0.496 + 0.788\, \Delta x^{0.869}\, .\label{eq:xis_0ac_fit}
		\end{align}
	obtained from 41 data points with $n$ varying between $64$ and $2048$. The non-vanishing value of ${|\xi_\mathrm{s} ( t \rightarrow \infty )|\approx 0.496}$  has the effect that the $x$-derivatives of $v ( t, x )$ at ${x = 0}$ never change during the RG flow and ${\partial_x v ( t, x ) \big|_{x = 0} = 1}$ for all times $t$, while all higher $x$-derivatives vanish. Yet, these derivatives are in direct correspondence to the $1$PI-correlation functions $\Gamma^{(n)}$, which are extracted from $v ( t, x )$ in the IR at the physical point ${x = 0}$ by differentiation \textit{w.r.t.}\ $x$,
		\begin{align}
			N^{\frac{n - 1}{2}} \, \Gamma^{(n + 1)} = \partial_x^n v ( t,  x ) \big|_{t \rightarrow \infty, x = 0} \, .	\label{eq:1pi_ir}
		\end{align}
	Hence, although having highly non-linear dynamics involving the interaction shocks and rarefaction waves for ${|x| \gtrsim 0.496}$, the function $v ( t, x )$ never changed its shape for ${- 0.496 \lesssim x \lesssim 0.496}$ and always resembles a massive free QFT in this part of field space. Metaphorically speaking and to stay in the fluid dynamic picture: It is as if the physical point $x = 0$ in field space is ``sitting in the eye of a cyclone''.
	
	Increasing $a$ towards the critical threshold $a_\mathrm{c}$ one observes that the shock waves freeze closer and closer to $x = 0$. Considering the metaphor of the previous paragraph, as $a$ approaches $a_\mathrm{c}$ from below the radius of the eye of the cyclone vanishes. At $a = a_\mathrm{c}$ (middle panel Fig.~\ref{fig:frg_largeN_flows}) one still observes a freezing of the shock wave in the IR at $|x| \approx 0.095$, which however is an artifact of the finite spatial resolution $\Delta x$ of the numerical scheme. This effect can be removed by successively decreasing the finite-volume computational cells $\Delta x$. We find that for $a = a_\mathrm{c}$ the shock freezes at $x = 0$, because the shock position in the IR scales as follows with $\Delta x$ for this situation,
		\begin{align}
			|\xi_\mathrm{s} ( t \rightarrow \infty )| \approx |\xi_\mathrm{s} ( t = 60 )| = 0.983\, \Delta x^{0.413}\, ,\label{eq:xis_1ac_fit}
		\end{align}
	again obtained from a fit to 41 data points with the number of volume cells $n$ varying between $64$ and $2048$ while keeping $x_\mathrm{max}$ fixed.
	
	However, as soon as $a > a_\mathrm{c}$ (middle panel Fig.~\ref{fig:frg_largeN_flows}) the interplay of the rarefaction waves and the shock waves no longer hinders the shock waves to collide and annihilate at $x = 0$. In turn, this has two direct consequences: Firstly, in the hydrodynamic language, the additional interaction of two discontinuities (the annihilation of the shock waves) again unavoidably leads to a loss of information and an abstract production of entropy on the level of the PDE. This is discussed in App.~\ref{app:flow_equation_in_rho}. Secondly, in the quantum field theoretical picture the annihilation of the shock waves caused a change in the slope of $v ( t, x )$ at the physical point $x = 0$. This directly affects the $1$PI-correlation functions, which are again extracted in the IR via Eq.~\eqref{eq:1pi_ir}. Indeed, we find that our numeric calculations reproduce the exact results \eqref{eq:two-point_exact} and \eqref{eq:n-point_exact}.
	
	In summary and again metaphorically speaking, the slight change in the slope $a$ of the initial condition \eqref{eq:RP_vofx} at $t=0$ on the interval $x \in [ 2, 4 ]$ causes a tremendous change of the non-linear dynamics of the fluid $v ( t, x )$, also at other positions in field space and later RG times, which can be seen as a ``butterfly effect'' in a QFT. The small deviations in the initial condition in the UV -- in the metaphor the minor perturbations caused by a distant butterfly flapping its wings  -- have tremendous impact on the solution in the IR at the physical point -- whether or not the formed cyclone has an eye or not. This further supports the notion of a first-order phase transition at $a_\mathrm{c}$ and the corresponding mechanism discussed in Ref.~\cite{Grossi:2019urj}.\\
	
\begin{figure}
			\centering
			\includegraphics{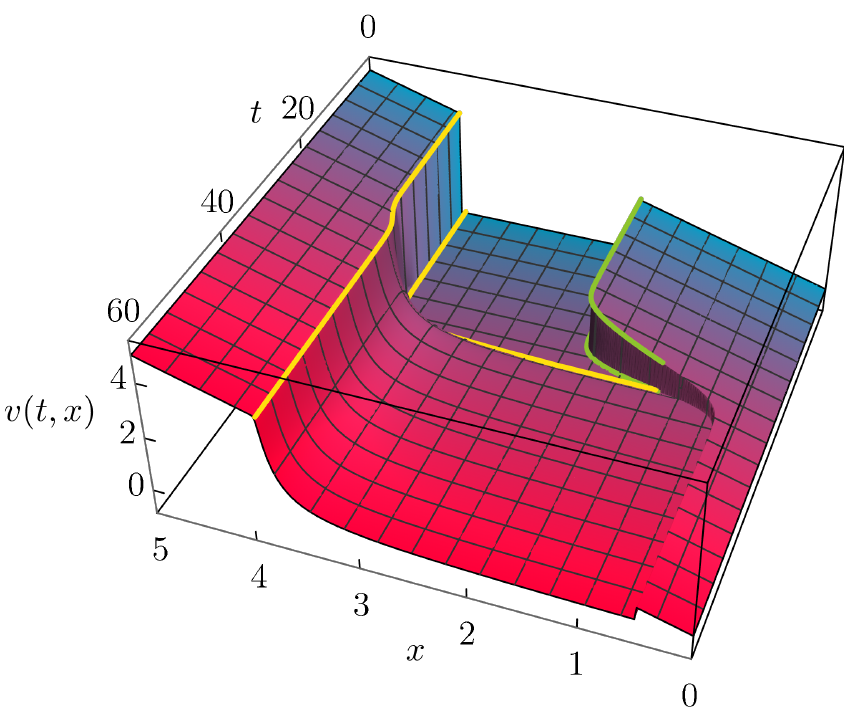}
			\caption{\label{fig:frg_largeN_0_flow3D}%
				RG flow of $v(t,x)$ for $a=0$ as 3D-plot corresponding to the flow displayed in the upper panel of Fig.~\ref{fig:frg_largeN_flows}. The left and right tips $( \xi_\mathrm{r}^\mp ( t ), t, v( t, \xi_\mathrm{r}^\mp ( t ) ) )$ of the rarefaction fan are plotted as {yellow} lines while the the shock $( \xi_\mathrm{s} ( t ), t, v ( t, \xi_\mathrm{s} ( t )^\pm ) )$ is marked with {green} lines. The left tip of the rarefaction fan and the shock are only marked up to $( t, x ) \approx ( 25.718, 1.115 )$, where they meet and the analysis based on the method of characteristics and Rankine-Hugoniot condition breaks down.
			}
		\end{figure}

\begin{figure}
			\centering
			\includegraphics{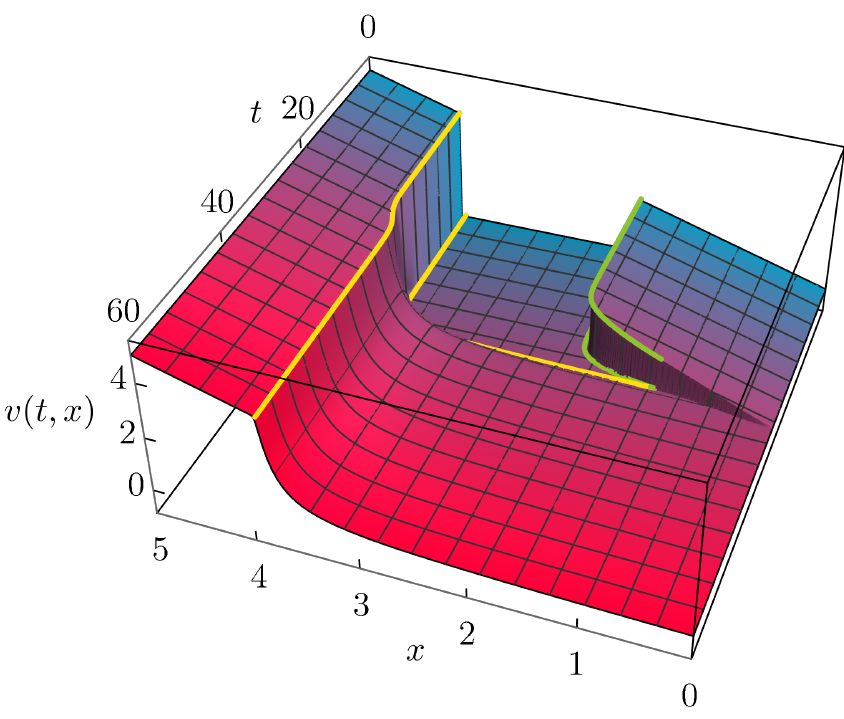}
			\caption{\label{fig:frg_largeN_2ac_flow3D}%
				RG flow of $v ( t, x )$ for $a = 2 a_\mathrm{c}$ as 3D-plot corresponding to the flow displayed in the lower panel of Fig.~\ref{fig:frg_largeN_flows}. The left and right tips $( \xi_\mathrm{r}^\mp ( t ), t, v ( t, \xi_\mathrm{r}^\mp ( t ) ) )$ of the rarefaction fan are plotted as {yellow} lines while the the shock $( \xi_\mathrm{s} ( t ), t, v ( t, \xi_\mathrm{s} ( t )^\pm ) )$ is marked with {green} lines. The left tip of the rarefaction fan and the shock are only marked up to $( t, x ) \approx ( 25.270, 1.146 )$, where they meet and the analysis based on the method of characteristics and Rankine-Hugoniot condition breaks down.
			}
		\end{figure}

	For better visualization of this dynamics, we present two supplemental 3D-plots for the RG flows of the upper and bottom panel of Fig.~\ref{fig:frg_largeN_flows}. The curves from Fig.~\ref{fig:frg_largeN_flows} are slices of constant intermediate times of the 3D-plots in Figs.~\ref{fig:frg_largeN_0_flow3D} and \ref{fig:frg_largeN_2ac_flow3D}. The color coding of all figures is identical.\\

	In addition to this rather qualitative discussion, we also provide explicit numerical errors, which can be used to judge to quality of the KT scheme \cite{KTO2-0} and our implementation in the context of RG flows. In Table~\ref{tab:KT1500errorsLargeN} we list the relative errors of the $1$PI-two point function $\Gamma^{(2)}$ extracted from the numerical RG flows of $v ( t, x )$ using Eq.~\eqref{eq:Gamma2vofx} using the exact results \eqref{eq:two-point_exact} and \eqref{eq:n-point_exact} as reference values.\\
		\begin{table}[b]
			\caption{\label{tab:KT1500errorsLargeN}%
				Relative numerical errors for the $1$PI-two-point function, see Eq.~\eqref{eq:Gamma2vofx}, for the results plotted in Fig.~\ref{fig:frg_largeN_flows} with corresponding exact reference values from the last row of Tab.~\ref{tab:Gamma2N}. The scaling of these errors with the number of volume cells can be found in Tab.~\ref{tab:frg_largeN_2ac_rates} for $a=2\,a_\mathrm{c}$.
			}
			\begin{ruledtabular}
				\begin{tabular}{l c c c}
					$N$			&	$a=0$					&	$a=a_\mathrm{c}$		&	$a = 2 a_\mathrm{c}$
					\\
					\colrule\addlinespace[0.25em]
					$\infty$	&	$7.994 \cdot 10^{-15}$	&	$1.199 \cdot 10^{-14}$	&	$3.333 \cdot 10^{-3}$
				\end{tabular}
			\end{ruledtabular}
		\end{table}
	
	We close our discussion on the analysis of the infinite-$N$ RG flows by noting that, in contrast to the $\tfrac{1}{N}$-saddle point expansion or perturbative methods, the FRG in its fluid dynamic framework is applicable and also produces reliable results in a highly non-perturbative regime. Furthermore, the FRG-fluid dynamic framework, naturally copes with different kinds of non-analyticities, while all kind of ``expansion-type'' methods\footnote{Also ``expansion-type'' schemes in the FRG framework, like a Taylor expansion of the local potential, tend to collapse \cite{Koenigstein:2021syz}, due to the Wilbraham-Gibbs phenomenon \cite{Wilbraham:1848,Gibbs:1898,Gibbs:1899,boyd2001chebyshev}, which is a well-known issue from the field of signal processing.} tend to collapse in the vicinity of relevant non-analytical physics that is only correctly described by fully fledged non-perturbative setups.
		
\subsection{RG flows at finite \texorpdfstring{$N$}{N} -- diffusion as a game changer}\label{subsec:FRGfiniteN}

	Next, we turn to the RG flows of our initial potential \eqref{eq:RP_Vofy} at finite $N$. To this end, we use the fluid-dynamic RG flow equation \eqref{eq:frg_flow_x} including advective and diffusive contributions by the pions and the $\sigma$-mode. As explained above, we cannot use Eq.~\eqref{eq:frg_flow_y} in the presence of diffusion, because the problem of diffusive influx at the $(y = 0)$-boundary, if formulated in $y$, is not settled yet \cite{Koenigstein:2021syz}. However, the KT scheme \cite{KTO2-0} with our FRG adapted boundary conditions in $x$ \cite{Koenigstein:2021syz,Koenigstein:2021rxj,Stoll:2021ori} can be directly applied to Eq.~\eqref{eq:frg_flow_x}. Thus, we can start our discussion without further reference to the numerical implementation.\\
	
	The main scope of this subsection is to demonstrate the astonishing role of the radial sigma mode in terms of highly non-linear and unconventional diffusion in RG flows of scale-dependent effective potentials $V ( t, x )$ or rather their derivatives $v ( t, x ) = \partial_x V ( t, x )$. To this end, let us again focus solely on the purely diffusive contribution of the RG flow equation \eqref{eq:frg_flow_x} and rewrite it in terms of a non-linear heat equation by executing the $\sigma$-derivative on the \textit{r.h.s.},
		\begin{align}
			\partial_t v ( t, x ) = \, & \frac{\mathrm{d}}{\mathrm{d} x} \bigg[ \ldots + \frac{1}{N} \, \frac{\tfrac{1}{2} \, \partial_t r ( t )}{r ( t ) + \partial_x v ( t, x )} \bigg] =	\vphantom{\bigg(\bigg)}
			\\
			= \, & \ldots + D [ t, \partial_x v ] \, \partial_x^2 v ( t, x ) \, ,	\vphantom{\bigg(\bigg)}	\nonumber
		\end{align}
	where we defined the manifestly positive diffusion coefficient (note the definition \eqref{eq:roft} of the regulator $r ( t )$),
		\begin{align}
			D [ t, \partial_x u ] \equiv - \frac{1}{N} \, \frac{\tfrac{1}{2} \, \partial_t r ( t )}{[ r ( t ) + \partial_x v ( t, x ) ]^2} \, .	\label{eq:diffusion_coefficient}
		\end{align}
	The identification of the sigma loop contribution with a non-linear version of the heat equation has sever numerical and conceptual implications.
	
	Numerically, one has to ensure that the spatial discretization scheme is able to handle non-linear diffusion, \textit{i.e.}\ parabolic type contributions in the PDE. This is the case for the KT scheme with diffusion fluxes \cite{KTO2-0} as was tested in various applications, also in the context of RG flow equations \cite{Koenigstein:2021syz,Koenigstein:2021rxj}.
	
	On a conceptual level, the diffusive contribution to the flow of $v ( t, x )$ clearly introduces a dissipative process into the RG flow and renders the RG flow manifestly irreversible right from the beginning -- as is absolutely natural and similar to all diffusive processes in our everyday life, \textit{e.g.}, heat conduction, diffusive mixture of fluids, \textit{etc.}. Furthermore, diffusion (dissipation) is a natural source of entropy and thereby introduces an abstract ``thermodynamic arrow of time'' \cite{Lebowitz:2008} into a system. For us, this ``arrow of time'' singles out the RG time $t\in[0,\infty)$, increasing from the UV to the IR, as a natural time-like parameter and the natural evolution parameter of the dissipative PDE \eqref{eq:frg_flow_x}. In part II of this series of publications \cite{Koenigstein:2021rxj}, we comment at length on this issue and draw direct connections between the manifestly dissipative character of the ERG equation \eqref{eq:wetterich-equation} and so-called $\mathcal{C}$-theorems \cite{Zamolodchikov:1986gt,Rosten:2010vm,Zumbach:1993zz,Zumbach:1994kc,Zumbach:1994vg,Banks:1987qs,Cardy:1988cwa,Osborn:1989td,Osborn:2011kw,Jack:1990eb,Komargodski:2011vj,Curtright:2011qg,Haagensen:1993by,Generowicz:1997he,Forte:1998dx,Codello:2013iqa,Codello:2015ana,Becker:2014pea,Becker:2016zcn}.
	
\begin{figure}
			\centering
			\includegraphics{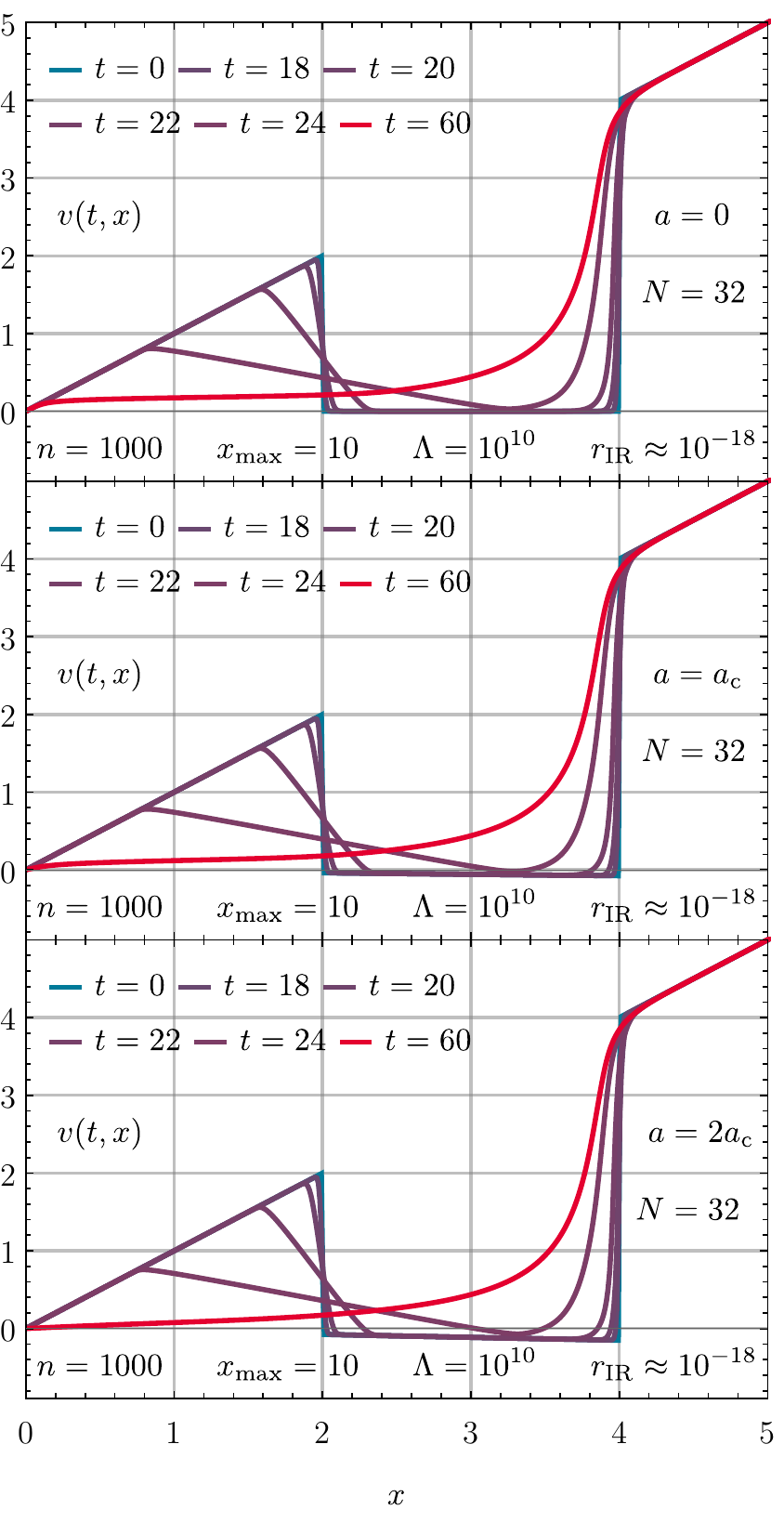}
			\caption{\label{fig:frg_N32_flows}%
				The RG flow of the derivative of the rescaled effective potential $v ( t, x )$ for the zero dimensional $O(N)$~model with $N = 32$ for the initial condition \eqref{eq:RP_vofx} with $a = 0$, $a = a_\mathrm{c}$, and $a = 2 a_\mathrm{c}$ in the upper, middle, and lower panel respectively. {Blue} curves represent the UV initial conditions at $t = 0$, {red} curves correspond to the IR potentials at $t = 60$ and the {violet} curves are at intermediate, selected RG times $t$.
			}
		\end{figure}

\begin{figure}
			\centering
			\includegraphics{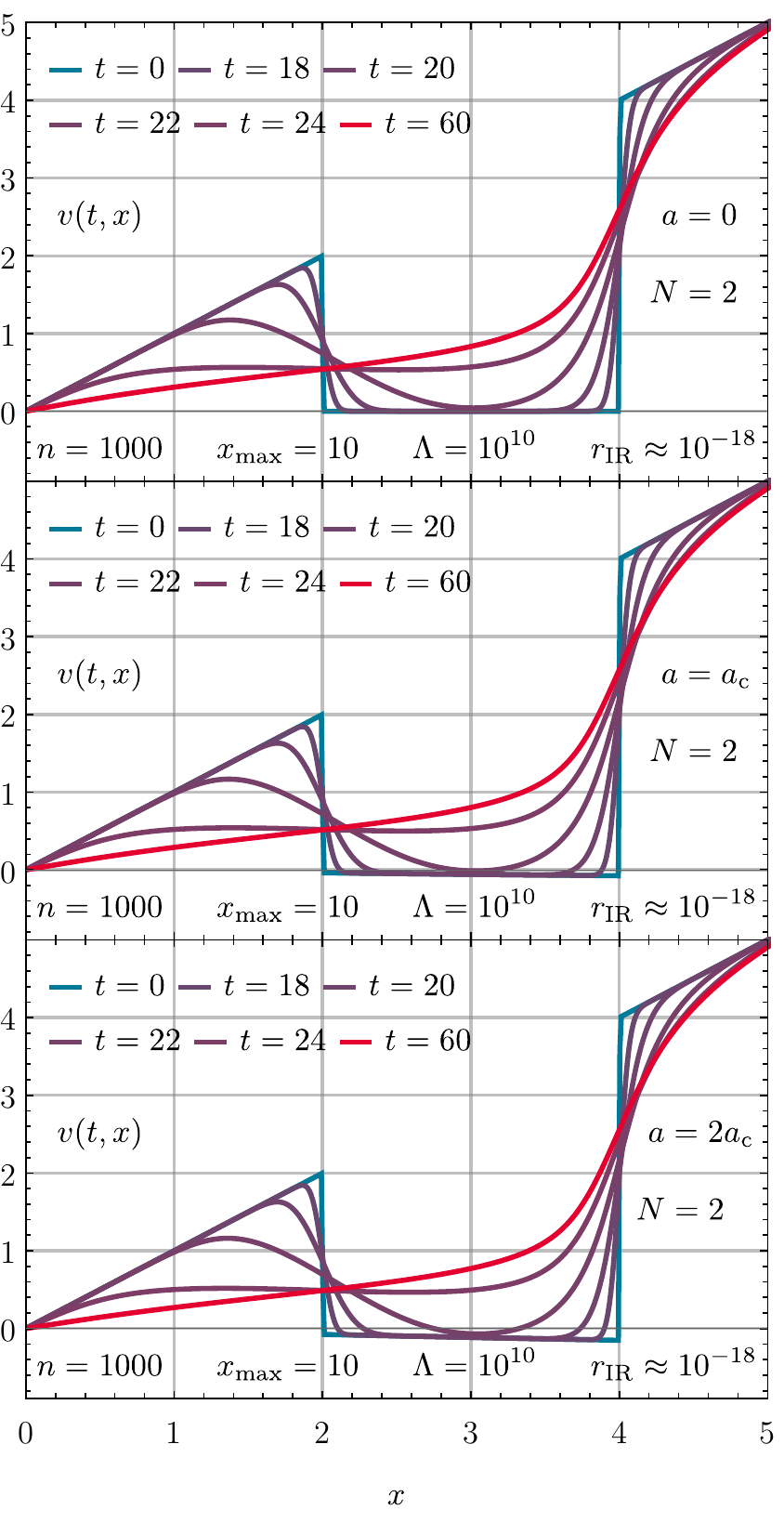}
			\caption{\label{fig:frg_N2_flows}%
				The RG flow of the derivative of the rescaled effective potential $v ( t, x)$ for the zero dimensional $O(N)$~model with $N = 2$ for the initial condition \eqref{eq:RP_vofx} with $a = 0$, $a = a_\mathrm{c}$, and $a = 2 a_\mathrm{c}$ in the upper, middle, and lower panel respectively. {Blue} curves represent the UV initial conditions at $t = 0$, {red} curves correspond to the IR potentials at $t = 60$ and the {violet} curves are at intermediate, selected RG times $t$.
			}
		\end{figure}

	In the context of this work, however, we are mainly interested in the influence of the non-linear diffusion on the explicit shape of $v ( t, x )$ and its drastic consequences for the reliability of $\tfrac{1}{N}$-expansions and the infinite-$N$ limit. To this end, we present our numerical solutions of Eq.~\eqref{eq:frg_flow_x} with initial condition \eqref{eq:RP_vofx} (see Fig.~\ref{fig:Vofx}) for two choices of $N$. \textit{W.l.o.g.}\ we choose $N = 2$ and $N = 32$ and present respective RG flows for $a = 0$, $a = a_\mathrm{c}$, and $a = 2 a_\mathrm{c}$ in Figs.~\ref{fig:frg_N32_flows} and \ref{fig:frg_N2_flows}. The figures are structured analogously to Fig.~\ref{fig:frg_largeN_flows} (for infinite $N$). For our numerical computations we used the same UV and IR cutoffs as for the infinite-$N$ case. Nevertheless, we had to change the size of the computational interval from $[ 0, 5]$ to $[ 0, 10 ]$ in order to exclude boundary effects due to the diffusion. Furthermore, it suffices to use $n = 1000$ computational finite volume cells on this interval, because it is no longer necessary to resolve the sharp shock fronts at extremely high resolution to obtain small numerical errors. For details on these two aspects, we refer to our lengthy and detailed discussion in part I of this series of publications \cite{Koenigstein:2021syz}, where we explicitly performed lots of tests for numerical parameters of finite $N$ computations.

	Qualitatively, we observe the following: Even though $N = 32$ seems to be rather large (especially in the context of the ``large-$N_\mathrm{color}$ or large-$N_\mathrm{flavor}$ discussions'' in the context of QCD, QCD-inspired models or holographic methods, where ``$N$'' is typically between $1$ and $6$) the RG flow of the $\tfrac{1}{N}$-rescaled $v ( t, x )$ entirely changes, if one compares corresponding panels of Figs.~\ref{fig:frg_N32_flows} and Fig.~\ref{fig:frg_largeN_flows} directly. Although the underlying shock, stemming from the still rather strong advective pion modes, dominates the overall shape of $v ( t, x )$ in Fig.~\ref{fig:frg_N32_flows} for all three choices of $a$, the diffusive character sets in rather early during the beginning of the RG flow and smears out the infinite negative slope of $v ( t, x )$ at the shock wave. Inspecting the non-linear diffusion coefficient \eqref{eq:diffusion_coefficient} this is expected for all finite $N$. Huge negative gradients $\partial_x v ( t, x )$ lower the difference $r ( t ) + \partial_x v ( t, x )$, which in turn drastically increases the diffusion coefficient leading, in combination with large $\partial_x^2 v ( t, x )$, to strong diffusion in regions where $v ( t, x )$ has large negative slopes, \textit{e.g.},\ next to the shock front. On the other hand, if $\partial_x v ( t, x )$ has large positive slope, as is the case close to the rarefaction fan, the diffusion coefficient is drastically suppressed, even if $\partial_x^2 v ( t, x )$ is large, such that the advection still dominates close to the rarefaction wave. For large $x \gg 5$ both, the $D [ t, \partial_x v ]$ and $\partial_x^2 v ( t, x )$ tend to zero (as is the case for $\tfrac{1}{x} \, v ( t, x )$ for the advection). For all other regions in $x$ we find complicated variations of these conceptual behaviors.
	
	Concerning the freezing or colliding of the shock wave, which was observed for infinite-$N$ in Fig.~\ref{fig:frg_largeN_flows}, we find that remnants of the freezing shocks are still visible in Figs.~\ref{fig:frg_N32_flows} (upper and middle panel). However, the gradient $\partial_x v ( t, x )$ no longer changes its sign at the right of the remnants of the freezing shock waves, such that overall the potential $V ( t, x )$ turns convex in the IR.
	
	Turning to Fig.~\ref{fig:frg_N2_flows} for the $N = 2$ scenario, where only one pion and one sigma mode are included in the calculation, we find that the overall the dynamics is very similar to the $N = 32$, but even more dominated by the diffusive $\sigma$-contribution to the RG flow. The freezing shock waves are no longer visible in the IR for $a = 0$ and $a = a_\mathrm{c}$ and the rarefaction wave is totally washed out. The latter effect is the reason, why the computational interval had to be increased.
	
	Before we turn to the overall interpretation of these findings, we remark that we also compared our numerical results for the $1$PI-two-point functions for all three choices of $a$ and $N = 2$ and $N = 32$ against exact results. In Table~\ref{tab:KT1500errors} we present the corresponding relative errors which are discussed further in App.~\ref{app:FRGnumericsFiniteN}.\\
		\begin{table}[b]
			\caption{\label{tab:KT1500errors}%
				Relative numerical errors for the $1$PI-two-point function $\Gamma^{(2)}$, see Eq.~\eqref{eq:Gamma2vofx}, for the results plotted in Figs.~\ref{fig:frg_N2_flows} and \ref{fig:frg_N32_flows} with corresponding exact reference values from the first two rows of Tab.~\ref{tab:Gamma2N}. The scaling of these errors with the number of volume cells can be found in Tabs.~\ref{tab:frg_N2_2ac_rates} and \ref{tab:frg_N32_2ac_rates} for $N=2$ and $N=32$ with $a=2\,a_\mathrm{c}$.
			}
			\begin{ruledtabular}
				\begin{tabular}{l c c c}
					$N$		&	$a=0$					&	$a = a_\mathrm{c}$		&	$a = 2 a_\mathrm{c}$
					\\
					\colrule\addlinespace[0.25em]
					$2$		&	$6.403 \cdot 10^{-5}$	&	$5.463 \cdot 10^{-5}$	&	$4.535 \cdot 10^{-5}$
					\\
					$32$	&	$4.227 \cdot 10^{-3}$	&	$6.405 \cdot 10^{-4}$	&	$8.521 \cdot 10^{-3}$
				\end{tabular}
			\end{ruledtabular}
		\end{table}
		
	In summary, we find that the radial $\sigma$-mode and the corresponding diffusion is a game changer in a QFT when switching from infinite to finite $N$. By directly comparing the infinite-$N$ and finite-$N$ results of the RG flows, we observe that for infinite-$N$ the $\tfrac{1}{N}$-rescaled potential $V ( t, x )$ does not turn convex in the IR and may still involve non-analyticities in terms of cusps. This is in direct opposition to the zero-dimensional version of the Coleman-Mermin-Wagner-Hohenberg theorem \cite{Coleman:1973ci,Mermin:1966,Hohenberg:1967,Koenigstein:2021syz,Moroz:2011thesis}, which states that the zero-dimensional IR potential has to be convex and smooth. On the other hand, we find that independent of the specific choice of $N$ -- as long as $N$ is finite -- the highly non-linear diffusion of the $\sigma$-mode restores convexity and smoothness of the IR potential. Depending on the specific choice of $N$ this may however happen at later times in the RG flow, respectively at lower RG scales, thus deeper in the IR\footnote{A similar effect is observed in a parallel study \cite{Stoll:2021ori} by the authors and their collaborators S. Rechenberger, J.~Stoll and N.~Zorbach on the Gross-Neveu(-Yukawa) model in $1+1$ dimensions, where, as long as as the number $N$ of fermions is finite and not infinite, the diffusion by the $\sigma$-mode completely changes the dynamics of the system, ensures convexity of the IR potential and even unavoidably restores the $\mathbb{Z}_2$-symmetry of the model at non-zero temperatures. Both effects are not present in the infinite-$N$ limit, see, \textit{e.g.} Refs.~\cite{Gross:1974jv,Wolff:1985av,Thies:2006ti}, where the diffusion in field space stemming from the bosonic quantum fluctuations of the $\sigma$-mode are totally suppressed.}. We conclude from these non-perturbative FRG studies, that calculations at infinite-$N$ and large-$N$, may lead to totally different results for certain aspects of a QFT.

\section{Conclusion and outlook}
	
	In the present work, we have discussed several fundamental aspects of and methods for QFTs using an simple $(0 + 0)$-dimensional $O(N)$-symmetric toy model that is exactly solvable, meaning that all correlation functions can be calculated analytically or numerically up to arbitrary precision as reference values. By inspecting a non-analytic piece-wise quadratic potential \eqref{eq:RP_Vofx} (see also Fig.~\ref{fig:Vofx}), we elucidated on the restricted applicability and validity of the large-$N$ expansion as well as the infinite-$N$ limit. Thereby we approached the task of calculating expectation values $\langle ( \vec{\phi}^{\, 2} )^n \rangle$ and the respective $1$PI-correlation functions $\Gamma^{(n)}$ with different methods.
	
	On the one hand, we studied the large-$N$ and infinite-$N$ limit within a saddle-point expansion of the partition function (path integral). On the other hand, we used the FRG and analyzed the same problem in terms of an exact untruncated RG flow equation. For our FRG analysis we made use of analytical and numerical tools from the field of computational fluid dynamics by interpreting the RG flow equation as a non-linear advection-diffusion equation involving different non-analyticities, like shock and rarefaction waves.
	
	Overall our result is that one should exercise great caution, when applying the large-$N$ expansion or large-$N$ limit, because of two main pitfalls. The first pitfall is the drastically limited applicability of the large-$N$ approximation within certain methods, like the saddle-point expansion, where analyticity of the expansion point needs to be guarantied (but is hardly ensured in higher-dimensional systems). The second pitfall is, that the infinite-$N$ limit (only retaining the zeroth order of the large-$N$ expansion) may alter fundamental aspects of a QFT, like the convexity of (effective) potentials, while other observables, like specific correlation functions, might not be totally off the exact results. Both effects as well as the exact results can be adequately resolved within our modern fluid dynamic formulation of the FRG.
	
	The major challenges arising in the application and generalization of our findings to higher $d$-dimensional QFTs are related to the issue of truncations necessary for practical FRG computations in $d>0$. Some further details can be found in parts I and II of this series of publications \cite{Koenigstein:2021syz,Koenigstein:2021rxj}, the planed part IV of this series \cite{Steil:partIV} and the set of publications \cite{Grossi:2019urj,Ihssen2020,Wink:2020tnu,Grossi:2021ksl,Stoll:2021ori} already using a fluid dynamic formulation of the FRG for computations in higher $d-$dimensional QFTs. Notwithstanding this early successes in $d>0$ a lot of research and development is required both on a conceptual as well as a practical level, when it comes to a fluid dynamic formulation of the FRG in non-zero space-time dimensions.
	
\begin{acknowledgments}
	A.K. and M.J.S. acknowledge the support of the \textit{Deutsche Forschungsgemeinschaft} (DFG, German Research Foundation) through the collaborative research center trans-regio  CRC-TR 211 ``Strong-interaction matter under extreme conditions''-- project number 315477589 -- TRR 211.
	
	A.K. acknowledges the support of the \textit{Friedrich-Naumann-Foundation for Freedom}.
	
	A.K. and M.J.S. acknowledge the support of the \textit{Giersch Foundation} and the \textit{Helmholtz Graduate School for Hadron and Ion Research}.
	
	We thank the co-authors of the first two parts of this series of publications \cite{Koenigstein:2021syz,Koenigstein:2021rxj}, J.~Braun, M.~Buballa, E.~Grossi, D.~H.~Rischke, and N.~Wink, for their collaboration, their support, and a lot of exceptionally valuable discussions. We are especially grateful to N.~Wink for pointing out some implications of the present work for first-order phase transitions and connections to details of their work \cite{Grossi:2019urj}.
	
	We thank L.~Pannullo, A.~Sciarra, and N.~Wink for useful comments on the manuscript.
	
	We further thank J.~Eser, F.~Divotgey, L.~Kurth, A.~Sciarra, J.~Stoll, M.~Winstel, and N.~Zorbach for very valuable discussions.
	
	All numerical numerical results as well as all figures in this work were obtained and designed using \texttt{Mathematica} \cite{Mathematica:12.2} including the following \textit{ResourceFunction}(s) from the \textit{Wolfram Function Repository}: \textit{PlotGrid} \cite{Lang:plotgrid}, \textit{PolygonMarker} \cite{Popkov:polygonmarker}, and \textit{MaTeXInstall} \cite{Horvat:matex}.
	
	The ``Feynman'' diagrams in Eqs.~\eqref{eq:advection_flux} and \eqref{eq:diffusion_flux} were generated via \texttt{Axodraw Version 2} \cite{Collins:2016aya}.
\end{acknowledgments}

\appendix

\section{Analytical solution for the instructive toy model}
\label{sec:RP_integrals}

	Within this appendix we present results for the integral $I_n^N [ V ]$ of Eq.~\eqref{eq:In_largeN} for the potential \eqref{eq:RP_Vofy},
		\begin{align}
			& I_n^N [ V ] \stackrel{\text{\eqref{eq:RP_Vofy}}}{=}	\vphantom{\bigg(\bigg)}	\label{eq:InNV_a0}
			\\
			= \, & N^{- ( \frac{N}{2} + n )} \, \Big( \Gamma \big( \tfrac{N}{2} + n \big) - \Gamma \big( \tfrac{N}{2} + n, 2 N \big) +	\vphantom{\bigg(\bigg)}	\nonumber
			\\
			& + \mathrm{e}^{6 N ( a + 1 )} \, \Gamma \big( \tfrac{N}{2} + n, 8 N \big) +	\vphantom{\bigg(\bigg)}	\nonumber
			\\
			& + ( - a )^{- ( \frac{N}{2} + n )} \, \mathrm{e}^{- 2 N ( a + 1 )} \times	\vphantom{\bigg(\bigg)}	\nonumber
			\\
			& \times \big[ \Gamma \big( \tfrac{N}{2} + n, - 2 N a \big) - \Gamma \big( \tfrac{N}{2} + n, - 8 N a \big) \big] \Big) \, ,	\vphantom{\bigg(\bigg)}	\nonumber
		\end{align}
	and in the special case $a = 0$,
		\begin{align}
			& I_n^N [ V ] \stackrel{\text{\eqref{eq:RP_Vofy}}|_{a = 0}}{=}	\vphantom{\bigg(\bigg)}	\label{eq:InNV}
			\\
			= \, & N^{- ( \frac{N}{2} + n )} \, \Bigg[ \Gamma \big( \tfrac{N}{2} + n \big) - \Gamma \big( \tfrac{N}{2} + n, 2 N \big) +	\vphantom{\bigg(\bigg)}	\nonumber
			\\
			& + \mathrm{e}^{6 N} \, \Gamma \big( \tfrac{N}{2} + n, 8 N \big) + \mathrm{e}^{-2 N} \, \frac{\big( 4^{\frac{N}{2} + n} - 1 \big) \, ( 2 N )^{\frac{N}{2} + n}}{\frac{N}{2} + n} \Bigg] \, ,  \vphantom{\bigg(\bigg)}	\nonumber
		\end{align}
	where 
		\begin{align}
			&	\Gamma ( a, z ) \equiv \int_{z}^{\infty} \mathrm{d} t \, t^{a - 1} \, \mathrm{e}^{-t} \, ,	&&	\Gamma ( z ) \equiv \Gamma ( z, 0 ) \, ,
		\end{align}
	is the (incomplete) gamma function. To determine the leading order contribution to $\langle ( \vec{\phi}^{\, 2} )^n \rangle$ in the limit ${N \rightarrow \infty}$ assuming finite $n$, we employ the asymptotic series, see, \textit{e.g.}, Secs.~6.1.41 and 6.5.32 of Ref~\cite{abramowitz+stegun} or Secs.~5.11 and 8.11 of Ref.~\cite{NIST:DLMF},
	\begin{align}
		\Gamma ( z ) = \, & \mathrm{e}^{-z} \, \sqrt{\tfrac{2 \pi }{z}} \, z^z \, \big( 1 + \tfrac{1}{12 z} + \tfrac{1}{288 z^2} + \ldots \big) \, ,	\vphantom{\bigg(\bigg)}
		\\
		\Gamma ( a, z ) = \, & \mathrm{e}^{-z} \, z^{a - 1} \, \big[ 1 + \tfrac{a-1}{z} + \tfrac{( a - 2 ) ( a - 1 )}{z^2} + \ldots \big] \, ,	\vphantom{\bigg(\bigg)}
	\end{align}
	valid for large real $z$ and in case of $\Gamma ( a, z )$ for $a \simeq \mathcal{O} ( z )$ \cite{Temme1975}.
	
	For $a = 0$ we find
		\begin{align}
			\lim_{N \rightarrow \infty} \tfrac{1}{N^n} \, \big\langle ( \vec{\phi}^{\, 2} )^n \big\rangle = \lim_{N \rightarrow \infty} \frac{2^n I_n^N [ V ]}{I_0^N [ V ]} \bigg|_{a = 0} = 1 \, ,
		\end{align}
	while for $a > 0$
		\begin{align}
			& \lim_{N \rightarrow \infty} \tfrac{1}{N^n} \, \big\langle ( \vec{\phi}^{\, 2} )^n \big\rangle =	\vphantom{\Bigg(\Bigg)}	\label{eq:RP_phi2_th}
			\\
			= \, & \lim_{N \rightarrow \infty} \frac{2^n I_n^N [ V ]}{I_0^N [ V ]} =	\vphantom{\Bigg(\Bigg)}	\nonumber
			\\
			= \, & \lim_{N \rightarrow \infty} \frac{17 \, \mathrm{e}^{6 N a} \, 16^{\frac{N}{2} + n} + 256 \, \sqrt{N \pi} \, \mathrm{e}^{\frac{n^2}{N} + \frac{3 N}{2}}}{17 \, \mathrm{e}^{6 N a} \, 4^N + 256 \, \sqrt{N \pi} \, \mathrm{e}^{\frac{3 N}{2}}} =	\vphantom{\Bigg(\Bigg)}	\nonumber
			\\
			= \, &
			\begin{cases}
				1		&	\text{for} \quad a \leq a_\mathrm{c} \, ,	\vphantom{\bigg(\bigg)}
				\\
				16^n	&	\text{for} \quad a_\mathrm{c} < a \, ,	\vphantom{\bigg(\bigg)}
			\end{cases}	\nonumber
		\end{align}
	where ${a_\mathrm{c}\equiv\tfrac{1}{4} - \tfrac{1}{3} \ln ( 2 ) \approx 0.018951}$. For $a > a_\mathrm{c}$ the first terms in the denominator and numerator of Eq.~\eqref{eq:RP_phi2_th} (third line) dominate, while for $a < a_\mathrm{c}$ the second terms dominate. For $a=a_\mathrm{c}$ Eq.~\eqref{eq:RP_phi2_th} (third line) can be simplified ultimately to $\mathrm{e}^{\frac{n^2}{N}}$ under the limit $N\rightarrow\infty$ and thus yielding $1$ in the limit.

\section{Saddle-point expansion at large-\texorpdfstring{$N$}{N}}
\label{sec:saddle_point_app}

	In this appendix we present the so-called saddle-point expansion for integrals of the type
		\begin{align}
			I^N [ f, g ] \equiv \int_{0}^{\infty} \mathrm{d} y \, g ( y ) \, \mathrm{e}^{- N f ( y )} \, .	\label{eq:SPapp_Idef}
		\end{align}
	Assuming that $f(y)$ has a unique global minimum at $y_0$ and further assuming analyticity (expandability to arbitrary order) of $f ( y )$ and also $g ( y )$ in $y_0$, it is possible to derive an asymptotic series of $I^N [ f, g ]$ for large $N$ if the series expansions of $f ( y )$ and $g ( y )$ around $y_0$ grow like polynomials. We focus here on the one-dimensional integral \eqref{eq:SPapp_Idef} see, \textit{e.g.}, Ref.~\cite{Arfken:2005} for further details and generalizations.

	For large $N$ the integrand of Eq.~\eqref{eq:SPapp_Idef} is peaked around $y_0$ and we therefore consider an expansion around $y_0$ using the computational coordinate $z$ defined by
		\begin{align}
			y = y_0 + \tfrac{z}{\sqrt{N}} \, .
		\end{align}
	We proceed with the computation of $I^N [ f, g ]$ at large $N$:
		\begin{widetext}
		\begin{align}
			& I^N [ f, g ] =	\vphantom{\Bigg(\Bigg)}
			\\
			= \, & \int_{0}^{\infty} \mathrm{d} y \, g ( y ) \, \mathrm{e}^{- N f ( y )} =	\vphantom{\Bigg(\Bigg)}	\nonumber
			\\
			= \, & \tfrac{1}{\sqrt{N}} \int_{- y_0 \sqrt{N}}^{\infty} \mathrm{d} z \, g \big( y_0 + \tfrac{z}{\sqrt{N}} \big) \, \exp \big[ - N f \big( y_0 +\tfrac{z}{\sqrt{N}} \big) \big] =	\vphantom{\Bigg(\Bigg)}	\label{eq:SPapp_Ishift}
			\\
			= \, & \tfrac{1}{\sqrt{N}} \int_{- y_0 \sqrt{N}}^{\infty} \mathrm{d} z \, g \big( y_0 + \tfrac{z}{\sqrt{N}} \big) \, \exp \Big[ - N f^{(0)} - \tfrac{1}{2} \, f^{(2)} \, z^2 - \tfrac{1}{6 \sqrt{N}} \, f^{(3)} \, z^3 - \tfrac{1}{24 N} \, f^{(4)} \, z^4 - \mathcal{O} ( z^5 ) \Big] \simeq	\vphantom{\Bigg(\Bigg)}
			\\
			\simeq \, & \mathrm{e}^{- N f^{(0)}} \tfrac{1}{\sqrt{N}} \int_{-\infty}^{\infty} \mathrm{d} z \, \mathrm{e}^{- \frac{1}{2} f^{(2)} z^2} \, g \big( y_0 + \tfrac{z}{\sqrt{N}} \big) \, \Big( 1 - \tfrac{1}{6 \sqrt{N}} \, f^{(3)} \, z^3 + \tfrac{1}{72 N} \, \big[ ( f^{(3)} )^2 \, z^6 - 3 \, f^{(4)} \, z^4 \big] + \mathcal{O} \big( N^{-\frac{3}{2}} \big) \Big) =	\vphantom{\Bigg(\Bigg)}	\nonumber
			\\
			= \, & \mathrm{e}^{- N f^{(0)}} \tfrac{1}{\sqrt{N}}\int_{-\infty}^{\infty} \mathrm{d} z \, \mathrm{e}^{-\frac{1}{2} f^{(2)} z^2} \Big( g^{(0)} + \tfrac{1}{\sqrt{N}} \, \big[ g^{(1)} - \tfrac{1}{6} \, g^{(0)} \, f^{(3)} \, z^2 \big] \, z +	\vphantom{\Bigg(\Bigg)}	\label{eq:SPapp_int}
			\\
			& \qquad \qquad \qquad \qquad \qquad \qquad \quad + \tfrac{1}{N} \, \big[ \tfrac{1}{2} \, g^{(2)} - \tfrac{1}{6} \, g^{(1)} \, f^{(3)} \, z^2 + \tfrac{1}{72} \, g^{(0)} \, ( f^{(3)} )^2 \, z^4 - \tfrac{1}{24} \, g^{(0)} \, f^{(4)} \, z^2 \big] \, z^2 + \mathcal{O} \big( N^{-\frac{3}{2}} \big) \Big) =	\vphantom{\Bigg(\Bigg)}	\nonumber
			\\
			= \, & \mathrm{e}^{- N f^{(0)}} \sqrt{\tfrac{2 \pi}{N f^{(2)}}} \, \sum_{i = 0}^{\infty} C_i [ f, g ] \, N^{-i} \, ,	\vphantom{\Bigg(\Bigg)}	\label{eq:SPapp_Isp}
		\end{align}
		\end{widetext}
	where we abbreviated $n$-th derivatives of $f$ and $g$ evaluated at $y_0$ with superscripts $(n)$. In the preceding set of equalities we first expanded the exponent in powers of $N$ after switching to the coordinate $z$. We split of the contributions of $\mathcal{O} ( N^1 )$ and $\mathcal{O} ( N^0 )$ in the exponent and then expanded the exponential in an asymptotic series in $N$, while shifting the lower integration bound\footnote{Since we are interested in an asymptotic power series for large $N$ shifting the lower integration bound in line \eqref{eq:SPapp_Ishift} is valid since contributions stemming from this shift decay exponentially and as such faster than any power.}. Afterwards, we continued by expanding $g$ and collecting terms of $\mathcal{O} ( N^{- \tfrac{n}{2}} )$. Ultimately, we were left with a sum over Gaussian integrals of $\mathcal{O} ( N^{-n} )$ and vanishing contributions of odd integrands of $\mathcal{O} ( N^{- \tfrac{2 n + 1}{2}} )$ in Eq.~\eqref{eq:SPapp_int} and performed those integrals, which left us with the desired power series \eqref{eq:SPapp_Isp} with coefficients $C_i [ f, g ]$ of $\mathcal{O} ( N^0 )$, \textit{e.g.},
		\begin{align}
			C_0 [ f, g ] = \, & g^{(0)} \, ,	\vphantom{\bigg(\bigg)}
			\\
			C_1 [ f, g ] = \, & \frac{g^{(2)}}{2 f^{(2)}} - \frac{g^{(1)} f^{(3)}}{2 ( f^{(2)} )^2} +\frac{5 g^{(0)} ( f^{(3)} )^2}{24 ( f^{(2)} )^3} - \frac{g^{(0)} f^{(4)}}{8 ( f^{(2)} )^2} \, .	\vphantom{\bigg(\bigg)}	\nonumber
		\end{align}
	The computation of higher order coefficients is straightforward and tedious by hand, but is easy to implement in computer algebra systems like \texttt{Mathematica} \cite{Mathematica:12.2}.
	
	The presented saddle-point expansion of $I^N [ f, g ]$ can be used in combination with Eq.~\eqref{eq:ON_expectation_value_largeN} for a large-$N$ expansion of the expectation values $\langle ( \vec{\phi}^{\, 2} )^n \rangle$
	\begin{widetext}
		\begin{align}
			\tfrac{1}{N^n} \, \langle ( \vec{\phi}^{\, 2} )^n \rangle = \, & \frac{2^n I^N [ V ( y ) - \tfrac{1}{2} \ln ( y ), y^{n - 1} ]}{I^N [ V ( y ) - \tfrac{1}{2} \ln ( y ), y^{-1} ]} =	\vphantom{\bigg(\bigg)}	\label{eq:SPapp_series}
			\\
			= \, & 2^n \, y_0^n + \frac{1}{N} \, \frac{n \, 2^n \, y_0^n \big[ 2 ( n - 3 ) \, y_0^2 \, V^{(2)} ( y_0 ) + n - 2 y_0^3 \, V^{(3)} ( y_0 ) - 1 \big]}{\big[ 2 y_0^2 \, V^{(2)} ( y_0 ) + 1 \big]^2} + \mathcal{O} ( N^{-2} ) \, ,	\vphantom{\bigg(\bigg)}	\nonumber
		\end{align}
	\end{widetext}
	which holds for $V(y)-\tfrac{1}{2}\log(y)$ which are analytic around their respective unique global minimum $y_0$. Corresponding expressions for the $1$PI-correlation functions can be derived using the relations between $\Gamma^{(n)}$ and $\langle ( \vec{\phi}^{\, 2} )^n \rangle$, see, \textit{e.g.}, Eqs.~(70)-(75) of Ref.~\cite{Koenigstein:2021syz} or Ref.~\cite{Keitel:2011pn}.

\section{Method of characteristics}
\label{app:method_of_characteristics}

	In this appendix we derive the expressions for the characteristic curves of Eqs.~\eqref{eq:frg_flow_Ninf_x} and \eqref{eq:frg_flow_Ninf_y} using the method of characteristics and to be specific the Lagrange–Charpit equations, see Ref.~\cite{Delgado2006Aug} for details or Refs.~\cite{polyanin2016handbook,LeVeque:2002} for a general overview.\\
	
	The quasilinear hyperbolic PDE of the form
		\begin{align}
			a ( t, z, v ) \, \partial_t v ( t, z ) + b ( t, z, v ) \, \partial_z v ( t, z ) = c ( t, z, v )	\label{eq:MoC_pde}
		\end{align}
	presents as an ordinary differential equation (ODE) along so called characteristic curves, which are given by the Lagrange–Charpit equations \cite{Delgado2006Aug} (also called characteristic equations) 
		\begin{align}
			\frac{\partial t ( \tau )}{\partial \tau} = \, & a ( t ( \tau ), z ( \tau ), v ( \tau ) ) \, ,	\vphantom{\bigg(\bigg)}	\label{eq:MoC_ode_1}
			\\
			\frac{\partial z ( \tau )}{\partial \tau} = \, & b ( t ( \tau ), z ( \tau ), v ( \tau ) ) \, ,	\vphantom{\bigg(\bigg)}	\label{eq:MoC_ode_2}
			\\
			\frac{\partial v( \tau )}{\partial \tau} = \, & c ( t ( \tau ), z ( \tau ), v ( \tau ) ) \, ,	\vphantom{\bigg(\bigg)}	\label{eq:MoC_ode_3}
		\end{align}
	with the curve-parameter $\tau$ and initial conditions
		\begin{align}
			t ( \tau = 0 ) = \, & t_0 \, ,	\vphantom{\bigg(\bigg)}
			\\
			z ( \tau = 0 ) = \, & z_0 \, ,	\vphantom{\bigg(\bigg)}
			\\
			v ( \tau = 0 ) = \, & v_0 ( t_0, z_0 ) \, ,	\vphantom{\bigg(\bigg)}
		\end{align}
	related to the original PDE~\eqref{eq:MoC_pde}. Solving this ODE system yields the functions $t(\tau)$, $z(\tau)$ and $v(\tau)$, which can be used to extract information about the actual solution of the PDE~\eqref{eq:MoC_pde} including in some cases the full solution itself.\\

	For the remainder of this appendix we will focus on the solution of the characteristic equations \eqref{eq:MoC_ode_1} - \eqref{eq:MoC_ode_3} for the FRG flow Eqs.~\eqref{eq:frg_flow_Ninf_x} and \eqref{eq:frg_flow_Ninf_y} of the zero dimensional $O(N)$~model in the limit $N \rightarrow \infty$. Since the equations in $x$ and $y$ are related by the coordinate transformation ${y = \tfrac{1}{2} \, x^2}$, the solutions and also characteristics curves are directly related. For simplicity we solve the characteristic equations for the flow equation \eqref{eq:frg_flow_Ninf_y} in the rescaled invariant $y$ and then compute the corresponding curves in $x$ using the coordinate transformation. A direct solution of the characteristic equations for the flow equation \eqref{eq:frg_flow_Ninf_x} in $x$ is also possible and shares a lot of computations with the slightly simpler computation in $y$.
	
	After performing the $y$ derivative in Eq.~\eqref{eq:frg_flow_Ninf_y} comparing coefficients with Eq.~\eqref{eq:MoC_pde} yields for the Eqs.~\eqref{eq:MoC_ode_1}-\eqref{eq:MoC_ode_3} explicitly
		\begin{align}
			\frac{\partial t ( \tau )}{\partial \tau} = \, & 1 \,,	\label{eq:MoC_ode_y_1}
			\\
			\frac{\partial y ( \tau )}{\partial \tau} = \, & - \frac{\Lambda \, \mathrm{e}^{- t ( \tau )}}{2 \, [ \Lambda \, \mathrm{e}^{- t ( \tau )} + v ( \tau ) ]^2} \, ,	\label{eq:MoC_ode_y_2}
			\\
			\frac{\partial v ( \tau )}{\partial \tau} = \, & 0 \, ,	\label{eq:MoC_ode_y_3}
		\end{align}
	with the UV initial conditions
		\begin{align}
			t ( \tau = 0 ) = \, & 0 \, ,	\vphantom{\bigg(\bigg)}
			\\
			y ( \tau = 0 ) = \, & y_0 \geq 0 \, ,	\vphantom{\bigg(\bigg)}
			\\
			v ( \tau = 0 ) = \, & v ( 0, y_0 ) \, ,	\vphantom{\bigg(\bigg)}
		\end{align}
	specifying the characteristic curves. The ODEs for $t(\tau)$ and $v(\tau)$ decouple and can be trivially integrated 
		\begin{align}
			t ( \tau )= \, & \tau \, ,	\vphantom{\bigg(\bigg)}	\label{eq:MoC_toftau}
			\\
			v ( \tau ) = \, & v ( 0, y_0 ) \, .	\vphantom{\bigg(\bigg)}	\label{eq:MoC_vyoftau}
		\end{align}
	Due to the direct equivalence of $t$ and $\tau$ we continue by using the RG time $t$ as the curve-parameter in the following. The ODE~\eqref{eq:MoC_ode_y_2} for $y(\tau)$ is independent of $y$ itself and can be integrated directly after inserting the solutions \eqref{eq:MoC_toftau} and \eqref{eq:MoC_vyoftau} for $t$ and $v$. The solution for $y ( t )$ follows as
		\begin{align}
			y ( t ) = \, & y_0 - \int_0^t \mathrm{d} \tau \, \frac{\Lambda \, \mathrm{e}^{- \tau}}{2 \, [ \Lambda \, \mathrm{e}^{- \tau} + v ( 0, y_0 ) ]^2} =	\vphantom{\Bigg(\Bigg)}	\label{eq:MoC_yoftau}
			\\
			= \, & y_0 - \frac{1}{2 \, [ \Lambda \, \mathrm{e}^{-t} + v ( 0, y_0 ) ]} + \frac{1}{2 \, [ \Lambda + v ( 0, y_0 ) ]} \, .	\vphantom{\Bigg(\Bigg)}	\nonumber
		\end{align}
	Using the coordinate transformation ${y = \tfrac{1}{2} \, x^2}$ and the associated relation for the first derivative ${\partial_y V ( t, y ) = \tfrac{1}{x} \, \partial_x V ( t, x )}$ we can compute the characteristic curves $x ( t )$ and $v ( t )$ for the flow Eq.~\eqref{eq:frg_flow_Ninf_x} from Eqs.~\eqref{eq:MoC_yoftau} and \eqref{eq:MoC_vyoftau},
		\begin{align}
			x ( t ) = \, & \pm \sqrt{2 y ( t )} =	\vphantom{\Bigg(\Bigg)}	\label{eq:MoC_xoftau}
			\\
			= \, & \pm \sqrt{x_0^2 - \frac{1}{\Lambda \, \mathrm{e}^{- t} + \tfrac{v ( 0, x_0 )}{x_0}} + \frac{1}{\Lambda + \tfrac{v ( 0, x_0 )}{x_0}} } \, ,	\vphantom{\Bigg(\Bigg)}	\nonumber
			\\
			v ( t ) = \, & \tfrac{v ( 0, x_0 )}{x_0} \, x ( t ) \, .	\vphantom{\Bigg(\Bigg)}	\label{eq:MoC_vxoftau}
		\end{align}
	A particularity of the flow equation in $x$ is that the conserved quantity $v ( t, x )$ (the derivative $\partial_x V ( t, x )$) is not constant along the characteristics, $\tfrac{\mathrm{d} v}{\mathrm{d} t} \neq 0$, due to the contribution stemming from $x ( t )$ in Eq.~\eqref{eq:MoC_vxoftau}.	
	
\section{Rankine-Hugoniot condition and shock position}
\label{app:rankine-hugoniot_condition_and_shock_position}
	
	The Riemann problems posed by the initial condition \eqref{eq:RP_vofy} with the flow Eq.~\eqref{eq:frg_flow_Ninf_y} include a shock discontinuity in the UV ($t = 0$) at $y = 2$, since $v ( 2^{-} ) > v ( 2^{+} )$ and $G [ t, v ] < 0$. For a discussion see Sub.Sec.~\ref{subsec:FRGlargeN}. This appendix is dedicated to the computation of the position of the shock as a function of flow time $t$ using the so called Rankine-Hugoniot condition \cite{Rankine:1870,Hugoniot:1887}, see, \textit{e.g.}, the textbooks \cite{Ames:1992,LeVeque:2002,Hesthaven2007} for a detailed discussion of this construction method. A computation in the invariant $y$ for a structurally identical flow equation and initial condition can be found in App.~C.1 of Ref.~\cite{Grossi:2019urj}. We present a derivation for the complementary problem (initial condition  \eqref{eq:RP_vofx} with the flow Eq.~\eqref{eq:frg_flow_Ninf_x}) in $x$ for the sake of completeness in the following.\\

	Assume that there is a single shock wave (discontinuity) at the position $\xi_\mathrm{s} ( t )$ between ${ x_\mathrm{L} ( t ) < \xi_\mathrm{s} ( t ) < x_\mathrm{R} ( t ) }$. Integration over the conservation law \eqref{eq:frg_flow_Ninf_x} yields
		\begin{align}
			& \int_{x_\mathrm{L} ( t )}^{x_\mathrm{R} ( t )} \mathrm{d} x \, \partial_t v ( t, x ) =	\vphantom{\bigg(\bigg)}
			\\
			= \, & - \int_{x_\mathrm{L} ( t )}^{x_\mathrm{R} ( t )} \mathrm{d} x \, \tfrac{\mathrm{d}}{\mathrm{d} x} \, F [ t, x , v ( t, x ) ] =	\vphantom{\bigg(\bigg)}	\nonumber
			\\
			= \, & - \big( F [ t, x_\mathrm{R} ( t ), v ( t, x_\mathrm{R} ( t ) ) ] - F [ t, x_\mathrm{L} ( t ), v ( t, x_\mathrm{L} ( t ) ) ] \big)	\vphantom{\bigg(\bigg)}	\, . \nonumber
		\end{align}
	For the \textit{l.h.s.}, we split the integral about the shock $\xi_\mathrm{s} ( t )$
		\begin{align}
			& \int_{x_\mathrm{L} ( t )}^{x_\mathrm{R} ( t )} \mathrm{d} x \, \partial_t v ( t, x ) =	\vphantom{\bigg(\bigg)}
			\\
			= \, & \int_{x_\mathrm{L} ( t )}^{\xi_\mathrm{s} ( t )} \mathrm{d} x \, \partial_t  v ( t, x ) + \int_{\xi_\mathrm{s} ( t )}^{x_\mathrm{R} ( t )} \mathrm{d} x \, \partial_t  v ( t, x ) =	\vphantom{\bigg(\bigg)}	\nonumber
			\\
			= \, & - v ( t, \xi_\mathrm{s} ( t) ) \, \partial_t \xi_\mathrm{s} ( t ) + v ( t, x_\mathrm{L} ( t ) ) \, \partial_t x_\mathrm{L} ( t ) +	\vphantom{\bigg(\bigg)}	\nonumber
			\\
			& + \tfrac{\mathrm{d}}{\mathrm{d} t} \int_{x_\mathrm{L} ( t )}^{\xi_\mathrm{s} ( t )} \mathrm{d} x \, v ( t, x ) -	\vphantom{\bigg(\bigg)}	\nonumber
			\\
			& - v ( t, x_\mathrm{R} ( t ) ) \, \partial_t x_\mathrm{R} ( t ) + v ( t, \xi_\mathrm{s} ( t) ) \, \partial_t \xi_\mathrm{s} ( t ) +	\vphantom{\bigg(\bigg)}	\nonumber
			\\
			& + \tfrac{\mathrm{d}}{\mathrm{d} t} \int_{\xi_\mathrm{s} ( t )}^{x_\mathrm{R} ( t )} \mathrm{d} x \, v ( t, x )	\vphantom{\bigg(\bigg)}	\, ,\nonumber
		\end{align}
	where we used Leibniz integral rule in the third line. Next, we study the limits ${x_\mathrm{L} ( t ) \rightarrow \xi_\mathrm{s}^- ( t )}$ and ${x_\mathrm{R} ( t ) \rightarrow \xi_\mathrm{s}^+ ( t )}$. We find that the two integrals with the total time derivatives vanish and by defining
		\begin{align}
			v_\mathrm{L} ( t ) = \, & \lim\limits_{x_\mathrm{L} ( t ) \rightarrow \xi_\mathrm{s}^- ( t )} v ( t, x_\mathrm{L} ( t ) ) \, ,	\vphantom{\bigg(\bigg)}
			\\
			F_\mathrm{L} ( t ) = \, & \lim\limits_{x_\mathrm{L} ( t ) \rightarrow \xi_\mathrm{s}^- ( t )} F [ t, x_\mathrm{L} ( t ), v ( t, x_\mathrm{L} ( t ) ) ] \, ,	\vphantom{\bigg(\bigg)}
			\\
			v_\mathrm{R} ( t ) = \, & \lim\limits_{x_\mathrm{R} ( t ) \rightarrow \xi_\mathrm{s}^+ ( t )} v ( t, x_\mathrm{R} ( t ) ) \, ,	\vphantom{\bigg(\bigg)}
			\\
			F_\mathrm{R} ( t ) = \, & \lim\limits_{x_\mathrm{R} ( t ) \rightarrow \xi_\mathrm{s}^+ ( t )} F [ t, x_\mathrm{R} ( t ), v ( t, x_\mathrm{R} ( t ) ) ] \, ,	\vphantom{\bigg(\bigg)}
		\end{align}
	the equation for the shock speed reads
		\begin{align}
			\partial_t \xi_\mathrm{s} ( t ) = \, & \frac{F_\mathrm{R} ( t ) - F_\mathrm{L} ( t )}{v_\mathrm{R} ( t ) - v_\mathrm{L} ( t )} \, ,\label{eq:RHeq}
		\end{align}
	and is refereed to as Rankine–Hugoniot (jump) condition for the shock.
	
	For the explicit problem under consideration the initial positions at $t = 0$ of the two shocks are $x = 2$ and $x = - 2$. \textit{W.l.o.g.}\ we consider the shock at $x=2$ since the discussion for the shock at $x=-2$ follows from the symmetry of the problem. Consider the characteristic curves \eqref{eq:MoC_xoftau} and $v ( t, x ( t ) )$, thus Eq.~\eqref{eq:MoC_vxoftau}, left and right of the shock we find
		\begin{align}
			v_\mathrm{L} ( t ) = \, & \frac{v_{\mathrm{UV}, \mathrm{L}}}{x_{\mathrm{UV}, \mathrm{L}}} \, x_\mathrm{L} ( t ) = \xi_\mathrm{s}^- ( t ) \, ,
			\\
			v_\mathrm{R} ( t ) = \, & \frac{v_{\mathrm{UV}, \mathrm{R}}}{x_{\mathrm{UV}, \mathrm{R}}} \, x_\mathrm{R} ( t ) = - a \, \xi_\mathrm{s}^+ ( t ) \, ,
		\end{align}
	and for the corresponding fluxes Eq.~\eqref{eq:frg_flow_Ninf_x} yields
		\begin{align}
			F_\mathrm{L} ( t ) = \, & - \frac{\frac{1}{2} \partial_t r ( t )}{r ( t ) + \frac{v_\mathrm{L} ( t )}{x_\mathrm{L} ( t )}} = - \frac{\frac{1}{2} \partial_t r ( t )}{r ( t ) + 1} \, ,
			\\
			F_\mathrm{R} ( t ) = \, & - \frac{\frac{1}{2} \partial_t r ( t )}{r ( t ) + \frac{v_\mathrm{R} ( t )}{x_\mathrm{R} ( t )}} = - \frac{\frac{1}{2} \partial_t r ( t )}{r ( t ) - a} \, .
		\end{align}
	Inserting those explicit results into the Rankine–Hugoniot (jump) condition \eqref{eq:RHeq} results in
		\begin{align}
			\partial_t \xi_\mathrm{s} ( t ) = \, & \frac{F_\mathrm{R} ( t ) - F_\mathrm{L} ( t )}{v_\mathrm{R} ( t ) - v_\mathrm{L} ( t )} =	\vphantom{\bigg(\bigg)}
			\\
			= \, & \frac{1}{\xi_\mathrm{s} ( t )} \frac{1}{a + 1} \, \bigg[ \frac{\frac{1}{2} \partial_t r ( t )}{r ( t ) - a} - \frac{\frac{1}{2} \partial_t r ( t )}{r ( t ) + 1} \bigg]	\vphantom{\bigg(\bigg)}\, ,	\nonumber
		\end{align}
	where we are allowed to set $\xi_\mathrm{s}^+ ( t ) = \xi_\mathrm{s}^- ( t ) = \xi_\mathrm{s} ( t )$. Using the monotonicity of the regulator shape function $r ( t )$, see Eq.~\eqref{eq:roft}, we find
		\begin{align}
			\partial_r \big( \xi_\mathrm{s}^2 ( r ) \big) = \frac{1}{a + 1} \, \bigg( \frac{1}{r - a} - \frac{1}{r + 1} \bigg)\, ,
		\end{align}
	which can be integrated from the UV ($r=\Lambda$) down to an arbitrary value $r(t)\geq 0$ yielding
		\begin{align}
			& \xi_\mathrm{s} ( t ) =	\vphantom{\Bigg(\Bigg)} \label{eq:xioft}
			\\
			= \, & \sqrt{ \xi_{\mathrm{s},\mathrm{UV}}^2 + \frac{1}{a + 1} \, \bigg[ \ln \bigg( \frac{r ( t ) - a}{\Lambda - a} \bigg) - \ln \bigg( \frac{r ( t ) + 1}{\Lambda + 1} \bigg) \bigg] }	\vphantom{\Bigg(\Bigg)}\, ,	\nonumber
		\end{align}
	with $\xi_{\mathrm{s},\mathrm{UV}}^2 =2^2=4$.
	
	For $a\geq 0$ (and $\Lambda\gg a$) we find $\xi_\mathrm{s} ( t_0 ) = 0$ for a finite $t_0 > 0$, which indicates that the shocks originating from $-2$ and $+2$ in the UV annihilate at $x=0$ at the RG time $t_0$ based on the discussion of this appendix. The applicability of the construction discussed in this appendix is however limited as outlined in Sub.Sec.~\ref{subsec:FRGlargeN}.

\section{The FRG flow equation formulated in the \texorpdfstring{$O(N)$}{O(N)}-invariant in the large-\texorpdfstring{$N$}{N} limit}
\label{app:flow_equation_in_rho}
	
	Before we discuss our numerical results for the flow equation~\eqref{eq:frg_flow_Ninf_y} formulated in the $\tfrac{1}{N}$-rescaled invariant $y \equiv \tfrac{1}{2} \, x^2$ in Sub.Sec.~\ref{app:RPandEntropy} of this App., we briefly introduce the employed numerical KNP scheme in the next Sub.Sec.~\ref{app:KNP}.

\subsection{First and second order KNP scheme}\label{app:KNP}
	
	For computations involving the flow equation in Eq.~\eqref{eq:frg_flow_Ninf_y} in the $\tfrac{1}{N}$-rescaled invariant $y$, the \textit{KNP scheme} introduced by A.~Kurganov, S. Noelle, and G. Petrova in Ref.~\cite{KTO2-1} has several advantages in the present context over the \textit{KT scheme} developed by A.~Kurganov and E.~Tadmor in Ref.~\cite{KTO2-0} and employed for the numerical computations in the main part of this work. We will discuss the advantages of the KNP scheme for the flow equation~\eqref{eq:frg_flow_Ninf_y} in the rescaled invariant $y$ in this appendix.\\
	
	The application of the KNP scheme, namely its numerical advection flux, to the flow equation \eqref{eq:frg_flow_Ninf_x} in $x$ is straightforward. While we present all formulas for the KNP scheme for the advection flux $G$ of Eq.~\eqref{eq:frg_flow_Ninf_y} in this appendix, directly replacing $G$ with $F$ and $x$ with $y$\footnote{Note that $F$ is explicitly position dependent and the advection fluxes $F$ and local speeds $a$ (consequently $\partial_u F$) have to be evaluated on the cell interfaces, \textit{cf.}\ Eq.~\eqref{eq:FV_KT_H} and part I of this series of publications~\cite{Koenigstein:2021syz}.} yields valid expressions for the application of the KNP scheme to Eq.~\eqref{eq:frg_flow_Ninf_x}. Conversely the direct application of the KT to the flow equation in $y$ is not straightforward even at infinite $N$ due to the left boundary at $y=0$. This is discussed at length Sec.~IV~D of part I of this series \cite{Koenigstein:2021syz}.\\
	
	The numerical advection flux of the KNP can be expressed in semi-discrete form
		\begin{align}
			\partial_t \bar{v}_j = - \tfrac{1}{\Delta y} \, \big( H_{j + \frac{1}{2}}^\mathrm{KNP} - H_{j - \frac{1}{2}}^\mathrm{KNP} \big) \, ,	\label{eq:FV_numericalAdvectionFlux}
		\end{align}
	for the volume cell $\bar{v}_j$ at continuous time $t$. The difference of numerical fluxes $H_{j \pm \frac{1}{2}}^\mathrm{KNP}$ at the cell interfaces $y_{j \pm \frac{1}{2}}$ depends in general on the five-point stencil ${ \{\bar{v}_{j - 2}, \bar{v}_{j - 1}, \bar{v}_j, \bar{v}_{j + 1}, \bar{v}_{j+2} \}}$. The explicit numerical flux of the KNP scheme is given by
		\begin{align}
			H_{j + \frac{1}{2}}^{\mathrm{KNP}} \equiv \, & \frac{a_{j + \frac{1}{2}}^+ G \big[ t, v_{j + \frac{1}{2}}^- \big] - a_{j + \frac{1}{2}}^- G \big[ t, v_{j + \frac{1}{2}}^+ \big]}{a_{j + \frac{1}{2}}^+ - a_{j + \frac{1}{2}}^-} -	\vphantom{\Bigg(\Bigg)}	\label{eq:FV_KNP_H}
			\\
			& - \frac{a_{j + \frac{1}{2}}^+ a_{j + \frac{1}{2}}^-}{a_{j + \frac{1}{2}}^+ - a_{j + \frac{1}{2}}^-}  \, \big( v_{j + \frac{1}{2}}^{+} - v_{j + \frac{1}{2}}^{-}\big) \, .	\vphantom{\Bigg(\Bigg)}	\nonumber
		\end{align}
	where $a_{j + \frac{1}{2}}^\pm$ are right- and left-sided local speeds and $v_{j + \frac{1}{2}}^{\pm}$ are reconstructed function values at the cell interface $y_{j + \frac{1}{2}}$ \cite{KTO2-1}. For the second order accurate KNP scheme we use the same piecewise linear, \textit{total variation diminishing} (TVD) MUSCL reconstruction \cite{LeVeque:1992,LeVeque:2002,HARTEN1983357} employed in the KT scheme \cite{KTO2-1}
		\begin{align}
			v_{j + \frac{1}{2}}^{-} = \, & \bar{v}_j + \tfrac{\Delta y}{2} \, ( \partial_y v )_j \, ,	\vphantom{\bigg(\bigg)}	\label{eq:FV_up12m}
			\\
			v_{j + \frac{1}{2}}^{+} = \, & \bar{v}_{j+1} -\tfrac{\Delta y}{2}  \, ( \partial_y v )_{j+1} \, ,	\vphantom{\bigg(\bigg)}	\label{eq:FV_up12p}
		\end{align}
	in which the slopes $( \partial_y v )_j$ are approximated from cell averages using 
		\begin{align}
			( \partial_y v )_j = \frac{\bar{v}_{j+1} - \bar{v}_{j}}{\Delta y} \, \phi \bigg( \frac{\bar{v}_{j} - \bar{v}_{j-1}}{\bar{v}_{j+1} - \bar{v}_{j}} \bigg) \, ,	\label{eq:FV_uxjn}
		\end{align}
		with with a TVD limiter $\phi(r)$. Here, we follow  Ref.~\cite{KTO2-0} and use the so-called \textit{minmod} limiter \cite{MinModRoe}
					\begin{align}
			\phi ( r ) = \, & \max[ 0, \min( 1, r )] \, .	\label{eq:FV_minmod}
		\end{align}
	The right- and left-sided local speeds at the cell interface $y_{j + \frac{1}{2}}$ are used to estimate the maximal propagation of a possible discontinuity in the case $v_{j + \frac{1}{2}}^+\neq v_{j + \frac{1}{2}}^-$ \cite{KTO2-1} and are given for a scalar advection equation in one spatial dimension by
		\begin{align}
			a_{j + \frac{1}{2}}^+ &\, \equiv \max \bigg\{ \frac{\partial G}{\partial v} \Big[t, v_{j + \frac{1}{2}}^{+} \Big] , \frac{\partial G}{\partial v} \Big[t, v_{j + \frac{1}{2}}^{-} \Big] , 0 \bigg\} \, ,	\vphantom{\Bigg(\Bigg)}	\label{eq:FV_KNP_ajp12p}
			\\
			a_{j + \frac{1}{2}}^- &\, \equiv \min \bigg\{ \frac{\partial G}{\partial v} \Big[t, v_{j + \frac{1}{2}}^{+} \Big] , \frac{\partial G}{\partial v} \Big[t, v_{j + \frac{1}{2}}^{-} \Big] , 0 \bigg\} \, .	\vphantom{\Bigg(\Bigg)}	\label{eq:FV_KNP_ajp12m}
		\end{align}
	with the partial derivatives $\frac{\partial G}{\partial v}$ as the eigenvalues of a trivial $1\times1$ Jacobian, \textit{cf.} Eq.~(3.2) of Ref.~\cite{KTO2-1}, for a scalar advection equation in one spatial dimension.

	Boundary conditions for the advection flux of the KNP (and KT) scheme are readily implemented by means of so-called ghost cells. For the left computational boundary at $y=0$ those ghost cells would be located at negative $y$ and a formulation of physically meaningful boundary conditions in this point is not obvious, for more details see Sec.~IV~D of part I of this series of publications \cite{Koenigstein:2021syz}. At the right boundary located at a finite $y_\mathrm{max}$ ghost cells can computed using linear extrapolation without any practical problems as long as $y_\mathrm{max}$ is large enough \cite{Koenigstein:2021syz,Stoll:2021ori}. Coming back to the problematic left boundary we recall from the discussion surrounding the flow Eq.~\eqref{eq:frg_flow_Ninf_y} that
		\begin{align}
			\frac{\partial G}{\partial v}= -\frac{1}{2} \frac{\Lambda\mathrm{e}^{-t}}{(\Lambda\mathrm{e}^{-t} + v)^2}
		\end{align}
	is manifest negative for all $y\in\mathbb{R}^+$ and $t\in\mathbb{R}^+$ for all valid initial conditions/UV initial scales realizing $\Lambda \, \mathrm{e}^{-t} + v>0$ in the UV, \textit{cf.}\ part I of this series of publications \cite{Koenigstein:2021syz}. This however implies in Eq.~\eqref{eq:FV_KNP_ajp12p} a vanishing right sided local speed $a_{j + \frac{1}{2}}^+=0$. Physically this means that the fluid is only propagated to the left which simplifies the expression \eqref{eq:FV_KNP_H} for the numerical flux of the KNP scheme immensely
		\begin{align}
			H_{j + \frac{1}{2}}^{\mathrm{KNP}} \big|_{a_{j + \frac{1}{2}}^+ = 0} = G [ t, v_{j + \frac{1}{2}}^+ ]
		\end{align}
	resulting in the numerical upwind advection flux for the KNP scheme
		\begin{align}
			\partial_t \bar{v}_j = \tfrac{1}{\Delta y} \, \big( G [ t, v_{j - \frac{1}{2}}^+] - G [ t, v_{j + \frac{1}{2}}^+ ] \big) \, .	\label{eq:FV_KNP_monotone}
		\end{align}
	It is this reduction to an upwind scheme in regions with directed local speeds equivalent to monotonic advection fluxes with either $a_{j + \frac{1}{2}}^+ = 0$ or $a_{j + \frac{1}{2}}^- = 0$, which has lead the authors of Ref.~\cite{KTO2-1} to call their scheme a central-upwind scheme. Note that Eq.~\eqref{eq:FV_KNP_monotone} does no longer include the left-sided local speed $a_{j + \frac{1}{2}}^-$ and only contains advection terms evaluated at $v_{j\pm\frac{1}{2}}^+$ involving the reconstructions from the cells to the right, \textit{cf.} Eqs.~\eqref{eq:FV_up12p} and \eqref{eq:FV_uxjn}. As a result the numerical flux of Eq.~\eqref{eq:FV_KNP_monotone} is based on a right-leaning four point stencil ${\{\bar{v}_{j-1},\bar{v}_j,\bar{v}_{j+1},\bar{v}_{j+2}\}}$, where $\bar{v}_{j-1}$ is required together with $\bar{v}_{j}$ and $\bar{v}_{j+1}$ to compute $( \partial_y v )_j$. For the numerical flux of the first volume cell $j=0$ which we choose to span over $y_{-\frac{1}{2}}=0$ to $y_{\frac{1}{2}}=\Delta y$ we require ${\{\bar{v}_{-1},\bar{v}_0,\bar{v}_{1},\bar{v}_{2}\}}$, where only $\bar{v}_{-1}$ is a ghost cell. Since it only appears in the flux limiting procedure, see \eqref{eq:FV_uxjn}, it is arguably not a ghost point related to physical boundary conditions but rather a computational one necessary to ensure formal second order accuracy of the MUSCL reconstruction while preventing spurious oscillations around discontinuities -- TVD time steps. 
	
	Two naive strategies for a practical choice of $\bar{v}_{-1}$ come to mind. The first one would be switching from a central reconstruction to a right-sided reconstruction. Constructing a right-sided TVD reconstruction or searching for one in literature seemed unappealing for the brief discussion of Sub.Sec.~\ref{app:RPandEntropy}. The second option is much simpler and related to the fact, that the KNP scheme with the position independent advection flux $G$ of Eq.~\eqref{eq:frg_flow_Ninf_y} has a meaningful first order reduction. Switching from a piecewise linear to a piecewise constant reconstruction in Eqs.~\eqref{eq:FV_up12m} and \eqref{eq:FV_up12p}
		\begin{align}
			v_{j + \frac{1}{2}}^{-} = \, & \bar{v}_j + \mathcal{O} ( \Delta y ) \, ,	\vphantom{\bigg(\bigg)}	\label{eq:FV_up12m1st}
			\\
			v_{j + \frac{1}{2}}^{+} = \, & \bar{v}_{j+1} + \mathcal{O} ( \Delta y ) \, ,	\vphantom{\bigg(\bigg)}	\label{eq:FV_up12p1st}
		\end{align}
	results in a first order accurate (in $\Delta y$) semi-discrete upwind scheme \cite{KTO2-1,10.2307/2157317,10.2307/2030019}
		\begin{align}
			\partial_t \bar{v}_j = \tfrac{1}{\Delta y} \, \big( G[t,\bar{v}_{j}] - G[t,\bar{v}_{j + 1}] \big) \, , \label{eq:FV_KNP_monotone1st}
		\end{align}
	valid for monotone advection fluxes with $\partial_u G<0$. The first order accurate KNP scheme is in this context equivalent to the so-called Godunov upwind scheme \cite{10.2307/2157317,10.2307/2030019}. Application of such first order upwind-schemes within the FRG framework are discussed and presented in Refs.~\cite{Wink:2020tnu,WinkHirschegg}. To avoid the ghost cell $\bar{v}_{-1}$ altogether we always use the first-order accurate KNP scheme \eqref{eq:FV_KNP_monotone1st} in the first volume cell $\bar{v}_0$ and either stick to the first-order accurate scheme or use the second-order accurate KNP scheme \eqref{eq:FV_KNP_monotone} for all other cells.
		
	In the following we will denote the first-order accurate KNP scheme \eqref{eq:FV_KNP_monotone1st} with KNP $\mathcal{O} ( \Delta y^1 )$ and the second-order accurate scheme of Eq.~\eqref{eq:FV_KNP_monotone} with KNP $\mathcal{O} ( \Delta y^2 )$. In the first volume cell we always use the KNP $\mathcal{O} ( \Delta y^1 )$ scheme to avoid complications with the boundary at $y=0$, while we use linear extrapolation ($\bar{v}_n = 2 \, \bar{v}_{n - 1} - \bar{v}_{n - 2}$ as well as $\bar{v}_{n + 1} = 3 \, \bar{v}_{n - 1} - 2 \, \bar{v}_{n - 2}$ when using the $\mathcal{O} ( \Delta y^2 )$ scheme) at the right computational boundary $y_\mathrm{max}$.\\
		
	We conclude this subsection with a brief remark on the KT scheme. Using the conservative, equal sided estimate ${a_{j + \frac{1}{2}}^+ = -a_{j + \frac{1}{2}}^-=a_{j + \frac{1}{2}}}$ for the right- and left-sided local speeds $a_{j + \frac{1}{2}}^\pm$, the numerical advection flux \eqref{eq:FV_KNP_H} of the KNP scheme reduces to the advection flux the KT scheme
		\begin{align}
			H_{j + \frac{1}{2}}^{\mathrm{KT}} \equiv \, & \frac{G \big[ t, v_{j + \frac{1}{2}}^+ \big] + G \big[ t, v_{j + \frac{1}{2}}^- \big]}{2} -	\vphantom{\Bigg(\Bigg)}	\label{eq:FV_KT_H}
			\\
			& - a_{j + \frac{1}{2}} \, \frac{v_{j + \frac{1}{2}}^{+} - v_{j + \frac{1}{2}}^{-}}{2} \, ,	\vphantom{\Bigg(\Bigg)}	\nonumber
		\end{align}
	with
		\begin{align}
			 a_{j + \frac{1}{2}} \equiv \max \bigg\{ \bigg| \frac{\partial G}{\partial v} \Big[ t, v_{j + \frac{1}{2}}^{+} \Big] \bigg| , \bigg| \frac{\partial G}{\partial v} \Big[ t,v_{j + \frac{1}{2}}^{-} \Big] \bigg| \bigg\} \, . \label{eq:FV_KT_ajp12}
		\end{align}
	In the first volume cell $\bar{v}_{-1}$ appears outside of the flux limiting procedure and is also present in the first-order accurate reduction of the KT scheme \eqref{eq:FV_KT_H} using Eqs.~\eqref{eq:FV_up12m1st} and \eqref{eq:FV_up12p1st} since the latter is based on a central scheme based on the stencil ${\{\bar{v}_{j-1},\bar{v}_j,\bar{v}_{j+1}\}}$. Lacking the more refined estimates for the right- and left-sided local speeds $a_{j + \frac{1}{2}}^\pm$ of the KNP scheme it is not obvious how to deal with the ghost cell at $\bar{v}_{-1}$. This is, why we chose the KNP scheme for our numerical computations in $y$. The advection flux of the KNP scheme is also suited for the position depended advection flux $F$ of the flow equation \eqref{eq:frg_flow_Ninf_x}in $x$. We have performed some heuristic tests with the KNP scheme and the flow equation \eqref{eq:frg_flow_Ninf_x} in $x$ and we come to the preliminary conclusion that it is in terms of accuracy and performance on par with the KT scheme in this scenario. Nevertheless, further detailed tests might be of interest for upcoming challenges in the context of FRG problems in dimensions $d > 0$ with more sophisticated truncations.
	
\subsection{Riemann problems and (numerical) entropy}\label{app:RPandEntropy}
	
\subsubsection{Riemann problems}
	
\begin{figure}
			\centering
			\includegraphics{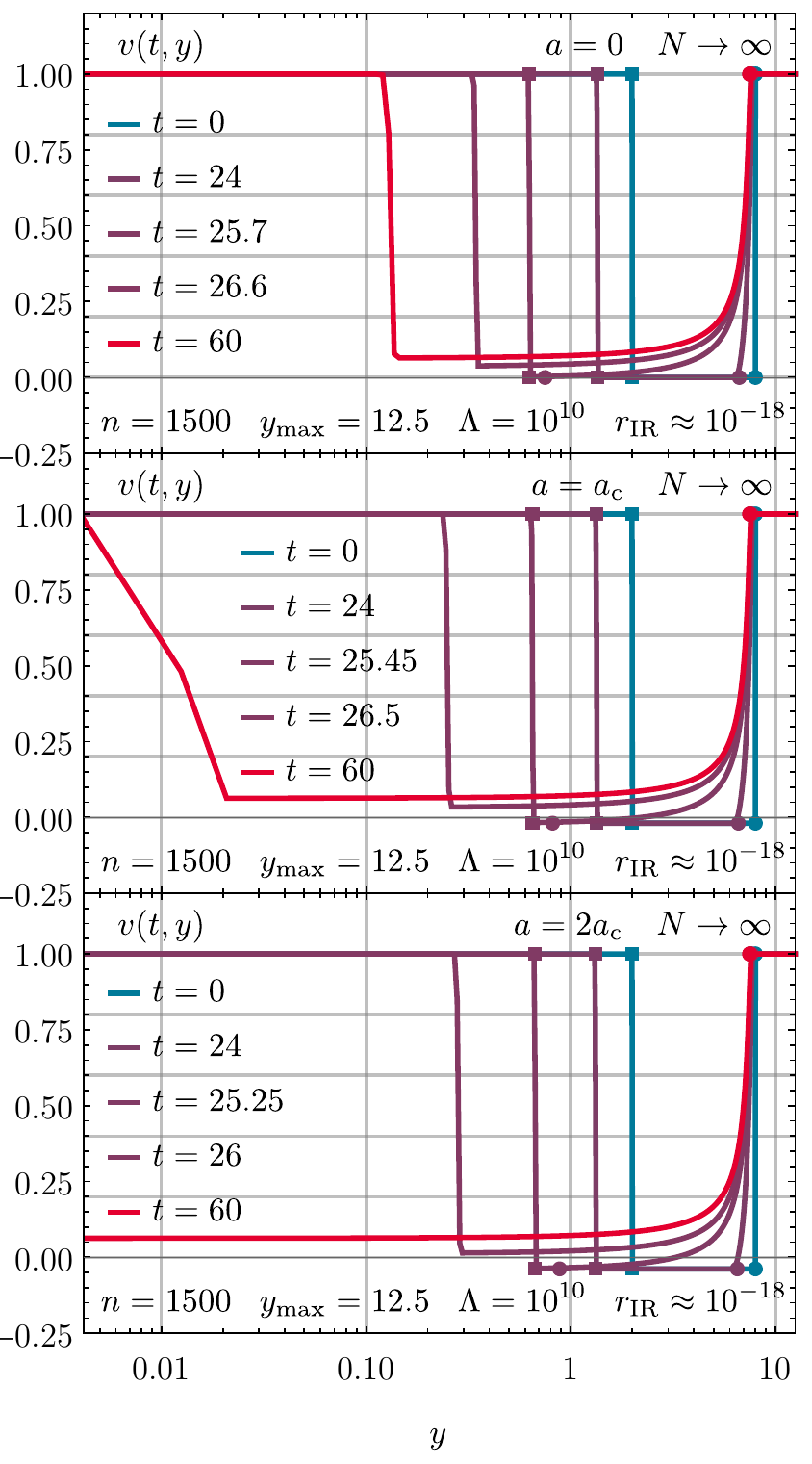}
			\caption{\label{fig:frg_largeN_KNPO1_flows}%
				The RG flow of the derivative of the rescaled effective potential $v ( t, y )$ for the zero dimensional $O(N)$~model in the limit $N \rightarrow \infty$ for the Riemann problems of Eq.~\eqref{eq:RP_vofy} with $a = 0$, $a = a_\mathrm{c}$ and $a = 2 a_\mathrm{c}$ in the upper, middle and lower panel respectively. Computations are equivalent to those presented in Fig.~\ref{fig:frg_largeN_flows} but in the invariant $y$ using the KNP scheme of $\mathcal{O} ( \Delta y^1 )$. The markings, color coding and parameters are the same as in Fig.~\ref{fig:frg_largeN_flows}. We choose a logarithmic scale for the $y$-axis for better visibility around $y = 0$ which is particularly useful for the visualization of the freezing shocks in the IR for $a = 0$ and $a = a_\mathrm{c}$.
			}
		\end{figure}

	In Fig.~\ref{fig:frg_largeN_KNPO1_flows} we present numerical results for the RG flow in the rescaled invariant $y$ using the flow equation \eqref{eq:frg_flow_Ninf_y} with the piecewise constant initial condition of Eq.~\eqref{eq:RP_vofy} obtained with the KNP $\mathcal{O} ( \Delta y^1 )$ scheme discussed in the previous subsection of this appendix. The flow equation \eqref{eq:frg_flow_Ninf_y} with the piecewise constant initial condition of Eq.~\eqref{eq:RP_vofy} constitutes two Riemann problems as outlined in the beginning of this paper in Sec.~\ref{chap:introduction}. The RG flows in $y$ displayed in Fig.~\ref{fig:frg_largeN_KNPO1_flows} are equivalent to the ones in $x$ presented in Fig.~\ref{fig:frg_largeN_flows} hence we will not repeat the qualitative discussion of the results but rather refer to Sub.Sub.Sec.~\ref{subsubsec:infiniteNflows}. In the following we will instead focus on certain aspects and problems inherent to the formulation and solution in the rescaled invariant $y$.\\
	
	For small RG times $t \lesssim 25$ the RG flows present as typical Riemann problems with a moving shock wave and an rarefaction fan. In Fig.~\ref{fig:frg_largeN_KNPO1_flows} the evolution for $t \lesssim 25$ is for all $a$ under consideration similar to the dynamics studied in Fig.~2~(a) of Ref.~\cite{Grossi:2019urj}, which originally motivated the chosen initial condition in this work. Beyond $t \approx 25$ the shock wave and the left tip of the rarefaction fan start interacting leading to a freeze-out of the shock wave for $a = 0$ with ${v ( t = 0, y = 0 ) = 1 = \partial_x v ( t = 0, x ) \big|_{x = 0}}$. For $a=2a_\mathrm{c}$ the shock moves out of the computational domain at $y=0$ and we recover ${v(t=0,y=0)=\frac{1}{16}=\partial_x v(t=0,x)\big|_{x = 0}}$. So far in complete agreement with the corresponding results in $x$ of Sub.Sub.Sec.~\ref{subsubsec:infiniteNflows}. For $a=a_\mathrm{c}$ we observe the remnant of the shock wave in the computational interval but the shock is strongly deformed by numerical(!) diffusion/the finite resolution of the computation. We will come back to this issue after briefly commenting on the numerical errors of the computation.
		
	The relative numerical errors for the $1$PI-two-point function $\Gamma^{(2)}$ corresponding to the IR results displayed in Fig.~\ref{fig:frg_largeN_KNPO1_flows} are presented for both the KNP $\mathcal{O} ( \Delta y^1 )$ and the KNP $\mathcal{O} ( \Delta y^2 )$ schemes in Tab.~\ref{tab:KNP1500errorsLargeN}. We choose to plot the results of the KNP $\mathcal{O} ( \Delta y^1 )$ scheme for their slightly better accuracy for $a=a_\mathrm{c}$. When comparing the errors to the ones of Tab.~\ref{tab:KT1500errorsLargeN} for the KT scheme and the flow equation in $x$ the only notable difference is in fact at $a=a_\mathrm{c}$. The error obtained with the KNP scheme is significantly worse by more than twelve orders of magnitude for both KNP $\mathcal{O} ( \Delta y^1 )$ and KNP $\mathcal{O} ( \Delta y^2 )$.
	
	The situation at $a=a_\mathrm{c}$ can be understood quite easily. The presented numerical computations use ${n=1500}$ volume cells equidistantly distributed in the interval ${y \in [ 0, 12.5 ]}$ resulting in ${\Delta y = \frac{1}{120} \simeq 8.33 \cdot 10^{-3}}$. Consequently the first two volume cells are centered at ${y_0 = \frac{1}{240} \simeq 4.17 \cdot 10^{-3}}$ and ${y_1 = \frac{1}{80} = 1.25 \cdot 10^{-2}}$. Those two volume cells are are clearly visible in the middle panel of Fig.~\ref{fig:frg_largeN_KNPO1_flows} and contain the frozen shock for ${a = a_\mathrm{c}}$. Form our computation in $x$ we found with the fit \eqref{eq:xis_1ac_fit} that the shock for $a = a_\mathrm{c}$ approaches $x = 0$ with $0.983 \, \Delta x^{0.413}$. For ${n = 1500}$ volume cell this amounts to a numerical shock position of ${| x | \approx 0.095}$ and consequently ${y \approx 4.513 \cdot 10^{-3}}$ which is for a computation in $y$ with ${n = 1500}$ retaining ${x_\mathrm{max} = 5 \Leftrightarrow y_\mathrm{max} = 12.5}$ approximately at the center of the first volume cell. Having no volume cell to the right of the shock makes it numerically impossible to resolve ${v(t=0,y=0)=1=\partial_x v(t=0,x)\big|_{x = 0}}$ accurately. Using the fit \eqref{eq:xis_1ac_fit} we can extrapolate that having the shock centered in the second or third cell would already require an extensive amount of volume cells namely ${n=3.7\cdot 10^5}$ or ${n=6.4\cdot 10^6}$ respectively while maintaining ${y_\mathrm{max}=12.5}$. Computations with $10^5$ and more volume cells overtax our current implementation and computational capacities, see App.~\ref{app:FRGnumerics} for details.
	
	Resolving dynamics at small $x$ with an equidistant grid of volume cells in $y = \tfrac{1}{2} \, x^2$ is in general difficult because equidistant cells in $y$ have a poor resolution around $x = \sqrt{2 y} = 0$. A drastic example is the freezing shock for $a = a_\mathrm{c}$ at $x = 0$, where the scaling $\propto \Delta x^{0.413}$ is already challenging. A situation with a scaling $\propto\Delta x^{p}$ with $p\geq\frac{1}{2}$ is also conceivable. Such a scenario would be impossible to resolve with an equidistant grid in the rescaled invariant $y = \frac{1}{2} \, x^2$. To improve or in some cases even facilitate computations at all around $x=0$ in the rescaled invariant $y$ a non-uniform mesh in $y$ seems necessary. The generalization of the KT and KNP scheme to non-uniform grids is straightforward in one spatial dimension, see, \textit{e.g.}, Ref.~\cite{Kurganov2020}, but will not be discussed in this work. 
	
	Anyhow, non-uniform grids and potentially adaptive mesh refinement techniques are of increasing importance when considering flow equations with more than one spatial domain. These improvements will for sure become important within the next years of FRG computations, \textit{e.g.}\ in low-energy effective models of QCD see, \textit{e.g.}, Refs.~\cite{Strodthoff:2011tz,Mitter:2013fxa,Rennecke:2016tkm,Lakaschus:2020caq}, with more than one condensate (quark-anti-quark and di-quark).\\
		\begin{table}[b]
			\caption{\label{tab:KNP1500errorsLargeN}%
				Numerical relative errors for the $1$PI-two-point function $\Gamma^{(2)}$, see Eq.~\eqref{eq:Gamma2vofy}, for the results plotted in Fig.~\ref{fig:frg_largeN_KNPO1_flows} and equivalent results computed with the second order accurate KNP scheme with corresponding exact reference values from the last row of Tab.~\ref{tab:Gamma2N}. The scaling of these errors with the number of volume cells can be found in Tabs.~\ref{tab:KNPac} and \ref{tab:KNP2ac} for $a = a_\mathrm{c}$ and $a = 2 a_\mathrm{c}$.
			}
			\begin{ruledtabular}
				\begin{tabular}{l c c c}
					Scheme								&	$a = 0$					&	$a = a_\mathrm{c}$		&	$a = 2 a_\mathrm{c}$
					\\
					\colrule\addlinespace[0.25em]
					KNP $\mathcal{O} ( \Delta y^1 )$	&	$2.220 \cdot 10^{-16}$	&	$1.700 \cdot 10^{-2}$	&	$8.255\cdot 10^{-3}$
					\\
					KNP $\mathcal{O} ( \Delta y^2 )$	&	$2.887 \cdot 10^{-15}$	&	$5.695 \cdot 10^{-1}$	&	$2.787\cdot 10^{-3}$
				\end{tabular}
			\end{ruledtabular}
		\end{table}

\subsubsection{Entropy and irreversibility}

	We now turn to the discussion of the (numerical) entropy associated with the purely advective RG flows in the rescaled invariant $y$ at infinite $N$ which is the main motivation for this whole appendix. In part II of this series of publications \cite{Koenigstein:2021rxj} we discussed the concept of (numerical) entropy of RG flows and its relation to the inherent irreversibility of RG flows in detail. We further argued for a connection between the (numerical) entropy of RG flows and Zamolodchikov's \cite{Zamolodchikov:1986gt} or more recent \cite{Codello:2013iqa,Codello:2015ana} formulations of the $\mathcal{C}-$function.
	
	In Ref.~\cite{Koenigstein:2021rxj} we focused on the limiting case $N=1$ of the zero-dimensional $O(N)$~model discussing numerical entropy and irreversibility of the purely diffusive RG flow equations in this case. The focus of this paper is the opposite limit of $N\rightarrow\infty$ yielding purely advective flow equations. While a (numerical) entropy production is almost intuitively understood for diffusive problems the present situation might seem less obvious for a non-expert reader. It is however well known from the study of non-linear advection equations, see, \textit{e.g.}, the textbooks \cite{Lax1973,Ames:1992,LeVeque:1992,LeVeque:2002,Hesthaven2007,Toro2009,RezzollaZanotti:2013}, that there is a meaningful notion of numerical entropy and that its increase is linked to the appearance and/or interaction of discontinuities like shocks and rarefaction waves. An increase in numerical entropy signals the irreversibility of the underlying flow, see, \textit{e.g.}, Refs.~\cite{LeVeque:2002,RezzollaZanotti:2013} for this in the context of non-linear (especially hyperbolic) conservation laws.\\

	Defining or constructing an explicit numerical entropy functional for general non-linear conservation laws is a difficult task especially when source terms are involved, \textit{cf.} Refs.~\cite{Monthe:2001,Beneito2008,Chen2011May,Bessemoulin:2012} and references therein. 
	
	When considering the flow equations \eqref{eq:frg_flow_x} and \eqref{eq:frg_flow_y} in $x$ or $y$ respectively, we note that the formulation in $x$ ($y$) involves a position dependent advection term (diffusion term). When executing the $x$-derivative in Eq. \eqref{eq:frg_flow_x} we can differentiate between three contributions in the resulting flow equation in primitive form: a parabolic diffusion term $\propto \partial_x^2 v(t,x)$ with a non-linear diffusion coefficient, a hyperbolic advection term $\propto \partial_x v(t,x)$ with a non-linear, position dependent advection velocity $\partial_v F$ and a non-linear, position dependent internal source term $\propto v(t,x)$ stemming from the product rule. As a consequence of the latter term the \textit{r.h.s} of the flow eq. \eqref{eq:frg_flow_x} and hence $\partial_t v(t,x)$ is non-vanishing for $v(t,x)$ constant in $x$. Similarly the flow Eq.~\eqref{eq:frg_flow_y} in $y$ contains such a non-linear, position dependent internal source term $\propto v(t,y)$ arising form the derivative of the explicitly $y$-dependent second term in Eq.~\eqref{eq:frg_flow_y}. Those internal source terms, explicit $x$- or $y$-dependences before executing the derivatives, in the flow equations in primitive form make the construction of explicit numerical entropy functionals at finite $N > 1$ challenging.
	
	At $N=1$ the formulation in $x$ manifests as a pure diffusion equation with a position independent diffusion term in Eq.~\eqref{eq:frg_flow_x}. Using standard techniques the authors of this papers and collaborators were able to construct a class of numerical entropy functionals for the $N=1$ flow equation in $x$ in part II of this series of publications \cite{Koenigstein:2021rxj}. The so-called \textit{total variation} (TV) \cite{HARTEN1983357} -- which is simply the arc length of $v(t,x)$ -- is among the class of viable entropy functionals at $N=1$.
	
	Incidentally in the opposite limit $N\rightarrow\infty$ but using the flow Eq.~\eqref{eq:frg_flow_Ninf_y} in the rescaled invariant $y$ the total variation is again a viable entropy functional. This goes back to general properties of (weak) solutions of purely hyperbolic non-linear advection equations -- like our $N\rightarrow\infty$ flow Eq.~\eqref{eq:frg_flow_Ninf_y}. Among other general qualitative statements about monotonicity and convexity (weak) solutions of hyperbolic non-linear advection equations like Eq.~\eqref{eq:frg_flow_Ninf_y} have a decreasing arc length -- they are 
	\textit{total variation diminishing} (TVD) or more precisely \textit{total variation non-increasing} (TVNI) \cite{HARTEN1983357,Lax1973} -- during time evolution when considered on a finite interval, for further details see also Refs.~\cite{LeVeque:1992,Toro2009} and especially Ref.~\cite{Redheffer1974Mar}. In terms of volume averages the arc length/total variation can computed using
		\begin{align}
			& \mathrm{TV} [ v ( t, y ) ] \equiv	\vphantom{\bigg(\bigg)}	\label{eq:TVdiscrete}
			\\
			\equiv \, & \mathrm{TV} [ \{ \bar{v}_i ( t ) \} ] \equiv \sum_{i = 0}^{n - 1} | \bar{v}_{i + 1} ( t ) - \bar{v}_{i} ( t ) | \, ,	\vphantom{\bigg(\bigg)}	\nonumber
		\end{align}
	and a corresponding entropy functional may be defined as
		\begin{align}
			\mathcal{C}\equiv \mathrm{TV}[v(t=0,y)]-\mathrm{TV}[v(t,y)]\,, \label{eq:Cfunction}
		\end{align}
	for further details see part II of this series of publications \cite{Koenigstein:2021rxj} and references therein. Since solutions of the underlying flow Eq.~\eqref{eq:frg_flow_Ninf_y} are TVNI ($\partial_t \mathrm{TV}[v(t,y)]\leq 0$) the entropy functional $\mathcal{C}$ is non-decreasing ($\partial_t\mathcal{C}\geq 0$).
	
	Solutions of the flow Eq.~\eqref{eq:frg_flow_x} in $x$ at $N>1$ are in general not TVNI. A fact we tested in numerical experiments with several initial conditions at various $N>1$ \cite{Koenigstein:2021syz,Koenigstein:2021rxj}. The loss of the TVNI property is most likely directly linked to the explicit position dependences in the flow equation manifesting as source terms when executing the $x$-derivatives of the \textit{r.h.s} of Eq.~\eqref{eq:frg_flow_x}. Formal results supporting this can be found in Ref.~\cite{Redheffer1974Mar}: non-linear parabolic differential equations of the type $0=\partial_t v-f(t,z,v,\partial_z v,\partial_{z}^2 v)$ have TVNI solutions if (among some other restrictions) the flux $f$ vanishes \textit{i.e.} $0=f(t,z,v,0,0)$ on constant solutions $0=\partial_z v=\partial_{z}^2 v$. The latter is not the case for flow equations in $x$ at $N>1$ and in $y$ for finite $N$ as discussed earlier in this subsection. It is intuitively obvious that source terms can increase the arc length of a (weak) solution and implications in the context of TVD schemes are discussed in, \textit{e.g.}, Refs.~\cite{Monthe:2001,Beneito2008,Chen2011May,Bessemoulin:2012}.
	
\begin{figure}
			\centering
			\includegraphics{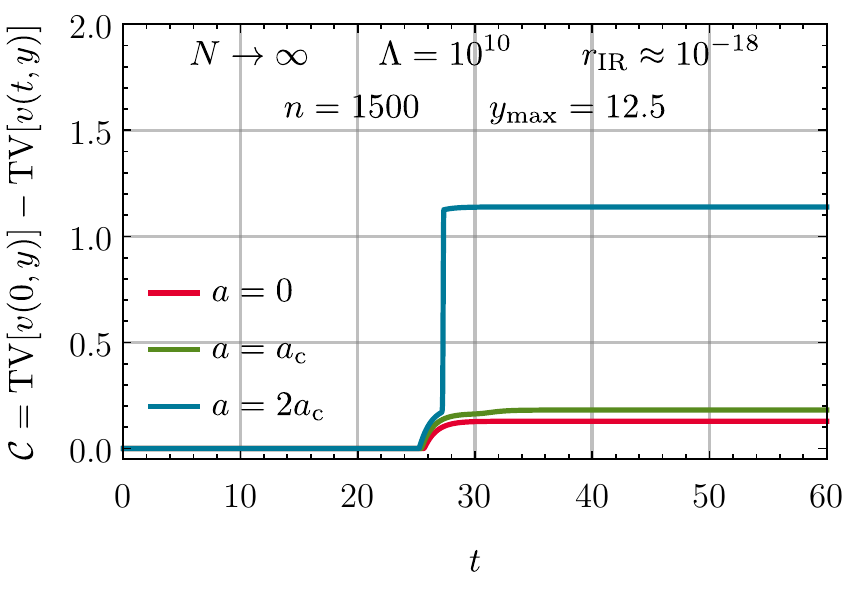}
			\caption{\label{fig:frg_largeN_TVD}%
				The RG flow of the $\mathcal{C}$-function, see Eq.~\eqref{eq:Cfunction}, for the zero dimensional $O(N)$~model in the limit ${N \rightarrow \infty}$ for the Riemanns problem of Eq.~\eqref{eq:RP_vofy} with $a = 0$, $a = a_\mathrm{c}$, and $a = 2 a_\mathrm{c}$ obtained with the KNP scheme of $\mathcal{O} ( \Delta y^1 )$. We observe plateaus in the UV and IR. The IR plateaus end for the individual values of $a = 0$, $a = a_\mathrm{c}$, and $a = 2 a_\mathrm{c}$ at the RG times when the shock wave and rarefaction fan intersect namely at $t \approx 25.718$, $25.469$, and $25.270$ respectively. The second jump in the curves for $a \geq a_\mathrm{c}$ is due to the collision of the shock waves at $x = 0$.
			}
		\end{figure}

	For $N \rightarrow \infty$ solutions in $x$ are still not TVNI but a reformulation in $y$ eliminates the explicit position dependence in the advection flux and the resulting source term. The solutions of the flow Eq.~\eqref{eq:frg_flow_Ninf_y} in $y$ are TVNI. A fact we tested numerically in this appendix, see Fig.~\ref{fig:frg_largeN_TVD}, for the Riemann problems posed by the initial condition~\eqref{eq:RP_vofy} with the flow Eq.~\eqref{eq:frg_flow_Ninf_y} and which is theoretically well established \textit{cf.}~Refs.~\cite{HARTEN1983357,Lax1973}.\\
		
	We conclude this appendix with a qualitative discussion of the numerical entropy for the Riemann problems posed by the initial condition~\eqref{eq:RP_vofy} with the flow Eq.~\eqref{eq:frg_flow_Ninf_y} for different $a$. The numerical entropies associated to the flows presented in Fig.~\ref{fig:frg_largeN_KNPO1_flows} are plotted in Fig.~\ref{fig:frg_largeN_TVD}.
	
	The numerical entropy stays constant in the UV up until the point where the shock wave and rarefaction fan intersect namely at $t \approx 25.718$, $25.469$, and $25.270$ for $a = 0$, $a = a_\mathrm{c}$, and $a = 2 a_\mathrm{c}$ respectively. Since both shock and rarefaction wave are already present in the initial condition $v(t=0,y)$ their simple advection does not increase the numerical entropy of Eq.~\eqref{eq:Cfunction}. The flow in the UV is therefore arguable reversible, which can be seen from the analytic solutions via the method of characteristics, but practical computations involving a finite resolution $\Delta y$ and finite precision during time evolution prevent a reversion by numerically integrating up in time $t$.
	
	Between $t \approx 25$ and $t \approx 35$ we observe an increase in numerical entropy related to the interaction of the shock and the rarefaction fan. For $a\leq a_\mathrm{c}$ the rise in entropy is rather small related to only marginal changes in arc length/total variation during the flow, see upper and middle panel of Fig.~\ref{fig:frg_largeN_KNPO1_flows}. For $a>a_\mathrm{c}$ namely $a=2a_\mathrm{c}$ we observe a steep rise in entropy at $t\approx 27.275$, which is the RG time at which the shock leaves the computational domain for $a=2a_\mathrm{c}$. Without the shock the arc length/total variation decreases dramatically leading to the observed rise in numerical entropy.
	
	In the IR for $t\gtrsim 35$ we again observe a plateau in the numerical entropy, related to the fact, that $k(t)$ for $t\gtrsim 35$ is sufficiently below the internal model scales of the problem under consideration meaning that all relevant fluctuations are already included. The plateaus in the numerical entropy in the UV and IR are indicators of RG consistency and sufficiently small numerical IR cutoffs respectively.

\section{(Numerical) parameters of the FRG computations}
\label{app:FRGnumerics}
	In this appendix we present further details on numerical parameters used for the numerical FRG computations with the KT and KNP scheme. To numerical integrate the RG flow equations in $x$ at finite and infinite $N$, see Eqs.~\eqref{eq:frg_flow_x} and \eqref{eq:frg_flow_Ninf_x} respectively, we employ the KT scheme in its semi-discrete form, see Sec.~IV~C of part I of this series of publications \cite{Koenigstein:2021syz} and the original publication \cite{KTO2-0} for details. To numerical integrate the RG flow equations in $y$ infinite $N$, see Eq.~\eqref{eq:frg_flow_Ninf_y}, we employ the first and second order KNP scheme in its semi-discrete form, see App.~\ref{app:KNP} and the original publication \cite{KTO2-1} for details.\\
	
	The semi-discrete KT and KNP scheme require an external numerical time stepper for evolution in $t$ -- the solution of the ODE system for the cell averages $\{\bar{v}_i(t)\}$. For this purpose we employ the default numerical ODE-solver \textit{NDSolve} of \texttt{Mathematica} \cite{Mathematica:12.2} with a \textit{PrecisionGoal} and \textit{AccuracyGoal} of 10.
	
	Given a finite volume solution $\{\bar{v}_{i}(t)\}$ for the $x$-derivative of the rescaled potential computed with the KT scheme, the $1$PI-two-point function can be computed by means of numerical differentiation using the first order difference coefficient\footnote{Due to the anti-symmetry of $v ( t, x )$ the first order and second order finite difference stencils are identical.}
		\begin{align}
			\Gamma^{(2)}_\mathrm{KT} = \, &\frac{v \big( t_\mathrm{IR}, x_1 \big) - v \big( t_\mathrm{IR}, x_0 = 0 \big)}{\Delta x} = \frac{\bar{v}_{1} ( t_\mathrm{IR} )}{\Delta x} \, ,	\label{eq:Gamma2vofx}
		\end{align}
	where $x_1 = \Delta x$ and we used the fact that $\bar{v}_0 ( t ) = 0$ due to anti-symmetry and the cell average of the second cell ${\bar{v}_{1}(t_\mathrm{IR})}$. Higher order finite difference coefficients can also be used, see part I of this series of publications~\cite{Koenigstein:2021syz}. When computing in the invariant $y$, thus in the $y$-derivative of the rescaled potential, the $1$PI-two-point function can be extracted directly from the volume averages $\{\bar{v}_{\mathrm{S},i}(t)\}$
		\begin{align}
			\Gamma^{(2)}_\mathrm{S}= \, &v_\mathrm{S}(t_\mathrm{IR}, y=0) = \bar{v}_{\mathrm{S},0}(t_\mathrm{IR})\, , \label{eq:Gamma2vofy}
		\end{align}
		with the cell average of the first cell ${\bar{v}_{\mathrm{S},0}(t_\mathrm{IR})}$ computed with the scheme $\mathrm{S}$. Relative numerical errors for a solution computed with the scheme $\mathrm{S}$ and related convergence rates are given by
		\begin{align}
			\epsilon_\mathrm{S} \equiv \, & \bigg| \frac{\Gamma^{(2)}_\mathrm{S}}{\Gamma^{(2)}} - 1 \bigg| \, ,	\vphantom{\Bigg(\Bigg)}	\label{eq:epsilonS}
			\\
			r_\mathrm{S} \equiv \, & \frac{\ln \big( \tfrac{\epsilon_{\mathrm{S}, i}}{\epsilon_{\mathrm{S}, i - 1}} \big)}{\ln \big( \tfrac{n_{i - 1}}{n_{i}} \big)} \, ,	\vphantom{\Bigg(\Bigg)}	\label{eq:rateS}
		\end{align}
	where we compare to the exact reference values $\Gamma^{(2)}$ computed from the integral \eqref{eq:ON_expectation_value} of the instructive toy model discussed in Sub.Sec.~\ref{subsec:RP} and presented in Tab.~\ref{tab:Gamma2N}. The convergence rate $r_\mathrm{S}$ compares errors obtained for computations involving a differing number of volume cells $n_{i-1}>n_i$ with otherwise unchanged numerical parameters.\\
	
	All numerical computations have been performed on an Intel\textsuperscript{\tiny\textcopyright} Core{\texttrademark} i7-8750H processor running up to 6 threads simultaneously. The wall times displayed in this appendix are not averaged over multiple runs and are given here to allow a comparison of computational cost between numerical computations at finite and infinite $N$ with the KT and KNP schemes. The total single thread wall time for all numerical computations discussed and displayed in this paper is approximately 60 hours.

\subsection{Computations in the large-\texorpdfstring{$N$}{N} limit}
\label{app:FRGnumericsLargeN}

	We now turn to the discussion of spatial resolution $\Delta x$ and $\Delta y$ for computations using the KT and KNP scheme in the limit $N\rightarrow\infty$. For the purely advective problems a computational extend with $x_\mathrm{max} = 5$ and equivalently $y_\mathrm{max} = \tfrac{1}{2} \, x_\mathrm{max}^2 = 12.5$ have proven sufficient.\\
	
\begin{figure}
			\centering
			\includegraphics{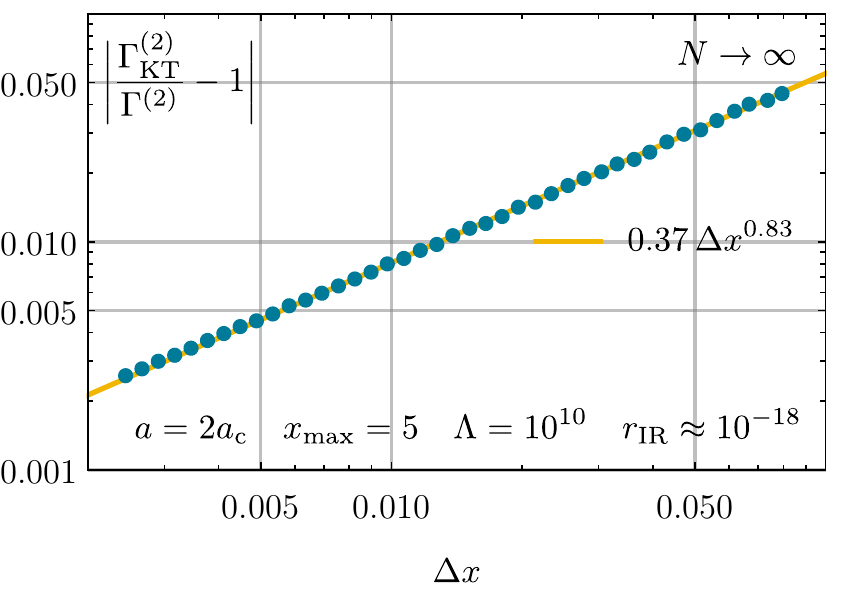}
			\caption{\label{fig:frg_largeN_2ac_Deltax}%
				The scaling of the relative error with decreasing grid spacing $\Delta x$ of the numerical results ({blue dots}) from the finite-volume KT scheme for the two-point function $\Gamma^{(2)}$ for the zero dimensional $O(N)$~model in the limit ${N \rightarrow \infty}$ for the initial condition \eqref{eq:RP_vofx} with $a = 2 a_\mathrm{c}$. The numerical derivatives at $x = 0$ of $v ( t_\mathrm{IR} = 60, x )$ were calculated using the first order difference coefficient \eqref{eq:Gamma2vofx}. The {yellow} straight line is included for optical guidance and represent a scaling with $\mathcal{O} ( \Delta x^{0.83} )$.
			}
		\end{figure}

	In the limit ${N \rightarrow \infty}$ computation with the KT scheme, \textit{cf.} Sub.Sub.Sec.~\ref{subsubsec:infiniteNflows}, converge rapidly towards the exact results in the IR for $a=0$ and $a = a_\mathrm{c}$: only approximately 120 (290) volume cells are required to reach relative errors on the level of machine/double precession ($\approx 10^{-15}$) for $a = 0$ ($a = a_\mathrm{c}$) and $\Lambda=10^{10}$ with $x_\mathrm{max}=5$. Corresponding errors and convergence rates for $a = 2 a_\mathrm{c}$ can be found in Tab.~\ref{tab:frg_largeN_2ac_rates} as well as Fig.~\ref{fig:frg_largeN_2ac_Deltax}. For $a = 2 a_\mathrm{c}$ (and other $a>a_\mathrm{c}$) we observe a convergence rate of the KT scheme in the number/size of volume cells of $\approx 0.8$. For the problem under consideration the theoretical scaling of $\Delta x^2$ and respective convergence rate of $2$ for the KT scheme is not archived practically. Reduced practical convergence rates are however not uncommon for involved equation systems and also depend strongly on the monitor used to compute them \cite{KTO2-0,KTO2-1}.\\

		\begin{table}[b]
			\caption{\label{tab:frg_largeN_2ac_rates}%
				Relative error $\epsilon_\mathrm{KT}$, corresponding rate of convergence $r_\mathrm{KT}$, see Eqs.~\eqref{eq:epsilonS} and \eqref{eq:rateS}, and wall time in seconds for a selected set of computations with varying number of grid points $n$ from Fig.~\ref{fig:frg_largeN_2ac_Deltax}. The average rate of convergence of $0.822$ is compatible with the trend of $0.83$ displayed in Fig.~\ref{fig:frg_largeN_2ac_Deltax}.
			}
			\begin{ruledtabular}
				\renewcommand{\arraystretch}{1.15}
				\begin{tabular}{l c c c}
					$n$			&	$\epsilon_\mathrm{KT}$	&	$r_\mathrm{KT}$	&	$t_\mathrm{W} ( \mathrm{s} )$
					\\
					\colrule\addlinespace[0.25em]
					$64$		&	$4.472 \cdot 10^{-2}$	&	$-$		&	$1.3 \cdot 10^{+0}$
					\\
					$128$		&	$2.474 \cdot 10^{-2}$	&	0.854	&	$4.1 \cdot 10^{+0}$
					\\
					$256$		&	$1.419 \cdot 10^{-2}$	&	0.802	&	$1.2 \cdot 10^{+1}$
					\\
					$512$		&	$8.004 \cdot 10^{-3}$	&	0.826	&	$6.1 \cdot 10^{+1}$
					\\
					$1024$		&	$4.502 \cdot 10^{-3}$	&	0.830	&	$4.1 \cdot 10^{+2}$
					\\
					$2048$		&	$2.586 \cdot 10^{-3}$	&	0.800	&	$5.6 \cdot 10^{+3}$
				\end{tabular}
			\end{ruledtabular}
		\end{table}

	We turn to the discussion of the KNP scheme in this context. The numerical challenges of a formulation in $y$ using the KNP scheme for $a$ approaching $a_\mathrm{c}$ related to the freezing of the shock at $x=y=0$ were discussed at length in Sub.Sec.~\ref{app:RPandEntropy}. Numerical errors, convergence rates and runtimes for the KNP scheme for $a=a_\mathrm{c}$ and $a=2a_\mathrm{c}$ can be found Tab.~\ref{tab:KNPac} and \ref{tab:KNP2ac} respectively. For $a=a_\mathrm{c}$ the rate of convergence improves with the number of volume cells as the freezing shock gets resolved better for larger $n$. The first-order accurate version of the KNP scheme has a better numerical error and rate of convergence for a given $n$ when compared to the second-order KNP scheme for $a=a_\mathrm{c}$. We observe that the KNP $\mathcal{O} ( \Delta y^1 )$ scheme runs marginally faster than KNP $\mathcal{O} ( \Delta y^2 )$ scheme.
	
	For $a=2a_\mathrm{c}$ the KNP $\mathcal{O} ( \Delta y^2 )$ scheme performs very similar to the KT scheme, \textit{cf.} Tab.~\ref{tab:KNP2ac} and \ref{tab:frg_largeN_2ac_rates}, in terms of numerical errors and convergence rates while running faster (twice as fast for small number of volume cells up to more than thirty times faster for 2048 volume cells) than KT scheme. The KNP $\mathcal{O} ( \Delta y^1 )$ scheme has a slower rate of convergence and larger numerical errors at a similar runtime when compared to the KNP $\mathcal{O} ( \Delta y^2 )$ scheme.	
	
	For $a=0$ we again observe rapid convergence against the exact results approximately 680 (740) volume cells are required to reach relative errors on the level of double precession for the KNP scheme of $\mathcal{O} ( \Delta y^2 )$ ($\mathcal{O} ( \Delta y^1 )$). Again we observe the cost of the relatively low resolution in $x=\sqrt{2y}$ at small $y$ when compared to a formulation in $x$ using the KT scheme, where only 120 cells are necessary to reach double precession for $a = 0$ in the $N \rightarrow \infty$ limit.\\

\begin{figure}
			\centering
			\includegraphics{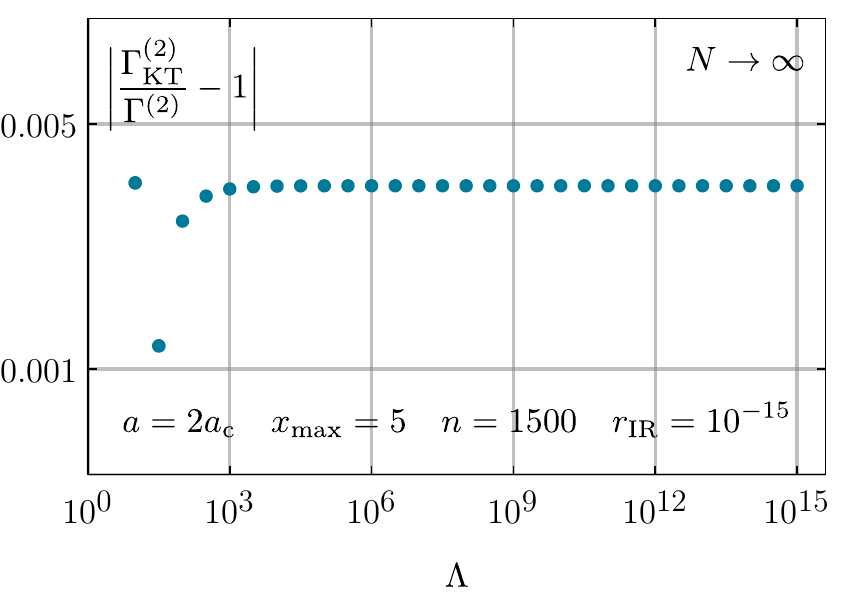}
			\caption{\label{fig:frg_largeN_2ac_Lambda}%
				The UV initial scale $\Lambda$ dependence of the relative error of the numerical results ({blue dots}) from the finite-volume KT scheme for the two-point function $\Gamma^{(2)}$ for the zero dimensional $O(N)$~model in the limit ${N \rightarrow \infty}$ with the initial condition \eqref{eq:RP_vofx} with $a = 2 a_\mathrm{c}$. The numerical derivatives at $x = 0$ of $v ( t_\mathrm{IR}, x )$ with fixed $r(t_\mathrm{IR})=10^{-15}$ were calculated using the difference coefficient \eqref{eq:Gamma2vofx}.
			}
		\end{figure}

	A discussion of RG consistency related to sufficiently high UV initial scales $\Lambda$ along the lines of part I of this series of publications \cite{Koenigstein:2021syz} is not very illuminating for the toy model under consideration. In Fig.~\ref{fig:frg_largeN_2ac_Lambda} we plot the numerical errors of the two-point function in the IR for KT scheme runs with $n=1500$ volume cells at varying UV initial scales. We hit a plateau beginning at $\Lambda\approx 3.2\cdot10^{3}$ where a further increase of the UV initial scale yields no improvement of the numerical error since it is dominated by the one related to the finite spatial resolution, \textit{cf.} Tabs.~\ref{tab:KT1500errorsLargeN} and \ref{tab:frg_largeN_2ac_rates}. The UV initial scale $\Lambda=10^{10}$ used for all other computations of this work, lies deep in the plateau implying that errors related to violations of RG consistency are several orders of magnitude smaller than the error related to the spatial discretization used in this work.

\subsection{Computations at finite \texorpdfstring{$N$}{N}}
\label{app:FRGnumericsFiniteN}
	
	In Tabs.~\ref{tab:frg_N2_2ac_rates} and \ref{tab:frg_N32_2ac_rates} we present results for numerical errors, convergence rates and runtimes for the finite $N$ KT computations of Sub.Sec.~\ref{subsec:FRGfiniteN} for completeness sake only. A detailed discussion of numerical parameters, errors and RG consistency at finite $N$ can be found in part I of this series of publications \cite{Koenigstein:2021syz}.
	
	At this point we only briefly comment on the numerical errors and related convergence rates. For finite $N$ the KT scheme operates on average with higher practical convergence rates than for infinite $N$. The results for small $N$, namely $N=2$ in Tab.~\ref{tab:frg_N2_2ac_rates}, are in agreement with the results of part I of this series of publications \cite{Koenigstein:2021syz}, \textit{cf.} especially test case I of Sub.Sec.~V~A. The oscillations in convergence rate and errors are related to the finite volume discretization of discontinuous initial conditions \cite{Koenigstein:2021syz}. The overall numerical errors at $N=2$ and $a=2a_\mathrm{c}$ are up to three orders of magnitude lower than the corresponding ones at $N\rightarrow\infty$ even though the resolution $\Delta x$ at $N=2$ is double the one at $N\rightarrow\infty$ for a given number of volume cells $n$ since we use $x_\mathrm{max}=10$ at $N=2$ instead of $x_\mathrm{max}=5$ used at $N\rightarrow\infty$. Comparing the results at $N=2$ and $N=32$ in Tabs.~\ref{tab:frg_N2_2ac_rates} and \ref{tab:frg_N32_2ac_rates} respectively we note that an increase of $N$ comes at a cost in both, runtime and numerical error. Computations at larger finite $N$ are numerically more challenging.

\onecolumngrid

		\begin{table}[!htpb]
			\caption{\label{tab:KNPac}
				Relative errors $\epsilon$, corresponding rates of convergence $r$, see Eqs.~\eqref{eq:epsilonS} and \eqref{eq:rateS}, and wall times in seconds for varying number $n$ of volume cells for the RG flows of the toy model in the limit ${N \rightarrow \infty}$ with $a = a_\mathrm{c}$ computed with the second and first order KNP scheme formulated in the invariant $y$. The corresponding KNP $\mathcal{O} ( \Delta y^1 )$ flow including parameters are displayed in the middle panel of Fig.~\ref{fig:frg_largeN_KNPO1_flows}.
			}
			\begin{ruledtabular}
				\renewcommand{\arraystretch}{1.15}
				\begin{tabular}{l c c c c c c}
							&	\multicolumn{3}{c}{KNP $\mathcal{O} ( \Delta y^2 )$}								&	\multicolumn{3}{c}{KNP $\mathcal{O} ( \Delta y^1 )$ }
					\\
							\cmidrule{2-4}																			\cmidrule{5-7} 
					$n$		&	$\epsilon_\mathrm{KNPO2}$	&	$r_\mathrm{KNPO2}$	&	$t_\mathrm{W} (\mathrm{s})$	&	$\epsilon_\mathrm{KNPO1}$	&	$r_\mathrm{KNPO1}$	&	$t_\mathrm{W}\ (\mathrm{s})$
					\\
					\colrule\addlinespace[0.25em]
					$64$	&	$7.069 \cdot 10^{-1}$		&	$-$					&	$4.746 \cdot 10^{-1}$		&	$4.610 \cdot 10^{-1}$		&	$-$					&	$2.896 \cdot 10^{-1}$
					\\
					$128$	&	$6.857 \cdot 10^{-1}$		&	$0.044$				&	$1.400 \cdot 10^{+0}$		&	$3.461 \cdot 10^{-1}$		&	$0.414$				&	$1.020 \cdot 10^{+0}$
					\\
					$256$	&	$6.593 \cdot 10^{-1}$		&	$0.057$				&	$4.495 \cdot 10^{+0}$		&	$2.055 \cdot 10^{-1}$		&	$0.752$				&	$3.490 \cdot 10^{+0}$
					\\
					$512$	&	$6.276 \cdot 10^{-1}$		&	$0.071$				&	$1.713 \cdot 10^{+1}$		&	$8.831 \cdot 10^{-2}$		&	$1.220$				&	$1.225 \cdot 10^{+1}$
					\\
					$1024$	&	$5.904 \cdot 10^{-1}$		&	$0.088$				&	$5.403 \cdot 10^{+1}$		&	$3.249 \cdot 10^{-2}$		&	$1.440$				&	$4.681 \cdot 10^{+1}$
					\\
					$2048$	&	$5.477 \cdot 10^{-1}$		&	$0.108$				&	$1.603 \cdot 10^{+2}$		&	$8.733 \cdot 10^{-3}$		&	$1.900$				&	$1.310 \cdot 10^{+2}$
				\end{tabular}
			\end{ruledtabular}
		\end{table}
		
		\begin{table}[!htpb]
			\caption{\label{tab:KNP2ac}
				Relative errors $\epsilon$, corresponding rates of convergence $r$, see Eqs.~\eqref{eq:epsilonS} and \eqref{eq:rateS}, and wall times in seconds for varying number $n$ of volume cells for the RG flows of the toy model in the limit ${N \rightarrow \infty}$ with $a = 2 a_\mathrm{c}$ computed with the second and first order KNP scheme formulated in the invariant $y$. The corresponding KNP $\mathcal{O} ( \Delta y^1 )$ flow including parameters are displayed in the lower panel of Fig.~\ref{fig:frg_largeN_KNPO1_flows}.
			}
			\begin{ruledtabular}
				\renewcommand{\arraystretch}{1.15}
				\begin{tabular}{l c c c c c c}
							&	\multicolumn{3}{c}{KNP $\mathcal{O} ( \Delta y^2 )$}								&	\multicolumn{3}{c}{KNP $\mathcal{O} ( \Delta y^1 )$}
							\\
							\cmidrule{2-4}																			\cmidrule{5-7}
					$n$		&	$\epsilon_\mathrm{KNPO2}$	&	$r_\mathrm{KNPO2}$	& $t_\mathrm{W}\ (\mathrm{s})$	&	$\epsilon_\mathrm{KNPO1}$	&	$r_\mathrm{KNPO1}$	&	$t_\mathrm{W}\ (\mathrm{s})$
					\\
					\colrule\addlinespace[0.25em]
					$64$	&	$4.102 \cdot 10^{-2}$		&	$-$					&	$4.888 \cdot 10^{-1}$		&	$9.271 \cdot 10^{-1}$		&	$-$					&	$3.147 \cdot 10^{-1}$
					\\
					$128$	&	$2.266 \cdot 10^{-2}$		&	$0.856$				&	$1.585 \cdot 10^{+0}$		&	$4.491 \cdot 10^{-2}$		&	$4.370$				&	$1.065 \cdot 10^{+0}$
					\\
					$256$	&	$1.252 \cdot 10^{-2}$		&	$0.856$				&	$4.713 \cdot 10^{+0}$		&	$2.731 \cdot 10^{-2}$		&	$0.717$				&	$3.749 \cdot 10^{+0}$
					\\
					$512$	&	$6.953 \cdot 10^{-3}$		&	$0.848$				&	$1.665 \cdot 10^{+1}$		&	$1.713 \cdot 10^{-2}$		&	$0.673$				&	$1.368 \cdot 10^{+1}$
					\\
					$1024$	&	$3.878 \cdot 10^{-3}$		&	$0.842$				&	$5.323 \cdot 10^{+1}$		&	$1.072 \cdot 10^{-2}$		&	$0.676$				&	$5.144 \cdot 10^{+1}$
					\\
					$2048$	&	$2.161 \cdot 10^{-3}$		&	$0.844$				&	$1.674 \cdot 10^{+2}$		&	$6.723 \cdot 10^{-3}$		&	$0.673$				&	$1.517 \cdot 10^{+2}$
				\end{tabular}
			\end{ruledtabular}
		\end{table}%

		\begin{table}[!htpb]
			\caption{\label{tab:frg_N2_2ac_rates}
				Relative errors $\epsilon_\mathrm{KT}$, corresponding rates of convergence $r_\mathrm{KT}$, see Eqs.~\eqref{eq:epsilonS} and \eqref{eq:rateS}, and wall times in seconds for varying number $n$ of volume cells for the RG flows of the toy model displayed in Fig.~\ref{fig:frg_N2_flows} for $N=2$.
			}
			\begin{ruledtabular}
				\renewcommand{\arraystretch}{1.15}
				\begin{tabular}{l c c c c c c c c c}
							&	\multicolumn{3}{c}{$a = 0$}														&	\multicolumn{3}{c}{$a = a_\mathrm{c}$}											&	\multicolumn{3}{c}{$a = 2 a_\mathrm{c}$}
					\\
							\cmidrule{2-4}																		\cmidrule{5-7}																					\cmidrule{8-10}
					$n$		&	$\epsilon_\mathrm{KT}$	&	$r_\mathrm{KT}$	&	$t_\mathrm{W} (\mathrm{s})$		&	$\epsilon_\mathrm{KT}$	&	$r_\mathrm{KT}$	&	$t_\mathrm{W} (\mathrm{s})$		&	$\epsilon_\mathrm{KT}$	&	$r_\mathrm{KT}$	&	$t_\mathrm{W} (\mathrm{s})$
					\\
					\colrule\addlinespace[0.25em]
					$64$	&	$7.959 \cdot 10^{-3}$	&	$-$				&	$2.8 \cdot 10^{-1}$				&	$5.082 \cdot 10^{-3}$	&	$-$				&	$3.0 \cdot 10^{-1}$				&	$2.133 \cdot 10^{-3}$	&	$-$				&	$3.9 \cdot 10^{-1}$
					\\
					$128$	&	$2.454 \cdot 10^{-3}$	&	$1.700$			&	$9.2 \cdot 10^{-1}$				&	$1.697 \cdot 10^{-3}$	&	$1.580$			&	$1.6 \cdot 10^{+0}$				&	$9.340 \cdot 10^{-4}$	&	$1.190$			&	$1.7 \cdot 10^{+0}$
					\\
					$256$	&	$4.591 \cdot 10^{-4}$	&	$2.420$			&	$1.7 \cdot 10^{+1}$				&	$2.456 \cdot 10^{-4}$	&	$2.790$			&	$2.3 \cdot 10^{+1}$				&	$3.221 \cdot 10^{-5}$	&	$4.860$			&	$3.1 \cdot 10^{+1}$
					\\
					$512$	&	$2.453 \cdot 10^{-4}$	&	$0.904$			&	$1.6 \cdot 10^{+2}$				&	$2.089 \cdot 10^{-4}$	&	$0.233$			&	$2.1 \cdot 10^{+2}$				&	$1.730 \cdot 10^{-4}$	&	$-2.430$		&	$2.6 \cdot 10^{+2}$
					\\
					$1024$	&	$3.425 \cdot 10^{-5}$	&	$2.840$			&	$8.2 \cdot 10^{+2}$				&	$2.224 \cdot 10^{-5}$	&	$3.230$			&	$9.3 \cdot 10^{+2}$				&	$1.035 \cdot 10^{-5}$	&	$4.060$			&	$1.2 \cdot 10^{+3}$
					\\
					$2048$	&	$8.510 \cdot 10^{-6}$	&	$2.010$			&	$2.4 \cdot 10^{+3}$				&	$5.530 \cdot 10^{-6}$	&	$2.010$			&	$3.0 \cdot 10^{+3}$				&	$2.574 \cdot 10^{-6}$	&	$2.010$			&	$3.0 \cdot 10^{+3}$
				\end{tabular}
			\end{ruledtabular}
		\end{table}%
		
		\begin{table}[!htpb]
			\caption{\label{tab:frg_N32_2ac_rates}
				Relative errors $\epsilon_\mathrm{KT}$, corresponding rates of convergence $r_\mathrm{KT}$, see Eqs.~\eqref{eq:epsilonS} and \eqref{eq:rateS}, and wall times in seconds for varying number $n$ of volume cells for the RG flows of the toy model displayed in Fig.~\ref{fig:frg_N32_flows} for $N=32$.
			}
			\begin{ruledtabular}
				\renewcommand{\arraystretch}{1.15}
				\begin{tabular}{l c c c c c c c c c}
							&	\multicolumn{3}{c}{$a = 0$}														&	\multicolumn{3}{c}{$a = a_\mathrm{c}$}											&	\multicolumn{3}{c}{$a = 2 a_\mathrm{c}$}
					\\
							\cmidrule{2-4}																		\cmidrule{5-7}																		\cmidrule{8-10}
					$n$		&	$\epsilon_\mathrm{KT}$	&	$r_\mathrm{KT}$	&	$t_\mathrm{W} (\mathrm{s})$	&	$\epsilon_\mathrm{KT}$	&	$r_\mathrm{KT}$	&	$t_\mathrm{W}\ (\mathrm{s})$	&	$\epsilon_\mathrm{KT}$	&	$r_\mathrm{KT}$	&	$t_\mathrm{W} (\mathrm{s})$
					\\
					\colrule\addlinespace[0.25em]
					$64$	&	$3.084 \cdot 10^{-1}$	&	$-$				&	$1.078 \cdot 10^{+0}$		&	$9.226 \cdot 10^{-2}$	&	$-$				&	$1.265 \cdot 10^{+0}$			&	$1.013 \cdot 10^{+0}$	&	$-$				&	$1.468 \cdot 10^{+0}$
					\\
					$128$	&	$1.363 \cdot 10^{-1}$	&	$1.180$			&	$9.857 \cdot 10^{+0}$		&	$1.206 \cdot 10^{-2}$	&	$2.930$			&	$1.455 \cdot 10^{+1}$			&	$3.724 \cdot 10^{-1}$	&	$1.440$			&	$1.387 \cdot 10^{+1}$
					\\
					$256$	&	$2.561 \cdot 10^{-2}$	&	$2.410$			&	$1.078 \cdot 10^{+2}$		&	$1.433 \cdot 10^{-2}$	&	$-0.248$		&	$1.515 \cdot 10^{+2}$			&	$1.260 \cdot 10^{-1}$	&	$1.560$			&	$1.588 \cdot 10^{+2}$
					\\
					$512$	&	$5.623 \cdot 10^{-3}$	&	$2.190$			&	$7.119 \cdot 10^{+2}$		&	$2.447 \cdot 10^{-3}$	&	$2.550$			&	$1.009 \cdot 10^{+3}$			&	$3.348 \cdot 10^{-2}$	&	$1.910$			&	$1.152 \cdot 10^{+3}$
					\\
					$1024$	&	$4.306 \cdot 10^{-3}$	&	$0.385$			&	$3.945 \cdot 10^{+3}$		&	$2.596 \cdot 10^{-3}$	&	$-0.085$		&	$5.318 \cdot 10^{+3}$			&	$1.042 \cdot 10^{-2}$	&	$1.680$			&	$5.709 \cdot 10^{+3}$
					\\
					$2048$	&	$1.248 \cdot 10^{-3}$	&	$1.790$			&	$1.657 \cdot 10^{+4}$		&	$4.792 \cdot 10^{-4}$	&	$2.440$			&	$2.091 \cdot 10^{+4}$			&	$2.539 \cdot 10^{-3}$	&	$2.040$			&	$2.222 \cdot 10^{+4}$
				\end{tabular}
			\end{ruledtabular}
		\end{table}
\FloatBarrier
\twocolumngrid

\bibliography{zero_dim_part_3} 

\end{document}